\documentclass[preprint,prd,aps,floatfix,preprintnumbers,superscriptaddress,bibnotes,nofootinbib]{revtex4-1}
\usepackage{graphicx}
\usepackage{pstricks}

\usepackage{subfigure}

\usepackage{version}    
\usepackage{amsmath}    
\usepackage{hyperref}
\usepackage{longtable}
\usepackage{ulem}
\usepackage{bm}  
\usepackage{color}
\usepackage{enumerate}  

\newcommand{\be}{\begin{equation}}
\newcommand{\ee}{\end{equation}}
\newcommand{\bea}{\begin{eqnarray}}
\newcommand{\eea}{\end{eqnarray}}
\newcommand{\bi}{\begin{itemize}}
\newcommand{\ei}{\end{itemize}}

\def\gammaO{\Gamma^O}

\begin{document}
\preprint{UTHEP-771, UTCCS-P-145, HUPD-2209}
\title{Nucleon isovector couplings in $N_f=2+1$ lattice QCD \\
at the physical point
}
%
\author{Ryutaro Tsuji\:}
\email[E-mail: ]{tsuji@nucl.phys.tohoku.ac.jp}
\affiliation{Department of Physics, Tohoku University, Sendai 980-8578, Japan}
\affiliation{RIKEN Center for Computational Science, Kobe 650-0047, Japan}
\author{Natsuki Tsukamoto\:}
\affiliation{Department of Physics, Tohoku University, Sendai 980-8578, Japan}
\author{Yasumichi Aoki\:}
\affiliation{RIKEN Center for Computational Science, Kobe 650-0047, Japan}
\author{Ken-Ichi~Ishikawa\:}
\affiliation{Core of Research for the Energetic Universe, Graduate School of Advanced Science and Engineering, Hiroshima University, Higashi-Hiroshima 739-8526, Japan}
\author{Yoshinobu~Kuramashi\:}
\affiliation{Center for Computational Sciences, University of Tsukuba, Tsukuba, Ibaraki 305-8577, Japan}
\author{Shoichi~Sasaki\:}
\email[E-mail: ]{ssasaki@nucl.phys.tohoku.ac.jp}
\affiliation{Department of Physics, Tohoku University, Sendai 980-8578, Japan}
\author{Eigo~Shintani\:}
\affiliation{Center for Computational Sciences, University of Tsukuba, Tsukuba, Ibaraki 305-8577, Japan}
\author{Takeshi~Yamazaki\:}
\affiliation{Faculty of Pure and Applied Sciences, University of Tsukuba, Tsukuba, Ibaraki, 305-8571, Japan}
\affiliation{Center for Computational Sciences, University of Tsukuba, Tsukuba, Ibaraki 305-8577, Japan}
\collaboration{PACS Collaboration}

\begin{abstract}

We present results for the scalar and tensor isovector-couplings ($g_S$ and $g_T$) 
of the nucleon measured at the physical point ($M_{\pi}=135$ MeV) with a single lattice 
spacing of $0.085\ \mathrm{fm}$ in 2+1 flavor QCD. 
Our calculations are carried out with two ensembles of gauge configurations 
generated by the PACS Collaboration with nonperturbatively ${\cal O}(a)$ 
improved Wilson quark action and Iwasaki gauge action 
on $(10.9\ {\rm fm})^4$ and $(5.5\ {\rm fm})^4$ lattices, where
the finite-size effect on the nucleon mass was not shown at the  
level of the statistical precision less than 0.5\%. 
We compute the nucleon three-point correlation functions in the vector, axial, scalar, and tensor channels. 
We confirm that our previous result of the nucleon axial coupling on the large spatial volume 
of $(10.9\ {\rm fm})^4$ has no finite-size effect at the level of the statistical precision of 1.9\%. 
For the renormalization, we first renormalize $g_S$ and $g_T$ nonperturbatively using
the RI/SMOM$_{(\gamma_\mu)}$ scheme, a variant of Rome-Southampton RI/MOM scheme 
with reduced systematic errors, as the intermediate scheme.
We evaluate our final results at the renormalization scale of 2 GeV in the $\overline{\rm MS}$ 
scheme through matching procedure between the RI/SMOM$_{(\gamma_\mu)}$ and $\overline{\rm MS}$ schemes
with the help of perturbation theory, and then obtain $g_S=0.927(83)_{\rm stat}(22)_{\rm syst}$ 
and $g_T=1.036(6)_{\rm stat}(20)_{\rm syst}$.
\end{abstract}

\maketitle

\section{Introduction}

Future and current precision {$\beta$}-decay measurements with cold and ultracold neutrons 
provide us an opportunity to study the sensitivity of the nucleon isovector matrix elements to new 
physics beyond the standard model (BSM)~\cite{Herczeg:2001vk}. The neutron life-time puzzle associated with the
nucleon axial-vector coupling ($g_A$) is one of such examples. 
The discrepancy between the results of beam experiments and storage experiments
remains unsolved. It is still an open question that deserves further investigation in terms of the ratio
of the nucleon axial and vector couplings ($g_A/g_V$)~\cite{Czarnecki:2018okw}. 
Although the vector and axial-vector channels dominate 
the weak decay of the neutron, there is no reason to forbid the other channel contributions
to the neutron decay if the BSM contributions are present. 
Therefore, the scalar, pseudo-scalar and tensor contributions, whose couplings are less known 
experimentally,
play important roles to constrain the limit of non-standard interactions mediated 
by undiscovered gauge bosons 
in the scalar, pseudo-scalar and tensor channels~\cite{{Bhattacharya:2011qm}, {Cirigliano:2013xha},{Gonzalez-Alonso:2018omy}}.

In general the isovector nucleon couplings $g_{O}$ for  
$O=V$ (vector), $A$ (axial-vector), $S$ (scalar), $P$ (pseudo-scalar) and $T$ (tensor)
are expressed by the neutron-proton transition matrix element with quark
charged (off-diagonal) currents
%
%
\begin{align}
\langle p(p,s)|\bar{u}\gammaO d |n(p,s)\rangle =g_O \bar{u}_p(p,s)\gammaO u_n(p,s),
\end{align}
where $\gammaO$ is a Dirac matrix appropriate for the channel $O$ ($O=V,A,S,P,T$).
Considering the $SU(2)$ Lie algebra associated with isospin, the isovector nucleon
matrix elements can be rewritten by the proton matrix elements of the diagonal isospin 
current
\begin{align}
\langle p|\bar{u}\gammaO d|n\rangle=
\langle p|\bar{u}\gammaO u|p\rangle - \langle p|\bar{d}\gammaO d|p\rangle
\end{align}
in the isospin limit. Therefore, the isovector couplings are related to the isovector charges of the
proton and they are rewritten by 
\begin{align}
g_O=g_O^u-g_O^d ,
\end{align}
where up and down quark contributions of the proton's isovector-charges $g_O^u$ and $g_O^d$
are defined by the proton forward matrix elements as 
$g_O^{f}=\langle p(p,s)|\bar{f}\gammaO f|p(p,s)\rangle$ 
with $f=u$ or $d$.
Recall that for the vector $g_V = 1$ is satisfied under the exact isospin symmetry.

The nucleon scalar isovector-coupling $g_S$, which is related to the size of the neutron-proton mass splitting induced solely by the light-quark mass difference in the absence of electromagnetic effects, is a phenomenologically interesting quantity~\cite{Gonzalez-Alonso:2013ura}. 
On the other hand, each flavor contribution of the proton's tensor charges $g_T^f$ 
is a weight factor for the contributions of the non-vanishing quark electric dipole moment (EDM) 
to the proton and neutron EDMs, since the tensor current has the same transformation properties under $P$ and $T$ discrete symmetries as the EDM current~\cite{Jackson:1957zz,Grossman:1997pa}. Furthermore, the nucleon tensor 
isovector-coupling $g_T$ itself is also an important information from the view-point 
of the proton spin structure~\cite{Courtoy:2015haa}.
The proton's tensor-charge is closely connected with the first Mellin moment of the proton's transversity 
parton distribution function (PDF), which has been recently studied in experiments such as semi-inclusive deeply inelastic scattering (SIDIS) and Drell-Yan measurements from HERMES at DESY, COMPASS at CERN and SoLID at Jefferson Lab. Recently, the first global analysis of the quark transversity distributions
with the available SIDIS data under the constraint of lattice estimates of $g_T$
has demonstrated significant reduction of the uncertainties 
on the resulting transversity PDFs~\cite{Lin:2017stx}.

As described previously, the knowledge of the nucleon scalar and tensor isovector-couplings 
is necessary for ongoing experimental researches for the BSM physics and also for the nucleon structure. 
Although the vector and axial-vector isovector-couplings ($g_V$ and $g_A$) are well measured in both experiment and lattice QCD, the scalar and tensor isovector-couplings ($g_S$ and $g_T$)
are so far not accessible in experiment. Of course, since the precise experiments are not yet available for 
$g_S$ and $g_T$, their accurate determination from lattice QCD is highly desired.
Lattice determinations of $g_S$ and $g_T$ have recently been performed by several groups~\cite{{Aoki:2021kgd},{Park:2021ypf},{Horkel:2020hpi},{Gupta:2018qil},{Bhattacharya:2016zcn},{Alexandrou:2019brg},{Hasan:2019noy},{Ottnad:2018fri},{Yamanaka:2018uud},{Bali:2014nma}}. As shown in FLAG Review~\cite{Aoki:2021kgd}, the most
precise determination of $g_S$ and $g_T$ is given by a single group (PNDME Collaboration)~\cite{{Gupta:2018qil},{Bhattacharya:2016zcn}} and their values are dominant in the FLAG average for
$N_f=2+1+1$. Their calculation had been performed with the mixed action simulations using the Wilson-clover valence quarks on the asqtad-improved gauge configurations with the highly improved staggered sea quarks (HISQ). This implies that their lattice QCD simulations are \textit{not} fully dynamical one, but rather \textit{partially quenched} one since they used the different action for the sea and valence quarks.
Even if we could not overcome the current precision given by the PNDME Collaboration, it is worth to reproduce their results in \textit{fully dynamical lattice QCD simulations}. 
Further comprehensive studies of the nucleon 
isovector-couplings including $g_S$ and  $g_T$ 
as well as $g_V$ and $g_A$ are still needed. In this study, we intend to determine $g_A$, $g_S$ and  $g_T$  
with high accuracy and high precision using the 2+1 flavor lattice QCD at the physical point on 
sufficiently large volume.

This paper is organized as follows. 
In Sec. II, we describe our method to calculate 
the nucleon isovector-couplings from the nucleon 2- and 3-point 
correlation functions in lattice simulations.
In Sec. III, we give a short outline of the renormalization procedure with 
the RI/SMOM and RI/SMOM$_{\gamma_\mu}$ intermediate schemes to connect 
the bare lattice operators to the continuum renormalized operators. 
The simulation parameters are described in Sec. III. We also
explain the error reduction technique employed in this study. 
The results of our lattice calculations for the nucleon isovector-couplings are 
presented in Sec. IV, which is divided into three subsections. We first
determine the bare values of the nucleon isovector-couplings 
for the vector, axial-vector, scalar and tensor channels in Sec. IV A.
Section IV B presents the results of the renormalization constants for 
the scalar and tensor. The results of the renormalized couplings are 
presented in the last subsection (Sec. IV C).
Finally, we close with summary in Sec. V. 
All dimensionful quantities are expressed in units of 
the lattice spacing throughout this paper, 
unless otherwise explicitly specified. A bold-faced variable represents a three-dimensional vector.

\section{Calculation method I: Bare coupling}

In this section we describe our method to calculate the nucleon isovector-couplings
from the nucleon 2- and 3-point correlation functions, which consist of the nucleon
source and sink interpolating operators (denoted as $\overline{N}(x)$ and $N(x)$, respectively), 
and the quark bilinear current operator $J(x)$.  

\subsection{Nucleon 2-point correlation function}

In this study, two types of smeared quark operator $q_S(t, {\mathbf x})$ are 
used for the construction of the nucleon interpolating operator as well as a local quark
operator $q(t, {\mathbf x})$. The smeared quark operator is given by a
convolution of the local quark operator with a smearing function $\phi(\mathbf{x},\mathbf{y})$
as
\begin{align}
    \label{eq:smear_qua}
    q_{S}(t,\mathbf{x})
    =
    \sum_{\mathbf{y}}
    \phi(\mathbf{x},\mathbf{y})
    q(t,\mathbf{y}),
\end{align}
where the color and Dirac indices are omitted. 

In the case of the exponentially smeared quark operator, a smearing function 
$\phi(\mathbf{x},\mathbf{y})$ is given by an isotropic function of $r=|\mathbf{x}-\mathbf{y}|$
in a linear spatial extent of $L$ as the following form
\begin{align}
    \label{eq:exp_smear_nuc}
    \phi(\mathbf{x},\mathbf{y})=\phi(r)=
\begin{cases}
    1 & (r=0) \\
    A\mathrm{e}^{-Br} & (r<L/2) \\
    0 & (r\ge L/2)
\end{cases}
\end{align}
with two smearing parameters $A$ and $B$. 
This procedure does not preserve the full gauge invariance of the 
hadron 2-point correlation functions consisting of the spatially smeared quark operators, so that 
the Coulomb gauge fixing is necessary. We have always adopted the exponentially smeared
quark operator as the spatially smeared method for the quark operator 
in our previous studies~\cite{{Ishikawa:2018rew},{Shintani:2018ozy},{Ishikawa:2021eut}}. 
In this study, we then adopt it as the primary smeared method in 
both two volumes ($L=128$ and $L=64$). In addition, 
the other type of the spatially smeared quark operator is also used 
in the larger volume ($L=128$) as supplementary result. 
For this purpose, we use gauge-covariant, approximately Gaussian-shaped smearing
method (denoted shortly as Gaussian smearing)~\cite{Gusken:1989qx}. 

To maintain the gauge covariance of the smeared quark operator through Eq.~(\ref{eq:smear_qua}),
we construct the following smearing function:
\begin{align}
\label{eq:gauss_smear_nuc}
\phi(\mathbf{x},\mathbf{y})\rightarrow
\phi(\mathbf{x},\mathbf{y}; U)=\delta_{\mathbf{x},\mathbf{y}}+\frac{w^2_{\mathrm{G}}}{4n_{\mathrm{G}}}\mathbf{\Delta}(\mathbf{x},\mathbf{y}; U),
\end{align}
where $n_{\mathrm{G}}$ and $w_{\mathrm{G}}$ represent two smearing parameters~\footnote{
In Ref.~\cite{Gusken:1989qx}, the smearing function is chosen as 
$$
\phi(\mathbf{x},\mathbf{y}; U)
=\delta_{\mathbf{x},\mathbf{y}}+\alpha\sum_{\mu=1}^{3}\left[
U_\mu(x)\delta_{\mathbf{x}+\hat{\mu},\mathbf{y}}+U_{\mu}^\dagger(x-\hat{\mu})\delta_{\mathbf{x}-\hat{\mu},\mathbf{y}}
\right].
$$
The parameter $\alpha$ that corresponds to the coupling strength of the nearest neighbors 
is related with our parameter set of $(n_{\mathrm{G}},w_{\mathrm{G}})$ as 
$\alpha=\frac{w_{\mathrm{G}}^{2}/4n_{\mathrm{G}}}{1-3w_{\mathrm{G}}^2/2n_{\mathrm{G}}}$.
The positivity of $\alpha$ requires the condition of $3w_{\mathrm{G}}^2<2n_{\mathrm{G}}$
in our parameter set.}.
Here, $\mathbf{\Delta}(\mathbf{x},\mathbf{y}; U)$ represents the three-dimensional covariant Laplacian, which is defined with the spatial components of the link variables $U_{\mu}(x)$ ($\mu=1,2,3$)~\footnote{In this study, six steps of stout-smearing are applied to the link variables.} as
\begin{align}
\mathbf{\Delta}(\mathbf{x},\mathbf{y}; U)=\sum_{\mu=1}^{3}\left[
U_\mu(x)\delta_{\mathbf{x}+\hat{\mu},\mathbf{y}}+U_{\mu}^\dagger(x-\hat{\mu})\delta_{\mathbf{x}-\hat{\mu},\mathbf{y}}-2\delta_{\mathbf{x},\mathbf{y}}
\right],
\end{align}
which ensures that the gauge covariance of the spatially extended quark operator is preserved.
Applying the same procedure recursively, an approximately Gaussian-shaped quark operator can be obtained through this iterative scheme. When the number of the iteration is chosen to be $n_{\mathrm{G}}$, $w_{\mathrm{G}}$ corresponds to the width of the Gaussian that results in $n_{\mathrm{G}}\rightarrow \infty$.

The nucleon 2-point function with the local nucleon sink operator $N_L(t, \mathbf{x})$ and smeared nucleon source operator $\overline{N}_S(t, \mathbf{x})$ is defined as
\begin{align}
 C^{LS}_{\mathrm{2pt}}(t - t_{\mathrm{src}}, \mathbf{x}-\mathbf{x}_{\mathrm{src}}) & = 
\frac{1}{4}\mathrm{Tr} \left[\mathcal{P}_+ \langle 0|N_L(t, \mathbf{x}) \overline{N}_S(t_{\mathrm{src}}, \mathbf{x}_{\mathrm{src}}) |0\rangle \right],
\end{align}
where $\mathcal{P}_{+}=(1+\gamma_{4})/2$~\footnote{
A projection operator $\mathcal{P}_{+}$ can eliminate contributions from opposite-parity state~\cite{{Sasaki:2001nf},{Sasaki:2005ug}}.
}, and the nucleon operator is given for the proton
state by
\begin{align}
    N_L(t, \mathbf{x})=
    \epsilon_{abc}u^{T}_{a}(t, \mathbf{x})C\gamma_5 d_{b}(t, \mathbf{x})u_{c}(t, \mathbf{x})
\end{align}
with $C=\gamma_4\gamma_2$, the up and down quark operators $u, d$ and $a, b, c$ being the color indices.  The smeared source operator $N_S(t, \mathbf{x})$ is the same as the local one 
$N_L(t, \mathbf{x})$, but all the quark operators $u, d$ are replaced by the smeared ones 
defined in Eq.~(\ref{eq:smear_qua}). The momentum projected 2-point function is
then given by
\begin{align}
\tilde{C}^{XS}_{\mathrm{2pt}}(t, \mathbf{p})=\sum_{\mathbf{r}}e^{-i\mathbf{p}\cdot\mathbf{r}}
C^{XS}_{\rm 2pt}(t, \mathbf{r}) 
\end{align}
with $X=L, S$ and a three-dimensional momentum $\mathbf{p}$.

\subsection{Nucleon 3-point correlation function}
\label{Sec2:ratio_method}
In order to calculate the isovector nucleon matrix element, we evaluate the nucleon 3-point functions, 
which are constructed with the spatially smeared source and sink operators of the nucleon 
and a given isovector current operator $J^{O}$ inserted at $t=t_{\mathrm{op}}$ being
subject to a range of $t_{\mathrm{sink}} > t_{\mathrm{op}} > t_{\mathrm{src}}$ as
\begin{align}
C_{O}^{k}(t_{\mathrm{op}} - t_{\mathrm{src}}, \mathbf{x}-\mathbf{x}_{\mathrm{src}})
 = \frac{1}{4}\sum_{\mathbf{y}}{\rm Tr}\left[
\mathcal{P}_k\langle 0|
N_{S}(t_{\mathrm{sink}}, \mathbf{y})J^{O}(t_{\mathrm{op}}, \mathbf{x})
\overline{N}_{S}(t_{\mathrm{src}}, \mathbf{x}_{\mathrm{src}}) |0\rangle \right],
\end{align}
where $\mathcal{P}_k$ is a projection operator, and $J^{O}$
is defined by the local bilinear operator as $J^{O}=
\overline{u} \gammaO u - \overline{d} \gammaO d$ with
a Dirac matrix $\gammaO$ appropriate for the channel $O$ 
$(O=V, A, S, P, T)$.
The momentum projected 3-point function is then given by 
\begin{align}
\tilde{C}_{O}^{k}(t, \mathbf{p})=\sum_{\mathbf{r}}e^{-i\mathbf{p}\cdot\mathbf{r}}
C_{O}^{k}(t, \mathbf{r}). 
\end{align}

A well-known procedure for determining the couplings is to calculate the following ratio of
the zero-momentum projected 3- and 2-point functions with a fixed source-sink separation ($t_{\mathrm{sep}}\equiv t_{\mathrm{sink}}-t_{\mathrm{src}}$) as
\begin{align}
    R(t_{\mathrm{op}}, t_{\mathrm{sep}})\equiv\frac{\tilde{C}_{O}^{k}(t_{\mathrm{op}}-t_{\mathrm{src}}, \mathbf{0})}{\tilde{C}^{SS}_{\mathrm{2pt}}(t_{\mathrm{sep}}, \mathbf{0})},
     \label{Eq:ratio}
\end{align}
where the nucleon 2-point function is given by the same smeared source and sink operators.
If the condition $t_{\mathrm{sep}} \gg t_{\mathrm{op}}-t_{\mathrm{src}}\gg 0$ is satisfied, 
the desired coupling $g_{O}$ relevant for the $O$ channel can be read off from an asymptotic plateau 
of the ratio (hereafter denoted as the plateau method), which is independent of a choice of $t_{\mathrm{op}}$ as
\begin{align}
R(t_{\mathrm{op}}, t_{\mathrm{sep}}) \xrightarrow[t_{\mathrm{sep}}\gg t_{\mathrm{op}}-t_{\mathrm{src}} \gg 0]{ }
g_{O} + {\cal O}(e^{-\Delta E(t_{\mathrm{op}}-t_{\mathrm{src}})}) + {\cal O}(e^{-\Delta E( t_{\mathrm{sep}}-t_{\mathrm{op}}+t_{\mathrm{src}}  )}),
\label{Eq:ratio_plateau}
\end{align}
where $\Delta E=E_1-E_0$ denotes a difference between the two energies of the ground state ($E_0$) and the lowest excited state ($E_1$). Narrower source-sink separation causes systematic uncertainties stemming from the excited-state
contamination represented by two terms of ${\cal O}(e^{-\Delta E(t_{\mathrm{op}}-t_{\mathrm{src}})})$ and
${\cal O}(e^{-\Delta E( t_{\mathrm{sep}}-t_{\mathrm{op}}+t_{\mathrm{src}}  )})$. 
The optimal choice of the smearing parameters can help to 
suppress as much as possible the coefficients of these terms due to the maximal overlap with the ground state.

Here, recall that the ratio vanishes unless $\gammaO=1(S)$, $\gamma_4(V)$, $\gamma_i\gamma_5(A)$, and $\sigma_{ij}(T)$ with $i,j=1,2,3$~\cite{Sasaki:2003jh}.
The projection operator $\mathcal{P}_k$ ($k=t, 5x, 5y, 5z$) is supposed to be chosen as below
\begin{align}
 \mathcal{P}_k = 
\begin{cases}
 \mathcal{P}_+ \gamma_5\gamma_1\ (k=5x) & {\rm for} \ \gammaO=\gamma_5\gamma_1(A), \sigma_{23}(T) \\
 \mathcal{P}_+ \gamma_5\gamma_2\ (k=5y) & {\rm for} \ \gammaO=\gamma_5\gamma_2(A), \sigma_{31}(T) \\
 \mathcal{P}_+ \gamma_5\gamma_3\ (k=5z) & {\rm for} \ \gammaO=\gamma_5\gamma_3(A), \sigma_{12}(T) \\
 \mathcal{P}_+ \gamma_4\ (k=t) & {\rm for}\ \gammaO=1(S), \gamma_4(V).
\end{cases}
\end{align}
In this study we focus on the axial ($A$), scalar ($S$) and tensor ($T$) channels.

In general, the nucleon 3-point functions have the connected and disconnected contributions, which correspond to the two types of Wick contraction topologies for all quark lines. 
The disconnected contributions corresponding to the sea quark effect are necessary for 
determining flavor diagonal charges of the nucleon, $g_O^q$ $(q = u, d, s, \cdots)$.
However, a computation of disconnected diagrams requires much-higher computational cost than that of 
connected diagrams. Recalled that in the case of 2+1 flavor lattice QCD, where the up and down quark masses are equal, 
the nucleon 3-point functions for the isovector current does not receive any contribution from the disconnected-type diagrams since they are canceled each other thanks to the exact isospin symmetry. 
Therefore, the isovector couplings $g_O=g_O^u-g_O^d$ can be determined only by the connected-type diagrams.

\section{Calculation method II: Renormalization}
\label{Renormalization}

The general idea for the nonperturbative renormalization of the quark operator is discussed 
in Ref.~\cite{Martinelli:1994ty}. Here, we briefly review the Rome-Southampton method~\cite{Martinelli:1994ty}. 
In this paper, we will calculate the renormalization constants for the quark bilinear 
operators
\begin{align}
\left(\overline{q}\gammaO q\right)_R & = {Z_O} \left(\overline{q} \gammaO q\right)_B ,
\end{align}
where $\gammaO$ denotes a Dirac gamma matrix appropriate for the channel $O$ ($O=V, A, S, P, T$).
The subscripts $R$ and $B$ represent ``renormalized'' and ``bare'' operators, respectively. 
Although we simply note $q$ and $\overline{q}$ for quark and anti-quark fields, only flavor non-singlet operators are considered.

In order to evaluate the renormalization constants ${Z_O}$, 
let us first define the one-side Fourier-transformed quark propagator as 
\begin{align}
 S(p|x_0) = \sum_{x^\prime} e^{-ip\cdot x^\prime}S(x^\prime, x_0) ,
\end{align}
where $S(x^\prime, x_0)$ represents a point-source quark propagator with a fixed source point $x_0$ in the Landau gauge.
The two-point Green's functions with the insertion of the quark bilinear operator $\overline{q} \gammaO q$ 
are given by 
\begin{align}
\langle G_O(p_1, p_2)\rangle= \frac{1}{n_{\mathrm{src}}}\sum_{x_0} \langle S(p_1|x_0)  \gammaO \gamma_5S(p_2|x_0)^\dagger\gamma_5\rangle ,
\end{align}
where two external quark lines carry momenta $p_1$ and $p_2$.
Here, recall that the explicit dependence of source location $x_0$ is averaged over
finite number of different source locations ($n_{\mathrm{src}}$). In this study,
$n_{\mathrm{src}}=4$ is chosen.

We then compute the bare vertex function (or amputated Green's function) as 

\begin{align}
\Lambda_O(p_1,p_2) & = \left[\frac{1}{n_{\mathrm{src}}}\sum_{x_0}\langle
S(p_1|x_0)\rangle\right]^{-1} \langle G_O(p_1, p_2)\rangle 
\left[\frac{1}{n_{\mathrm{src}}}\sum_{x_0}\langle \gamma_5 S(p_2|x_0)^\dagger \gamma_5 \rangle\right]^{-1}.
\end{align}

Following Ref.~\cite{Martinelli:1994ty}, we define the renormalized vertex function by
\begin{align}
\left(\Lambda_O\right)_R = \frac{Z_O}{Z_q} \left(\Lambda_O\right)_B,
\end{align}
where $Z_q$ denotes the quark field renormalization constant. 

In the regularization independent (RI) momentum-subtraction scheme, 
the $Z$ factors are fixed by imposing the following renormalization condition  
\begin{align}
    \left.\left(\Lambda_O\right)_R\right|_{\mu^2=\mu^2(p_1, p_2)}
        =  \left(\Lambda_O\right)_{\rm tree} = \gammaO , 
\end{align}
where the renormalization scale $\mu$ is introduced according to 
the subtraction kinematics with the external quark momenta
$p_1$ and $p_2$. 
In practice, the ratio of the renormalization constants $Z_O/Z_q$ 
is calculated from the projected vertex functions with 
the appropriate projection operator $\mathcal{P}_O$:
\begin{align}
    \left. \frac{Z_O}{Z_q}
        {\rm Tr}\left[ \left(\Lambda_O\right)_B \mathcal{P}_O\right]
        \right|_{\mu^2=\mu^2(p_1, p_2)}
    = {\rm Tr} \left[ \gammaO \mathcal{P}_O\right],
\end{align}
where the symbol ``Tr'' means a trace over color and Dirac indices. 
There is no unique choice of the projection operators and the momentum subtraction kinematics. 
There are some variants of the RI schemes as enumerated later
in Sec.~\ref{variants_of_RI_schemes}.

Although the field renormalization constant $Z_q$ can be solely evaluated, the value of $Z_O$ given by 
taking the product of $Z_O/Z_q$ with $Z_q$ is strongly affected by statistical fluctuations on $Z_q$. 
Recall that $Z_V$ and $Z_A$ are precisely determined using the Schr\"odinger functional (SF) scheme~\cite{Ishikawa:2015fzw} 
being independent of the renormalization scheme and scale. 
Therefore, we simply calculate the ratio of $Z_O/Z_q$ and $Z_V/Z_q$ (or $Z_A/Z_q$) 
in the RI scheme without the direct evaluation of $Z_q$ for the scalar and tensor channels ($O=S$ and $T$).
The values of $Z_{S}^{\rm RI}$ and $Z_{T}^{\rm RI}$ at a renormalization scale $\mu$
can be obtained with help of $Z_V^{\mathrm{SF}}$ or $Z_A^{\mathrm{SF}}$ in fully nonperturbative manner. 
For instance, in the scalar case with $Z_V^{\mathrm{SF}}$, $Z_{S}^{\rm RI}$ is evaluated
by
\begin{align}
    \label{eq:eq_renormalization}
    Z_{S}^{\rm RI}(\mu^2)
    =
    Z_{V}^{\rm SF} \times
    \frac{{\rm Tr} \left[\Gamma_{S} \mathcal{P}_{S} \right]}{{\rm Tr} \left[\Gamma_{V} \mathcal{P}_{V}\right]} \cdot
    \frac{\left. {\rm Tr} \left[\left(\Lambda_{V}\right)_B \mathcal{P}_{V}\right]\right|_{\mu^2=\mu^2(p_1, p_2)}}
        {\left. {\rm Tr} \left[\left(\Lambda_{S}\right)_B \mathcal{P}_{S}\right]\right|_{\mu^2=\mu^2(p_1, p_2)}},
\end{align}
where the renormalization scale is introduced by the external momenta $p_1$ and $p_2$ as $\mu=\sqrt{p_1^2}=\sqrt{p_2^2}$.

\subsection{Variants of the RI schemes}
\label{variants_of_RI_schemes}
\subsubsection{Momentum subtraction (MOM) scheme}
The regularization independent momentum-subtraction (RI/MOM) scheme is an intuitive and conventional approach 
to impose the renormalization condition with a specific kinematics where incoming and outgoing momenta 
are equal as $p_1= p_2$. In this kinematics, the projection operator $\mathcal{P}_O$ should 
be simply chosen to be equal to $\gammaO$, since the momentum inserted at the operator becomes zero 
as $q = p_1 - p_2 = 0$. However, as discussed in~Refs.~\cite{Aoki:2009alo,Aoki:2007xm}, the vanishing momentum $q$ enhances unwanted infrared effects such as the chiral symmetry breaking, which causes some large residual scale dependence in the infrared region of the renormalization scale $\mu$.  This, indeed, is a major source of systematic uncertainties when the renormalization constants, especially $Z_S$ and $Z_P$, are determined in the RI/MOM scheme.  

\subsubsection{Symmetric momentum (SMOM) scheme}
To avoid the unwanted infrared effects, one may choose
the symmetrical momentum configurations $p_1^2=p_2^2=q^2$ (denoted as the symmetric 
momentum (SMOM) scheme). 
In the RI/SMOM scheme \cite{Sturm:2009kb}, we can expect better infrared behavior than 
the conventional RI/MOM scheme 
owing to $q^2\neq 0$. 
The use of the non-exceptional momentum configuration, like the symmetric configuration, and its effect to suppress the unwanted infrared effects has first discussed in Ref.~\cite{Aoki:2007xm}.

The projection operators can be chosen as below~\cite{Sturm:2009kb}.
\begin{align}
\mathcal{P}_O = 
\begin{cases}
 \frac{q^\nu\gamma_\nu q_\mu}{q^2} &{\rm for \ } \gammaO = \gamma_\mu \\
 \frac{q^\nu\gamma_\nu q_\mu\gamma_5}{q^2} &{\rm for \ } \gammaO =
 \gamma_\mu\gamma_5 \\
 1 &{\rm for \ } \gammaO = 1 \\
 \gamma_5 &{\rm for \ } \gammaO = \gamma_5 \\
 \gamma_\mu\gamma_\nu &{\rm for \ } \gammaO = \gamma_\mu\gamma_\nu ,
\end{cases}
\label{eq:SMOM_proj}
\end{align}
where the vector and axial-vector cases can take the projection operators that are different from 
those used in the RI/MOM scheme, since they can maintain the Ward-Takahashi identities 
for the vector and axial-vector channels even for finite $q$. 

There is one alternative scheme called the RI/SMOM$_{\gamma_\mu}$ scheme \cite{Sturm:2009kb}, where the projection operators
are the same as in the RI/MOM scheme, while the symmetrical momentum configurations are adopted as in the RI/SMOM
scheme. 
The same effect on the suppression of the infrared effects as the RI/SMOM scheme is expected for this scheme as well. Adding another scheme will appear useful to assess the systematic uncertainty associated with the use of the perturbation theory, as different schemes may develop different convergence properties in the perturbative series.

The kinematics of momentum subtraction point and the choice of the projection operators for all three RI schemes are 
summarized in the Tab.~\ref{tab:RI_scheme}.

%
%
\begin{table*}
\caption{Summary of types of the RI schemes.
\label{tab:RI_scheme}}
\begin{ruledtabular}
 \begin{tabular}{lcc}
Type & Kinematics  & Projection operator \\\hline
RI/MOM & $p_1 = p_2,\; q = 0$ & equal to $\gammaO$ \\ 
RI/SMOM& $p_1^2 = p_2^2 = q^2$ & Eq.~(\ref{eq:SMOM_proj})\\
RI/SMOM$_{\gamma_\mu}$& $p_1^2 = p_2^2 = q^2$ & equal to $\gammaO$  \\
\end{tabular} 
\end{ruledtabular}
\end{table*}

\subsubsection{Conversion and evolution}
In order to relate operators renormalized in the RI and the $\mathrm{\overline{MS}}$ schemes,
we use the matching factor, $C^{\mathrm{RI}}_{O}(\mu_0)=Z^{\mathrm{\overline{MS}}}_{O}(\mu_0)/Z^{\mathrm{RI}}_{O}(\mu_0)$ at a matching scale $\mu_0$,
which is computed under the Landau gauge in the perturbation theory with help of the renormalization group as detailed in Appendix~\ref{app:pert_matching}.
The two-loop results for the matching factor have been already obtained for the RI/SMOM$_{(\gamma_\mu)}$ scheme
in Ref.~\cite{Almeida:2010ns}, while the three-loop result was known for the RI/MOM scheme~\cite{{Gracey:2000am},{Gracey:2003yr}}. 
In addition, the evolution factor $R_{O}(\mu, \mu_0)=Z_O^{\mathrm{\overline{MS}}}(\mu)/Z_O^{\mathrm{\overline{MS}}}(\mu_0)$ 
from the matching scale $\mu_0$ to a reference scale $\mu$ in the $\overline{\rm MS}$ scheme, 
is calculated with the four-loop beta function and the three-loop anomalous dimension. 
Combining the matching factor and the evolution factor, the conversion from $Z^{\mathrm{RI}}_{O}(\mu_0)$
to $Z^{\mathrm{\overline{MS}}}_{O}(\mu)$ can be achieved at 
 next-to-next-to-leading (NNLO)
accuracy for both the RI/MOM and RI/SMOM$_{(\gamma_\mu)}$ schemes as
\begin{align}
    \label{eq:eq_matching}
    Z_O^{\mathrm{\overline{MS}}} (\mu)
    =
    R_{O}(\mu, \mu_0)
    \cdot
    C^{x}_{O}(\mu_0)
    \times
    Z_O^{x}(\mu_0)
    \quad : \quad
    x \in
    \left\{
     \begin{array}{l}
         \mathrm{RI/MOM}  \\
         \mathrm{RI/SMOM}_{(\gamma_\mu)}
         \end{array}
     \right\} .
\end{align}
In this study, we determine the renormalization constants by mainly using the RI/SMOM$_{(\gamma_\mu)}$
scheme
that can suppress the unwanted infrared effects as will be described in Sec.~\ref{unwanted_infrared_divergenve}, 
while detailed comparisons between the RI/MOM and RI/SMOM$_{(\gamma_\mu)}$ schemes are provided in Appendix~\ref{app:mom_and_smom}.

\subsection{Residual scale dependence}
\label{residual_scale_dependence} 
In general, however, the value of $Z_O^{\mathrm{\overline{MS}}} (\mu)$ obtained 
from $Z_O^{\mathrm{RI}}(\mu_0)$ in Eq.(\ref{eq:eq_matching}) 
receives the residual dependence on the choice of the matching 
scale $\mu_0$. As explained previously, in the RI scheme, the renormalization scale $\mu_0$ is 
introduced by the external momentum $p$ as $\mu_0 = \sqrt{p^2}$.
For the higher matching scale $\mu_0$, the perturbation theory becomes relevant and
the correct scaling behavior of the renormalization constants is well predicted 
by the perturbation theory with help of the renormalization group. 
However, there is one issue at higher momenta $p$ due to the presence of the lattice
artifacts, which are associated with the discretization errors of ${\cal O}((ap)^{2n})$ 
in the lattice quark propagator $S(p|x)$. On the other hand, the error associated with the truncation of
the perturbative series at a given order grows as the momenta $p$ decrease. 
Furthermore, if some nonperturbative effect is introduced at lower
momenta $p$, the effect cannot be corrected by perturbation theory and fails the matching. 
This is the window problem, which arises as the renormalization scale $\mu_0=\sqrt{p^2}$ in the RI scheme.
{An ideal situation is realized if the window is sufficiently wide ($\Lambda_{\rm QCD} \ll \mu_0 \ll \mathcal{O}(a^{-1})$). However in practice one needs to deal with the both edges of the window, and then associated systematic uncertainties need to be carefully studied. }

In summary, there are three sources for the residual scale dependence as below. 
\begin{enumerate}
 \item  Lattice discretization artifacts at higher $\mu_0$. 
 \item  Unwanted infrared effects at lower $\mu_0$. 
 \item  Truncation errors in perturbation matching and evolution at lower $\mu_0$.
\end{enumerate}
The first and second points cause the peculiar behavior as a function of $\mu_0$.
The former functional form can be represented by a positive power series in $(a\mu_0)^2$, while
the latter point can be incorporated in the functional form that contains the negative power term of $\mu_0$.
Here, our aim is to extract the relevant $\mu_0$-independent 
value as the final result of $Z_O^{\mathrm{\overline{MS}}} (\mu)$. The truncation error, however, is not avoided 
if the perturbation theory is used. Therefore, we will also evaluate the systematic uncertainties associated with 
the third point after extracting $\mu_0$-independent value. 
In the following subsections, we describe how to reduce the systematic uncertainties 
associated with the residual $\mu_0$-dependence, which are mainly caused by
the first and second points, so as to extract the relevant $\mu_0$-independent value.

\subsubsection{Lattice artifacts}
The renormalization constants should have the lattice artifacts stemming from the lattice discretization. 
They are associated with the $H(4)$ hypercubic symmetry in the lattice discretization, which breaks $O(4)$ invariance. There are four types of invariants: $p^{[2n]}=\sum_{\mu}p_{\mu}^{2n}$ for $n=1, 2, 3, 4$,
under the $H(4)$ symmetry~\cite{Boucaud:2003dx}. The higher-order invariants $p^{[4]}$, $p^{[6]}$ and $p^{[8]}$ 
correspond to the non $O(4)$ invariants, whose presence calls the hypercubic artifacts~\cite{Boucaud:2003dx}. 
Therefore, the lattice quark propagator is no longer expressed by the single variable function of $p^2=p^{[2]}$, 
and then will form a  ``fish-bone structure''. 
It is empirically known that Wilson fermions show less-pronounced structure than chiral fermions, and thus develop less 
problematically~\cite{Boucaud:2003dx}.
In this study, we simply adopted the following ``democratic'' strategy. First, we impose a ``cylindrical'' cut on the values of momenta and then minimize the $O(4)$ variant hypercubic artifacts. We take an average over the selected momentum configurations, so that the single-valued function of $p^2$ can be obtained.

As described earlier, the renormalization scale $\mu_0$ is introduced according to 
the subtraction kinematics with the external quark momenta $p$ 
as $\mu_0=\sqrt{p^2}$ in determination of the renormalization constant from the vertex function.
The lattice artifacts can be described by the polynomial function 
of the dimensionless variable ($a\mu_0$) as
\begin{align}
Z_O(\mu_0) \approx \sum_{k > 0}^{k_{\mathrm{max}}} c_k (a\mu_0)^{2k} ,
\label{eq:ren_fit_poly}
\end{align}
where the odd power terms do not present under the momentum reflection symmetry ($p \rightarrow -p$) 
for the external momentum. We will adopt the above functional form to subtract the lattice discretization artifacts
caused by the finite lattice spacing $a$ at higher $\mu_0$ and then estimate the relevant $\mu_0$-independent value $c_0$. The selection of 
the range of $\mu_0$ and the choice of the number of the polynomial terms ($k_{\mathrm{max}}$)
may introduce systematic uncertainties. Therefore, we will make a detailed assessment of such the systematic 
uncertainties in order to determine the systematical errors on the final result of the $\mu_0$-independent value.

\subsubsection{Unwanted infrared divergence}
\label{unwanted_infrared_divergenve}
There is another major source of systematic uncertainties associated with the {\it infrared} 
physics scale $\Lambda_{\mathrm{IR}}$. 
In the quark sector, typical IR scales come from quark mass $m_q$ ($\sim \Lambda_{\mathrm{IR}}$), chiral condensate 
$\langle \bar{q}q \rangle$ ($\sim \Lambda_{\mathrm{IR}}^3$) and so on.
The infrared divergence
can remain even in the chiral limit due to nonperturbative vacuum condensates. 
It is demonstrated in the RI/MOM scheme \cite{Aoki:2007xm}, that $\langle\overline{q}q\overline{q}q\rangle$ condensate contributes at $O(\mu_0^{-2})$ by using the Weinberg's power counting (necessary dimensional adjustment is done using $\Lambda_{\mathrm{QCD}}$ as a typical low energy momentum scale triggering the condensate) ~\cite{Weinberg:1959nj}. This particular contribution is, however, removed by the use of SMOM schemes. Although large infrared effect is much reduced in SMOM schemes in this way, there is no guarantee that all the $O(\mu_0^{-2})$ effects are completely removed.
Indeed it is pointed out~\cite{Boucaud:2005rm} that the vertex functions also may receive from the dimension two condensate contribution ($\Lambda_{\mathrm{IR}}^2=\langle A^2 \rangle$) since $\langle A^2 \rangle$ condensate could couple to the quark propagator in the case of the Landau gauge, which is suggested in RI' scheme~\cite{Boucaud:2005rm} whose wavefunction renormalization is equivalent to RI/SMOM scheme
\cite{Sturm:2009kb}.
As the leading infrared effect we take a double pole with respect to $\mu_0$ 
\begin{align}
     Z_O(\mu_0) \approx \frac{c_{-1}}{(\Lambda_{\mathrm{IR}}^{-1}\mu_0)^2} ,
\label{eq:ren_fit_pole}
\end{align}
where a scale $\Lambda_{\mathrm{IR}}$ is responsible for characterizing infrared physics.

\subsubsection{Uncertainties from perturbation and others}
\label{error_pt}
As described earlier, in order to compare with the experimental values, we have to
evaluate the renormalized values in the $\overline{\mathrm{MS}}$ at the renormalization scale
of 2 GeV using the continuum perturbation theory for the scheme matching and the scaling evolution. 
Therefore, we would like to estimate possible truncation errors in perturbation matching and evolution.
To this end,  we utilize the two intermediate schemes where the conversion factors differ.
A difference of resulting renormalization factors through RI/SMOM and RI/SMOM$_{\gamma_\mu}$ schemes may be interpreted as a truncation error. 
We have four results of renormalization factor for each operator.
They are given through the four combinations with the choices of two types of the SMOM scheme, which purely care the truncation error, combined with two ways of estimating the wave function renormalization through the vector vertex $\Lambda_V$ or axial-vector vertex $\Lambda_A$ with help of $Z_V^{\mathrm{SF}}$ or $Z_A^{\mathrm{SF}}$.

It is worth remarking that there is no difference between $\Lambda_V$ and $\Lambda_A$ in the continuum perturbation theory.
Therefore, the difference that occurs only when choosing  
the vector or axial-vector with a fixed choice of the SMOM scheme does not originate from the truncation error and rather
arises from other sources of non-perturbative effects.
Indeed, such difference in the RI/MOM scheme becomes sizable  
due to unwanted infrared divergence associated with non-perturbative effects as discussed in Appendix~\ref{app:mom_and_smom}. 
The above-mentioned difference is already suppressed by means of the SMOM schemes, and then is expected to be subdominant.
In this study, we simply assume that a typical size of 
systematic uncertainties associated with the truncation error
can be explored by
the difference among the four combinations with the choices of the intermediate schemes (RI/SMOM or RI/SMOM$_{\gamma_\mu}$ schemes), 
combined with two inputs of $Z_V^{\mathrm{SF}}$ or $Z_A^{\mathrm{SF}}$.
We confirm that the estimated uncertainties are comparable with the naive truncation error
 $\mathcal{O}(\alpha_s^3)\sim3\%$
and this alternative estimation works, as detailed in Appendix~\ref{app:uncertainties_pert}.

\subsection{Renormalization constants}
\label{methods_ren}
In order to reduce the systematic uncertainties associated with the residual $\mu_0$-dependence,
let us introduce two types of fitting functional forms as a function of the matching scale $\mu_0$. 
The first functional form which includes both the infrared divergent contribution at lower $\mu_0$
and the discretization corrections at higher $\mu_0$ is given as
\begin{align}
    f_{\mathrm{Global}}(\mu_0)=\frac{c_{-1}}{(\Lambda_{\mathrm{IR}}^{-1}\mu_0)^2} + c_0 + \sum_{k > 0}^{k_{\mathrm{max}}} c_k (a\mu_0)^{2k}
\label{eq:ren_fit_global}
\end{align}
with $c_0$ being the $\mu_0$-independent value of $Z_O^{\mathrm{\overline{MS}}} (\mu)$ at the
renormalization scale $\mu=2$ GeV. Eq.~(\ref{eq:ren_fit_global}) is used for the global fit to a wide 
range of $\mu_0$ (denoted as ``Global" in subscript). 
In this fit model, the leading contribution at lower $\mu_0$ is simply introduced 
by a double pole with respect to a dimensionless variable of 
$\Lambda_{\mathrm{IR}}^{-1}\mu_0$, where a scale $\Lambda_{\mathrm{IR}}$ 
is responsible for characterizing infrared physics.
A model dependence that is mainly introduced by the double pole term should be addressed.

In this context, we use different fitting strategy to assess the systematic error associated with the model dependence of the fit form~(\ref{eq:ren_fit_global}). 
For the sake of this purpose, we use the second functional form
\begin{align}
    f_{\mathrm{IR-trunc.}}(\mu_0) =  c_0 + \sum_{k > 0}^{k_{\mathrm{max}}} c_k (a\mu_0)^{2k} ,
\label{eq:ren_fit_IRtru}
\end{align}
where only positive power series in $(a\mu_0)^2$ are included.This functional form is widely used in the previous studies~\cite{{Horkel:2020hpi},{Gupta:2018qil},{Bhattacharya:2016zcn},{Alexandrou:2019brg},{Hasan:2019noy}}.
We then apply the functional form of Eq.~(\ref{eq:ren_fit_IRtru}) for fitting the data in a restricted range $\mu_0\gg \Lambda_{\mathrm{IR}}$ 
(denoted as ``IR-trunc.'' in subscript) for avoiding unwanted infrared effects. However, the fit result would be sensitive to the fit range ($\mu_{\mathrm{min}} \le \mu_0 \le \mu_{\mathrm{max}}$), especially the choice of $\mu_{\mathrm{min}}$ that is supposed to 
satisfy $\mu_{\mathrm{min}} \gg \Lambda_{\mathrm{IR}}$. In this study, the fitting with Eq.~(\ref{eq:ren_fit_IRtru}) 
is only applied to the case of $\mu_{\mathrm{min}}\ge 2$ GeV, where the truncation errors in perturbation 
matching can be simultaneously reduced.

Using two functional forms of Eq.~(\ref{eq:ren_fit_global}) and Eq.~(\ref{eq:ren_fit_IRtru}), 
the appropriate fit range ($\mu_{\mathrm{min}} \le \mu_0 \le \mu_{\mathrm{max}}$) with a smaller value of $k_{\mathrm{max}}$ is carefully determined by means of uncorrelated fits according to the following guidelines:
\begin{enumerate}[I]
    \item $c_0$ is insensitive to the variation of the fit range.
    \item $k_{\mathrm{max}}$ ensures that inclusion of the higher power terms does not change the value of $c_0$.
    \item $\chi^2$/dof is close to unity as much as possible.
\end{enumerate}
If the above two fit procedures 
lead to compatible results, both the model dependence in the determination of the $\mu_0$-independent value 
with the usage of Eq.~(\ref{eq:ren_fit_global}) and the strong dependence 
of the fit range using Eq.~(\ref{eq:ren_fit_IRtru}) are reasonably well under control. 
A representative value is selected from them as the central value of the final result, while 
discrepancies among them can be estimated as systematic errors.

In this study, three types of systematic uncertainties for the determination of the renormalization constant
are examined as follows:  
\begin{enumerate}
    \item $(\Delta Z)_{\mathrm{fit}}$ is determined by the maximum difference among the fit results that
     are stable against the variation of the fit range and the choice of $k_{\mathrm{max}}$. 
    \item $(\Delta Z)_{\mathrm{scheme}}$ is determined from the maximum difference among
     the four combinations with the choices of the intermediate schemes (RI/SMOM or RI/SMOM$_{\gamma_\mu}$ schemes)
     combined with the SF inputs of $Z_V$ and $Z_A$ (hereafter denoted 
     as \{SF input, scheme\}=\{$Z_V$, SMOM\}, \{$Z_V$, SMOM$_{\gamma_\mu}$\}, 
     \{$Z_A$, SMOM\},  \{$Z_A$, SMOM$_{\gamma_\mu}$\}).
     \item $(\Delta Z)_{\mathrm{model}}$ is determined by discrepancy between
    results obtained from the global fit with the functional form of $f_{\mathrm{Global}}(\mu_0)$ 
    and the IR-truncated fit 
    with the functional form of $f_{\mathrm{IR-trunc.}}(\mu_0)$ 
    with the optimal combination of \{SF input, scheme\}.
\end{enumerate}
Furthermore, all three systematic uncertainties are added in quadrature to get 
\begin{align}
(\Delta Z)_{\mathrm{syst}} = \sqrt{((\Delta Z)_{\mathrm{fit}})^2 + ((\Delta Z)_{\mathrm{scheme}})^2 +
((\Delta Z)_{\mathrm{model}})^2}.
\end{align}
We then will quote the total error of the renormalization constant as 
\begin{align}
(\Delta Z)_{\mathrm{total}} = \sqrt{((\Delta Z)_{\mathrm{stat}})^2 + ((\Delta Z)_{\mathrm{syst}})^2},
\end{align}
which is given by both statistical and systematic uncertainties combined in quadrature.

\section{Simulation details}

In this paper, we use two ensembles of gauge configurations, which are
generated by the PACS Collaboration with $L^3 \times T=128^3 \times 128$ and $64^3 \times 64$ 
lattices using the six stout-smeared $O(a)$-improved Wilson quark action and the Iwasaki 
gauge action~\cite{Iwasaki:1983iya}
at fixed gauge coupling $\beta=1.82$ corresponding to the lattice spacing
of $a=0.08520(16)$ fm [$a^{-1} = 2.3162(44)$ GeV] determined
from the $\Xi$ baryon mass input~\cite{PACS:2019ofv}.
The stout-smearing parameter is set to $\rho=0.1$~\cite{Morningstar:2003gk}.
The improved coefficient, $c_{\mathrm{SW}}=1.11$, is nonperturbatively
determined using the Schr\"odinger functional (SF) scheme~\cite{Taniguchi:2012gew}. 
The PACS Collaboration has performed the finite-size study for light meson and
baryon sectors at the physical point in 2+1 flavor QCD with the two
lattices~\cite{{Ishikawa:2018jee},{PACS:2019ofv}}. 
A brief summary of the simulation parameters for the two lattice volumes is given in Tab.\ref{tab:simulation_details}.

As pointed out in Ref.~\cite{PACS:2019ofv}, the two volumes (linear spatial extents of 10.9 fm 
and 5.5 fm) are large enough for the finite-size effect on the nucleon mass to be negligible.
Even for the nucleon matrix elements, the finite-size effects are hardly detected under
the current statistical precision~\cite{Ishikawa:2021eut}.
Thanks to the large spatial volume ($L=128$), the small momentum transfer region
close to the zero momentum transfer, $q^2=0$ can be accessible in calculation 
of the various hadron form factors. Therefore, the $L=128$ lattice was
utilized for studies of the nucleon form factors~\cite{Ishikawa:2018rew} 
and also the $K_{l3}$ form factor~\cite{PACS:2019hxd}, both of which require
the $q^2$ interpolation (or extrapolation) to the zero momentum point.
In this work, we use 20 gauge configurations separated by 10 molecular dynamics trajectories for the $L=128$ lattice, while 100 gauge configurations separated by 20 molecular dynamics trajectories are used for the $L=64$ lattice.
Their statistical errors are estimated by the single elimination jackknife method.
Significant reduction of the computational cost is achieved by employing
the all-mode-averaging (AMA) method~\cite{Blum:2012uh,Shintani:2014vja,vonHippel:2016wid} with the deflated Schwartz Alternating Process  (SAP)~\cite{Luscher:2003qa} and 
Generalized Conjugate Residual (GCR) \cite{Luscher:2007se}
for the measurements as in our previous works~\cite{{Shintani:2018ozy},{Ishikawa:2021eut}}.
We compute the combination of the correlation function with high-precision $O^{(\mathrm{org})}$ 
and low-precision $O^{(\mathrm{approx})}$ as
\begin{align}
    O^{(\mathrm{AMA})} 
    =
    \frac{1}{N_{\mathrm{org}}}\sum^{N_{\mathrm{org}}}_{f\in G}
    \left(O^{(\mathrm{org})f} - O^{(\mathrm{approx})f} \right)
    +
    \frac{1}{N_{G}}\sum^{N_{G}}_{g\in G}O^{(\mathrm{approx})g},
\end{align}
where the superscripts $f, g$ denote the transformation under the lattice symmetry $G$. 
In our calculations, it is translational symmetry, e.g., changing the position of the source operator, and
changing the temporal direction of the configuration using its hypercubic symmetry as in Refs.~\cite{{Ishikawa:2018rew},{Ishikawa:2018jee},{PACS:2019ofv},{PACS:2019hxd}}. $N_{\mathrm{org}}$ and $N_{G}$ are the numbers for $O^{(\mathrm{org})}$ and $O^{(\mathrm{approx})}$, respectively. The numbers and the stopping conditions of the quark propagator for the high- and low-precision measurements are summarized in Tab.~\ref{tab:measurements}. 

In calculation of the nucleon 2- and 3-point functions, we use the same quark action 
as in the gauge configuration generation with the hopping parameter 
$\kappa=0.126117$ for the degenerated up-down quarks, the improved coefficient, $c_{\mathrm{SW}}=1.11$ and 
six steps of stout-smearing to the link variables.
The periodic boundary condition in all the temporal 
and spatial directions is adopted in the quark propagator calculation.

The quark propagator is calculated using the exponential smeared source (sink) with the Coulomb gauge fixing. 
The smearing parameters for the quark propagator 
defined in Eq.~(\ref{eq:exp_smear_nuc}) are chosen 
as $(A,B)=(1.2, 0.16)$ for the $L=128$ lattice~\cite{Shintani:2018ozy}
and $(A,B)=(1.2, 0.14),\ (1.2, 0.16)$ for the $L=64$ lattice~\cite{Ishikawa:2021eut}. 
In addition to the exponential source (sink), we also use the gauge-covariant, approximately Gaussian-shaped source (sink)~\cite{Gusken:1989qx} for the $L=128$ lattice with a single parameter set of 
$(n_{\mathrm{G}}, w_{\mathrm{G}})$=(110, 8.0), which is defined in Eq.~(\ref{eq:gauss_smear_nuc}),
so as to make sure that there is no dependence on the gauge fixing condition and 
the excited-state contamination is well controlled on our final results. 
All parameters of adopted sources are chosen to optimize 
the effective mass plateau for the smear-local case in each of $L=128$ and $L=64$ lattices.

The nucleon 3-point functions are calculated using the sequential source method with a fixed
source-sink separation~\cite{{Martinelli:1988rr},{Sasaki:2003jh}}. 
This method requires the sequential quark propagator for each choice of a projection operator
$\mathcal{P}_k$ regardless of the types of current $J^{O}$. 
To minimize the numerical cost for the larger volume calculation ($L=128$), 
we have to restrict ourselves to calculating a single 3-point function with the projection $\mathcal{P}_{5z}$,
while we calculate three 3-point functions with the projection $\mathcal{P}_{5j}$
in all the three spatial directions $j=x, y, z$ for the smaller volume calculation ($L=64$). 
Furthermore, in the case of the $L=64$ volume, we also take the average of the forward 
and backward 3-point functions to increase statistics.

As for the source-sink separation of $t_{\mathrm{sep}}$ (denoted as 
$t_{\mathrm{sep}}=t_{\mathrm{sink}}-t_{\mathrm{src}}$), we use four variations of $t_{\mathrm{sep}}=10,12,14,16$ 
with the exponential source~\footnote{For $t_{\mathrm{sep}}=16$, we include additional numerical simulations
which lead to a slightly larger number of measurements in comparison to our earlier work~\cite{Shintani:2018ozy}.}
and two variations of $t_{\mathrm{sep}}=13,16$ with the gauge-covariant Gaussian source for the $L=128$ lattice, while four variations of $t_{\mathrm{sep}}=11,12,14,16$ for 
the $L=64$ lattice as also summarized in Tab.~\ref{tab:measurements}.
We investigate the effects of the excited-state contamination by varying $t_{\mathrm{sep}}$ from 0.85 
to 1.36 fm in the standard plateau method that was explained in Sec.~\ref{Sec2:ratio_method}.

The renormalization constants for vector and axial-vector currents, $Z_{O}(O=V,A)$ are obtained with the SF scheme at vanishing quark mass as $Z_{V}=0.95153(76)(1487)$ and $Z_{A}=0.9650(68)(95)$, where each error represents statistical and systematic error stemming from the difference between two volumes~\cite{Ishikawa:2015fzw}. However,
the second errors are simply ignored in the later analysis, since we choose the larger volume~\cite{Ishikawa:2015fzw} to set the physical scale.
As for the renormalization of other quark bilinears, scalar and tensor currents, we employed the 
RI scheme for their intermediate scheme, using quark propagators which are nonperturbatively calculated on the $L=64$ lattice with the Landau gauge fixing. The details of them are described in Sec.~\ref{Renormalization}.
%
%
\begin{table*}[ht]
\caption{
Parameters of 2+1 flavor PACS ensembles with two different
lattice volumes. See Refs.~\cite{{Ishikawa:2018jee},{PACS:2019ofv}} for further details.
\label{tab:simulation_details}}
\begin{ruledtabular}
\begin{tabular}{ccccccc}
 $\beta$ &$L^3\times T$ & $a^{-1}$ [GeV] &  $\kappa_{ud}$ & $\kappa_{s}$& $M_\pi$ [GeV] \\
          \hline        
     1.82& $128^3\times 128$ &2.3162(44)& 0.126117 & 0.124902 &  0.135\\
          \hline
     1.82& $64^3\times 64$    &2.3162(44)& 0.126117  &0.124902  & 0.138 \\
\end{tabular}
\end{ruledtabular}
\end{table*}
%

%
%
\begin{table*}[ht]
    \caption{
    Details of the measurements: the spatial extent ($L$), 
    time separation ($t_{\mathrm{sep}}$), type of the smearing
    method, smearing parameters of the quark operator, the stopping condition of quark propagator in the high- and low-precision calculations ($\epsilon_{\mathrm{high}}$ and $\epsilon_{\mathrm{low}}$), the number of measurements for 
    the high- and low-precision calculations ($N_{\mathrm{org}}$ and $N_{G}$), the number of configurations ($N_{\mathrm{conf}}$) and the total number of the measurements ($N_{\mathrm{meas}}=N_{G}\times N_{\mathrm{conf}}$), respectively. 
    \label{tab:measurements}}
\begin{ruledtabular}
\begin{tabular}{ccccccccccc}
$L$ &  $t_{\mathrm{sep}}$ & Smearing-type & Smearing parameters & $\epsilon_\mathrm{high}$ & $\epsilon_\mathrm{low}$& $N_{\mathrm{org}}$
& $N_{G}$ & $N_{\mathrm{conf}}$ & $N_{\mathrm{meas}}$\\
\hline
    128 & 10& Exp.(I) &$(A,B)=(1.2,0.16)$ & $10^{-8}$& ---\footnotemark[1]\footnotetext[1]{As for the $L=128$ lattice, the low-precision calculations use 
    a fixed number of iterations for the stopping condition as
    $5$ GCR iterations using $8^4$ SAP domain size with $50$ deflation fields.} & 1& 128& 20& 2,560\\
                   & 12& &             & $10^{-8}$& ---\footnotemark[1] & 1& 256& 20& 5,120\\
                   & 14& &             & $10^{-8}$& ---\footnotemark[1] & 2& 320& 20& 6,400\\
                   & 16& &             & $10^{-8}$& ---\footnotemark[1] & 4& 512& 20& 10,240\\
                   \cline{2-10}
                   & 13& Gauss& $(n_{\mathrm{G}}, w_{\mathrm{G}})=(110, 8.0)$
                                             & $10^{-8}$& ---\footnotemark[1] & 1 &   128 &  20 & 2,560\\
                   & 16& &             & $10^{-8}$& ---\footnotemark[1] & 6 &   450 &  20 & 9,000\\
          \hline
    64 & 11& Exp.(I) &$(A,B)=(1.2,0.16)$ & $10^{-8}$& $0.005$& 4& 40& 50& 2,000\\
                  & 14& &                       & $10^{-8}$& $0.02$& 4& 64& 100& 6,400\\
                   \cline{2-10}
                  & 12& Exp.(II)&$(A,B)=(1.2,0.14)$ & $10^{-8}$& $0.005$& 4& 256& 100& 25,600\\
                  & 14& &             & $10^{-8}$& $0.005$& 4& 1,024& 100& 102,400\\
                  & 16& &             & $10^{-8}$& $0.002$& 4& 2,048& 100& 204,800\\
\end{tabular}
\end{ruledtabular}
\end{table*}

\clearpage
\section{Numerical results}

In this section we present the isovector couplings of the nucleon for the vector, axial, scalar and tensor channels, 
obtained from both $128^4$ and $64^4$ lattices. We will first show results of the bare couplings, which 
can be determined by the plateau measured in the ratio defined in Eq.(\ref{Eq:ratio}). 
For the renormalization of the scalar and tensor channels, we use the $\mathrm{RI/SMOM}_{(\gamma_\mu)}$ schemes 
as the intermediate scheme and then evaluate the renormalized values of the scalar and tensor couplings in the $\overline{\mathrm{MS}}$ scheme at the renormalization scale of 2 GeV with the help of perturbation 
theory of the matching between the intermediate and $\overline{\mathrm{MS}}$ schemes.

\subsection{Bare coupling}
The bare coupling can be read off from the ratio~(\ref{Eq:ratio}) as the asymptotic plateau that is independent of a choice of $t_{\mathrm{op}}$, with some moderate source-sink separation $t_{\mathrm{sep}}$ and the optimal choice of the smearing parameters. 
Since the excited-state contaminations could not be fully eliminated by tuning smearing parameters in practice, 
one should calculate the ratio~(\ref{Eq:ratio}) with several choices of $t_{\mathrm{sep}}$, and then makes sure
whether the evaluated value of the bare coupling does not change with a variation of $t_{\mathrm{sep}}$ within
a certain precision. 

\begin{figure*}
 \includegraphics[width=0.32\textwidth,bb=30 30 405 525,clip]{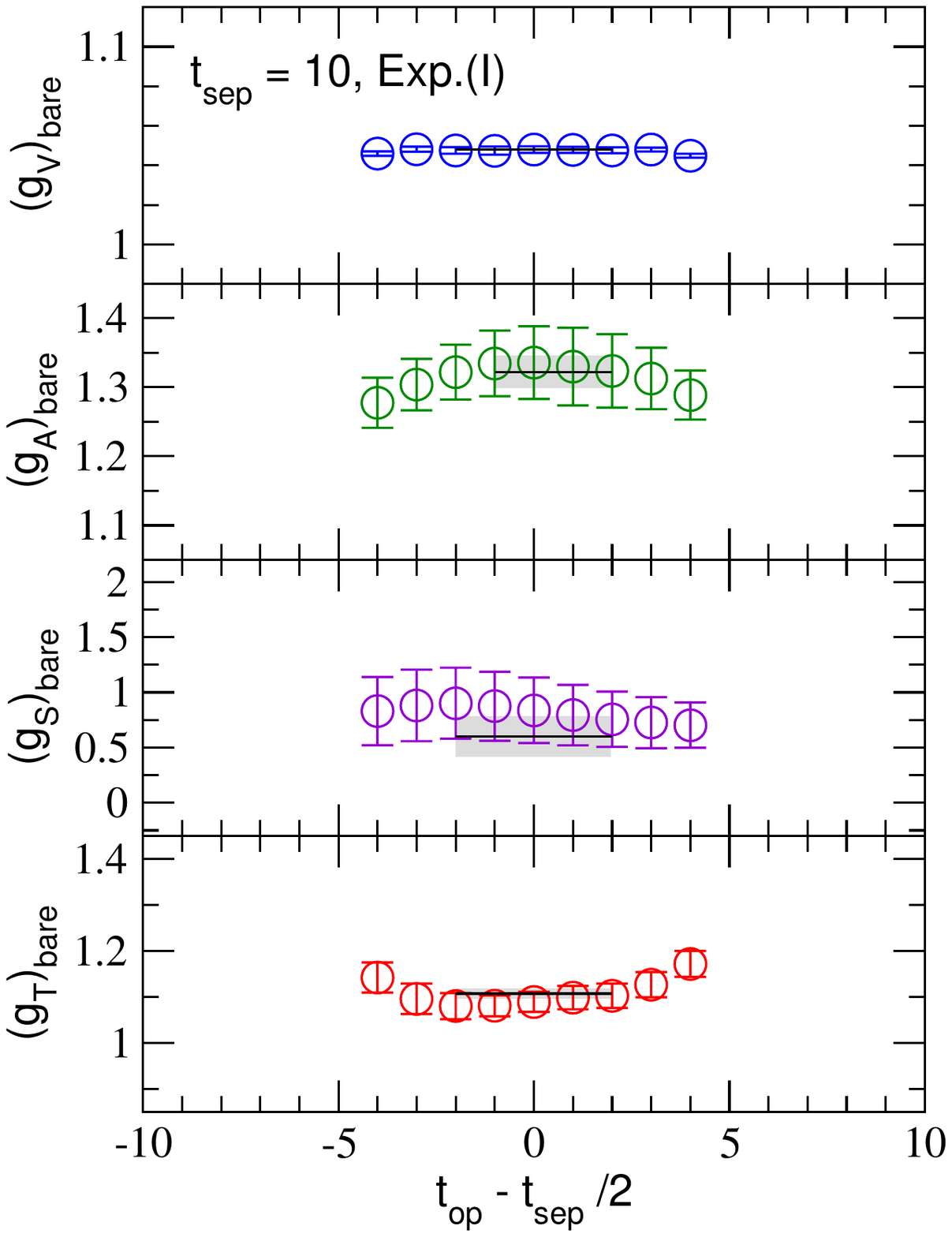}
 \includegraphics[width=0.32\textwidth,bb=30 30 405 525,clip]{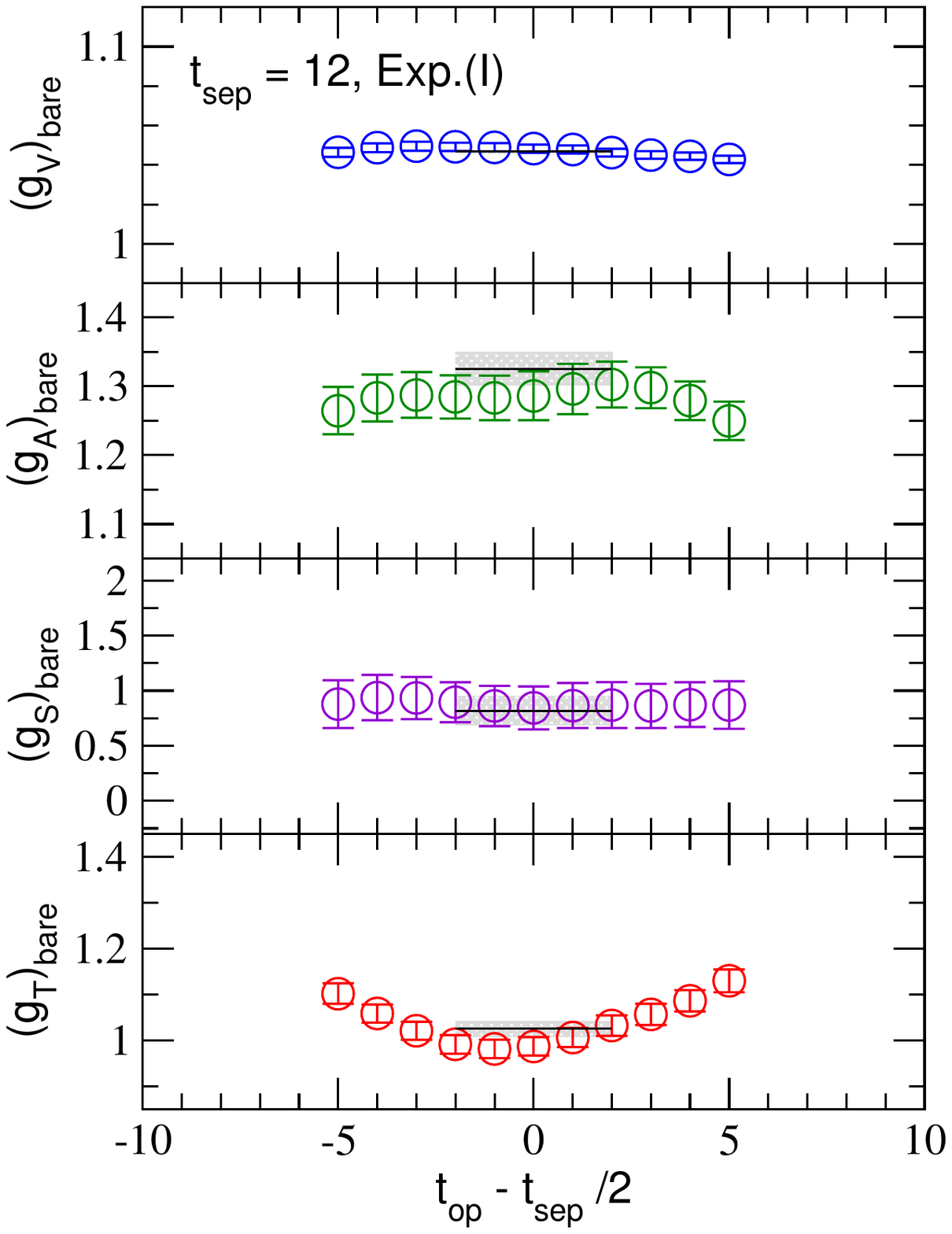}
 \includegraphics[width=0.32\textwidth,bb=30 30 405 525,clip]{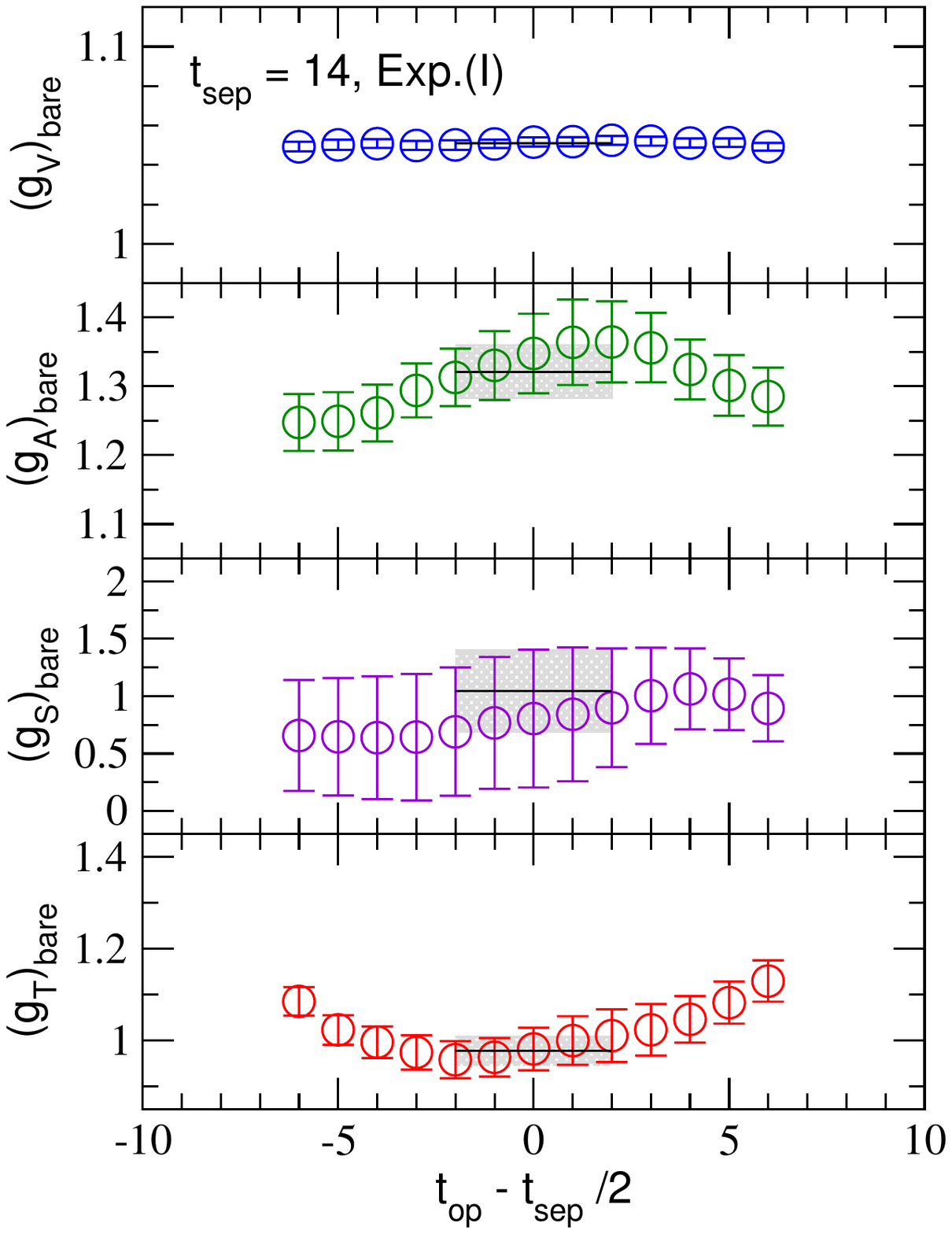}
 \includegraphics[width=0.32\textwidth,bb=30 30 405 525,clip]{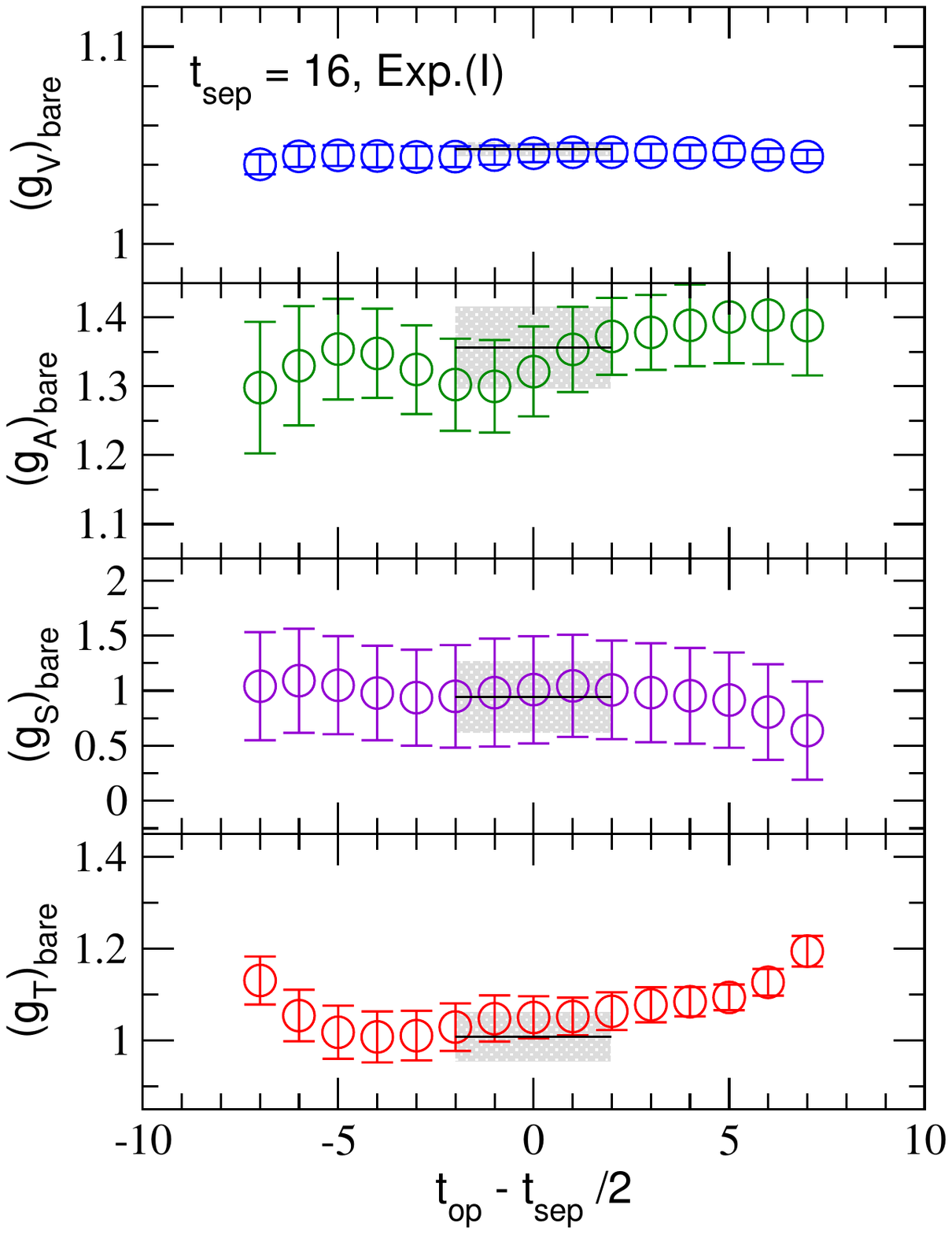}
 \includegraphics[width=0.32\textwidth,bb=30 30 405 525,clip]{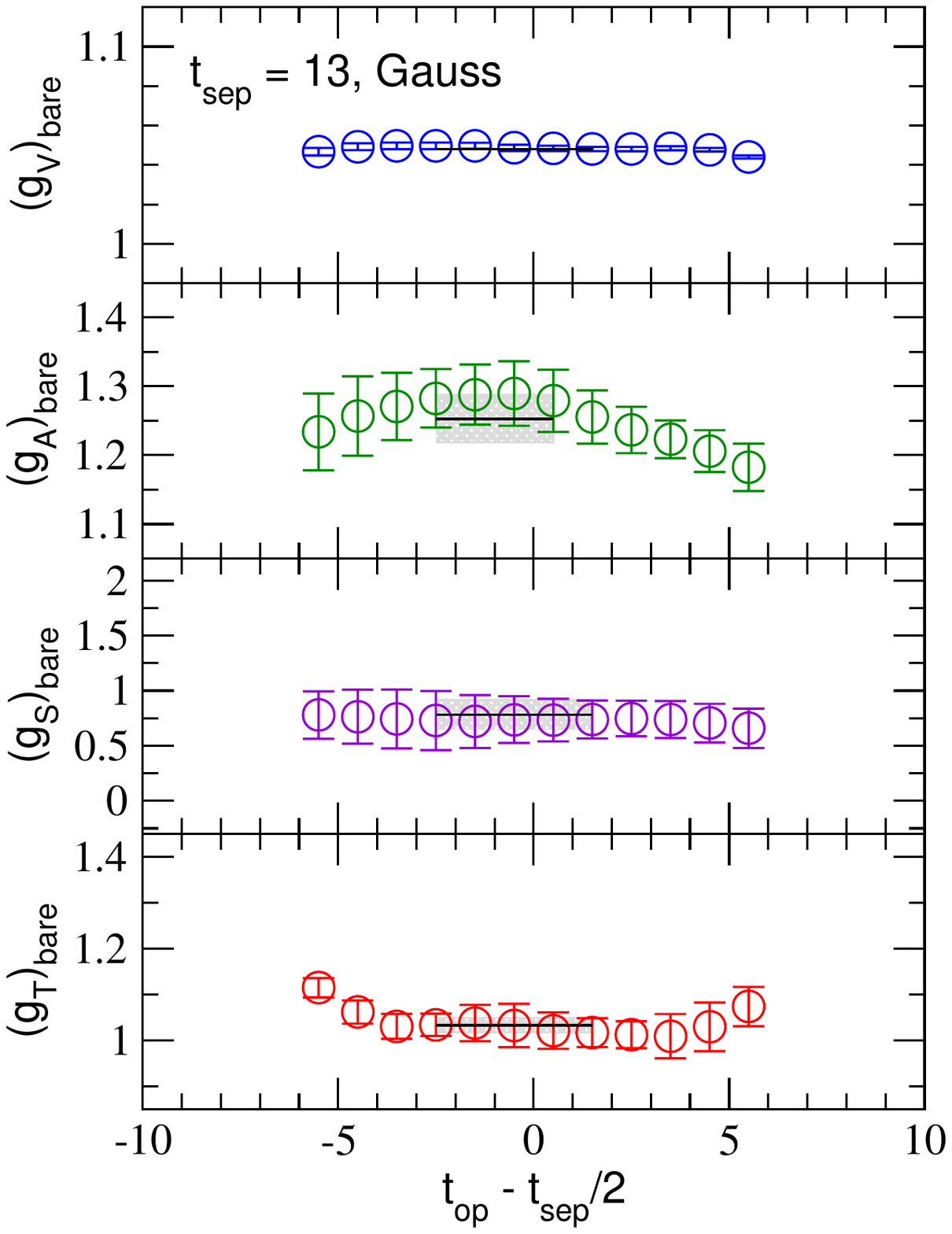}
 \includegraphics[width=0.32\textwidth,bb=30 30 405 525,clip]{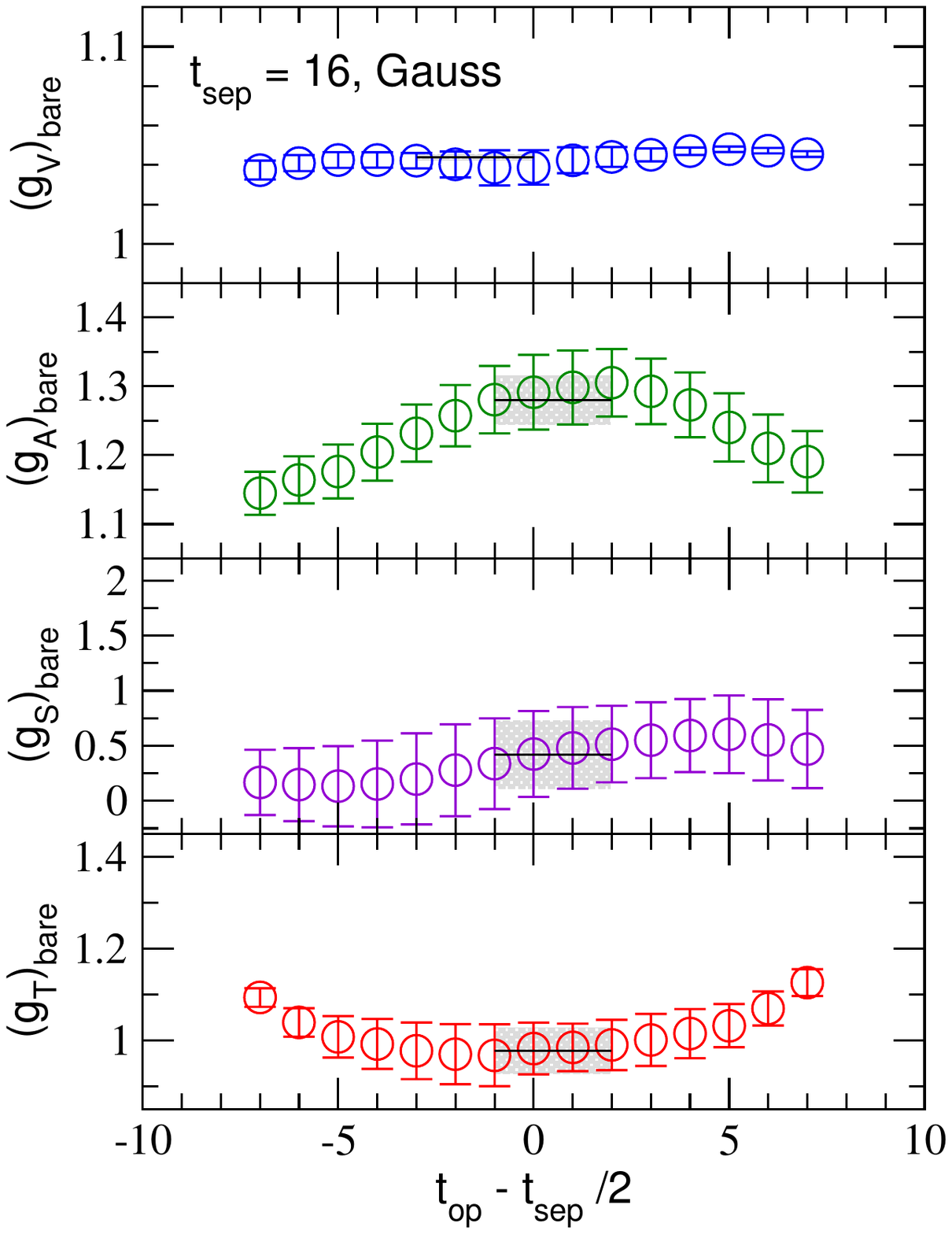}
 \caption{
Bare coupling $g_O$ for $O=V, A, S, T$ obtained from the $128^4$ lattice as a function of the current operator insertion point $t_{\mathrm{op}}$. In each panel, the results for the bare values of $g_V$, $g_A$, $g_S$ and $g_T$ are plotted from top to bottom with different choices of $t_{{\mathrm{sep}}}$ and smearing type.
For the sake of comparison, the interval of the $y$ axis in each channel is selected to be the same for all the panels. The solid line represents the average value, and shaded band displays the fit range and one standard deviation.
The value of $t_{\mathrm{sep}}$ and type of smearing are given in the legends.}
\label{fig:128_bare_all}
\end{figure*}

In Fig.~\ref{fig:128_bare_all}, we show results of the bare couplings
obtained from the $L=128$ lattice for different choices of $t_{\mathrm{sep}}$ and two kinds of the smearing-type. 
In each panel, the results for the bare values of $g_V$, $g_A$, $g_S$ and $g_T$ are plotted from top to bottom
as a function of the current operator insertion point $t_{\mathrm{op}}$. 
At a glance, plateau behaviors are observed for all four channels in the middle region between the source and sink points, thanks to our elaborate tuning of the smearing parameters. 

\begin{figure*}
\includegraphics[width=0.32\textwidth,bb=30 30 405 525,clip]{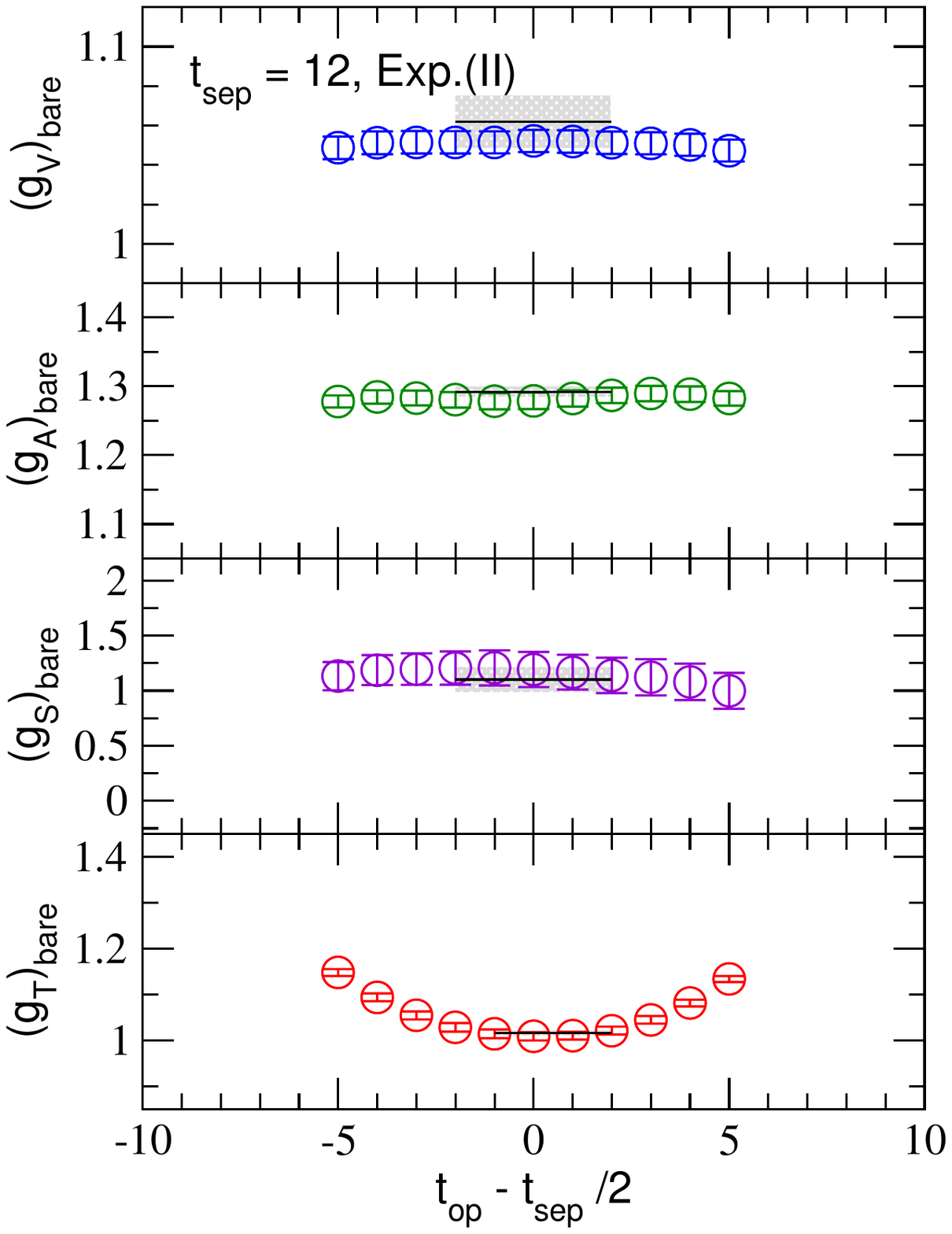}
\includegraphics[width=0.32\textwidth,bb=30 30 405 525,clip]{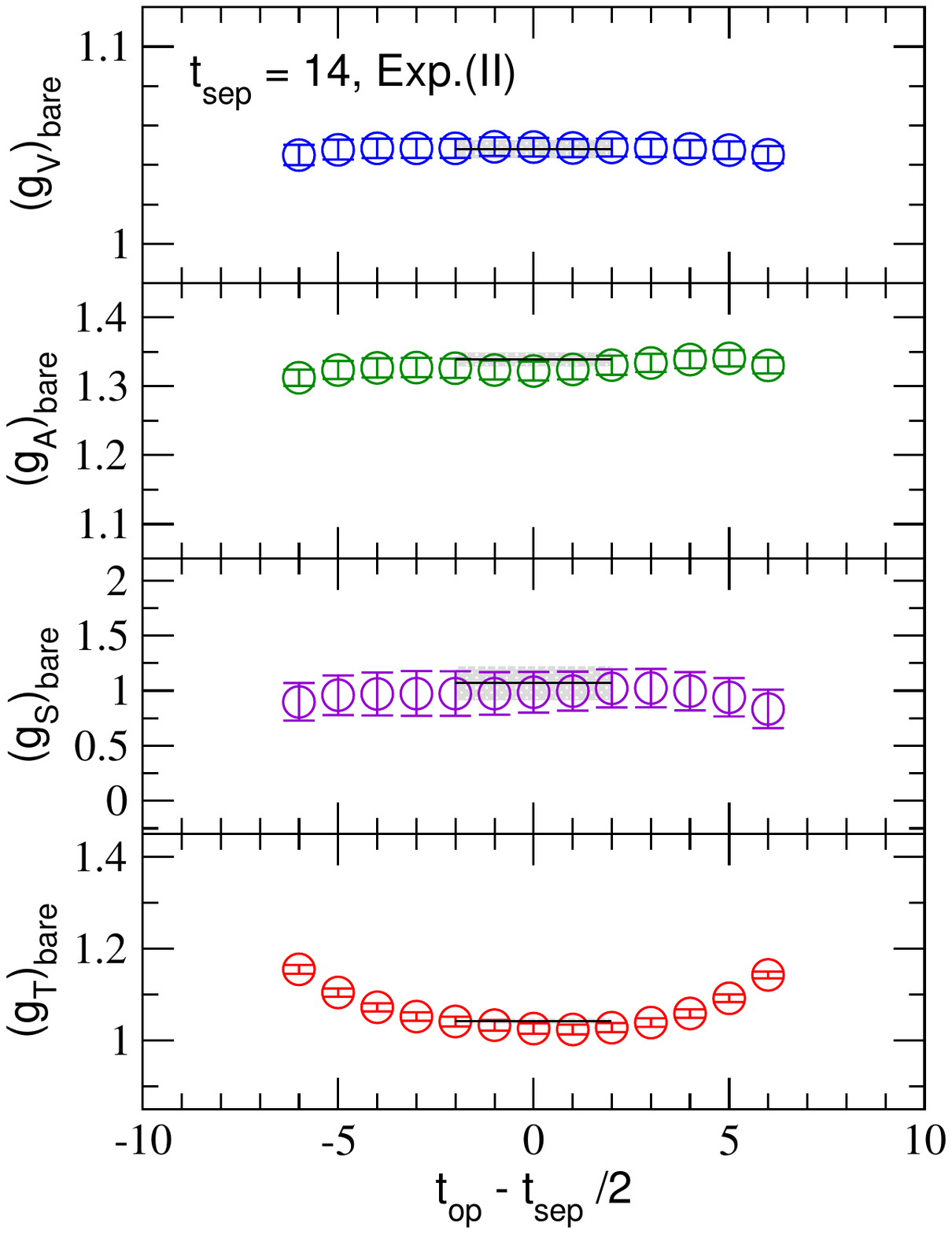}
\includegraphics[width=0.32\textwidth,bb=30 30 405 525,clip]{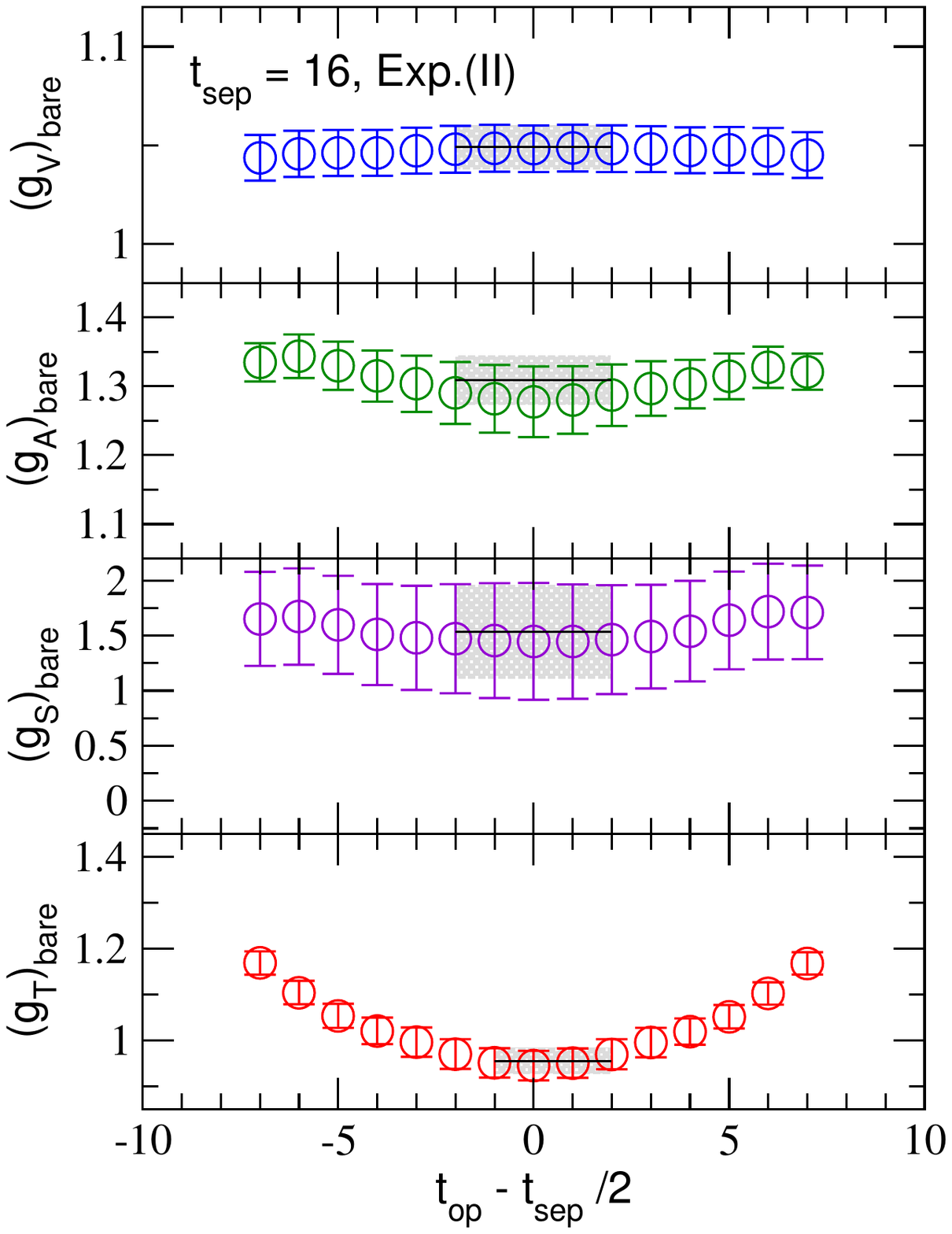}
\includegraphics[width=0.32\textwidth,bb=30 30 405 525,clip]{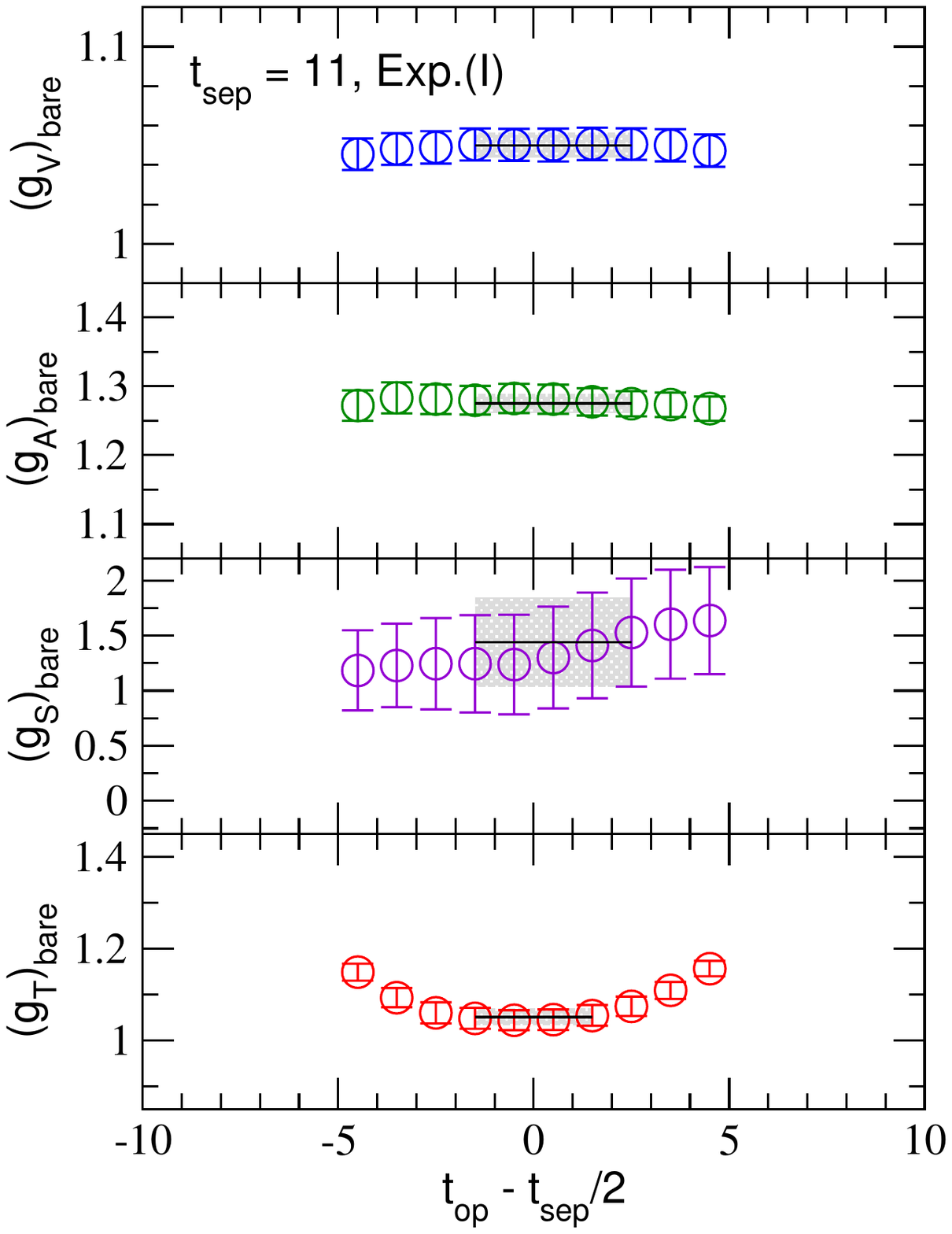}
\includegraphics[width=0.32\textwidth,bb=30 30 405 525,clip]{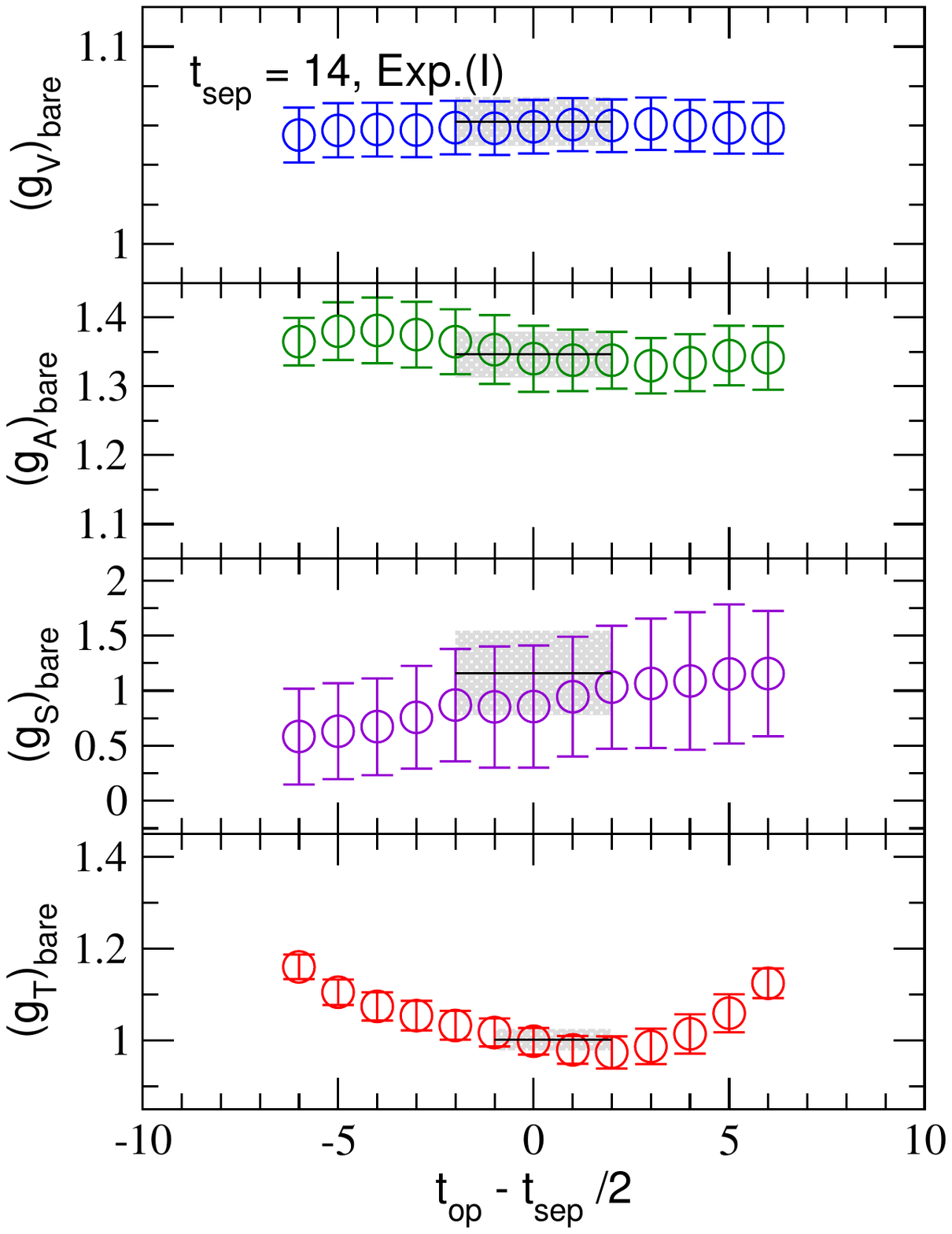}
 \caption{
Bare coupling $g_O$ for $O=V, A, S, T$ obtained from the $64^4$ lattice as a function of the current operator insertion point $t_{\mathrm{op}}$. The rest is the same as in Fig.~\ref{fig:128_bare_all}
}
 \label{fig:64_bare_all}
\end{figure*}

We also show the bare couplings obtained from the $L=64$ lattice in Fig.~\ref{fig:64_bare_all}. 
The exponential smearing is only used with two sets of smearing parameters, $(A,B)=(1.2, 0.14)$ and $(1.2, 0.16)$. 
The former set is obtained by retuning for the $L=64$ lattice (denoted as Exp.(II)), while
the latter set is the same as the one used in the previous $L=128$ calculation (denoted as Exp.(I)). 

Let us focus on the upper three panels in Fig.~\ref{fig:64_bare_all}, where the symmetric behavior 
with respect to $t_{\mathrm{op}}$ is clearly seen in all four channels. It is worth reminding that 
in the $L=64$ calculation, the time-reversal average was implemented in our averaging procedure 
on each configuration with multiple measurements, which includes both forward and backward 
propagations in time, as described early. More importantly, it is clearly observed that
there is some channel dependence. The vector, axial-vector and scalar channels show
the very long plateau behavior up to the vicinity of the source and sink points, while the tensor channel 
exposes the residual $t_{\mathrm{op}}$ dependence that would be originated from the excited-state
contaminations. This tendency is slightly seen in the $L=128$ calculations, where the symmetric behavior 
would be hindered by larger statistical fluctuations because each total number of the measurements
with the $L=128$ lattice is much smaller than the cases of Exp.(II) with $t_{\mathrm{sep}}=12, 14, 16$ 
in the $L=64$ calculations.

Although the asymptotic plateau behavior is relatively shortened in the tensor channel,
there is still room for extracting the value of the bare coupling from the plateau behavior appearing
in the middle region between the source and sink points even in the $L=64$ calculations. 
We simply use the correlated constant fit to extract the value of the bare coupling for all channels in
both $L=128$ and $L=64$ calculations.
In the $L=128$ calculations, we choose the same fit range 
that is determined by the axial-vector channel, while the narrower fit range is adopted for the tensor 
channel in the $L=64$ calculations where the small statistical fluctuations make it difficult to perform 
the constant fit with the same fit range used for the other channels. 
The fit ranges, which are carefully chosen to obtain a reasonable value of $\chi^2/{\mathrm{dof}}$, 
are denoted as gray-shaded areas in each panel of both figures.
The obtained values of the bare couplings are summarized in Tab.~\ref{tab:bare_couplings}.

%
%

%
%

%
%
\begin{table}[ht]
     \caption{
     Results of the bare couplings $g_{O}$ for $O=V, A, S, T$ obtained from both $128^4$ and $64^4$ lattices
     with different choices of the source-sink separation ($t_{\mathrm{sep}}$) and the smearing type. 
     The values of the bare couplings are extracted by the correlated constant fit on the relevant ratio~(\ref{Eq:ratio})
     with the fit range, which is denoted as gray-shaded area in each panel of Figs.~\ref{fig:128_bare_all} 
     and \ref{fig:64_bare_all}.
     }
     \label{tab:bare_couplings}
     \renewcommand{\arraystretch}{1.3}
     \begin{ruledtabular}
     \begin{tabular}{ccccccc}
        $L$ & $t_{\mathrm{sep}}$ & Smearing-type &  $g_V^{\mathrm{bare}}$& $g_A^{\mathrm{bare}}$& $g_S^{\mathrm{bare}}$& $g_T^{\mathrm{bare}}$\\
       \hline
       128 &10 & Exp.(I) & 1.048(1)& 1.322(25)& 0.599(197)& 1.107(14)\\
               &12 &                       & 1.047(2)& 1.325(23)& 0.817(147)& 1.025(21)\\
               &14 &                       & 1.051(2)& 1.321(38)& 1.044(353)& 0.977(36)\\
               &16 &                       & 1.048(4)& 1.356(61)& 0.943(336)& 1.008(57)\\
               \cline{2-7}
               &13 & Gauss 
                                     & 1.048(1)& 1.252(34)& 0.783(153)& 1.033(20)\\
               &16 &            & 1.044(2)& 1.280(38)& 0.418(325)& 0.977(54)\\
       \hline
       64  &11 & Exp.(I) & 1.050(7)& 1.275(16)& 1.442(417)& 1.051(21)\\
               &14 &                       & 1.062(13)& 1.346(35)& 1.159(395)& 1.002(26)\\
               \cline{2-7}
              &12 & Exp.(II)  & 1.062(14)& 1.292(9)& 1.102(127)& 1.016(8)\\
               &14 &                       & 1.048(4)& 1.339(12)& 1.068(168)& 1.042(9)\\
               &16 &                       & 1.049(12)& 1.309(38)& 1.535(436)& 0.955(32)\\
      \end{tabular}
     \end{ruledtabular}
   \end{table}

\subsection{Renormalization}
\label{Sec.5:renorm}
In order to compare with the experimental values or other lattice results, the bare couplings $g_S$ and $g_T$ should be renormalized with the renormalization constants in the $\overline{\mathrm{MS}}$ scheme, while the renormalization constants for vector and axial-vector current, $Z_{O} (O=V,A)$ are obtained with the SF scheme at vanishing quark masses as $Z^{\mathrm{SF}}_{V}=0.95153(76)$ and $Z^{\mathrm{SF}}_{A}=0.9650(68)$~\cite{Ishikawa:2015fzw}.
For the scalar and tensor cases, we use the RI scheme as the intermediate scheme 
in order to evaluate the renormalization constants $Z^{\mathrm{RI}}_{O}(O=S,T)$ in fully nonperturbative manner 
with the value of $Z^{\mathrm{SF}}_{O}(O=V,A)$ as shown in Eq.~(\ref{eq:eq_renormalization}).
The resulting renormalization constants are then converted to the $\overline{\mathrm{MS}}$ scheme at an intermediate scale
and evolved to the scale of 2 GeV using the perturbation theory as described in Eq.~(\ref{eq:eq_matching}).

The smaller volume ensemble ($L=64$) with 101 gauge configurations 
is used for computing the renormalization constants. Averaging the results of
the vertex functions over multiple sources would help 
to reduce the statistical uncertainties. 
The vertex functions can be constructed by the (Fourier-transformed) quark propagator 
on each gauge configuration. The quark propagator is calculated by using a point source since 
all modes in the momentum space are equally taken into account.  We then calculate the quark propagator 
with four different source locations at $x_0=(0,0,0,0)$, (16,16,16,16), (32,32,32,32), (48,48,48,48).
In the analysis, all four sets of vertex functions are folded together to create the single-vertex functions,
respectively.

\subsubsection{Test for the chiral symmetry}
Our calculations are performed with the stout-smeared $O(a)$ improved Wilson fermions, where 
the chiral symmetry is explicitly broken even at vanishing quark masses due to the discretization error.
This is indeed confirmed by the fact of $Z^{\mathrm{SF}}_{V} \neq Z^{\mathrm{SF}}_{A}$ for the local currents. 
In order to examine the explicit chiral symmetry breaking in the RI scheme,  
we introduce the following quantity
\begin{align}
   c_{V-A} & \equiv \frac{\tilde{\Lambda}_A^{x} - \tilde{\Lambda}_V^{x}}{(\tilde{\Lambda}_A^{x} + \tilde{\Lambda}_V^{x})/2} ,
\label{eq:CAV}
\end{align}
where $\tilde{\Lambda}_{V(A)}^{x}$ denotes the traced vertex function with an appropriate projection operator as 
$\tilde{\Lambda}_{V(A)}^{x}={\rm Tr}\left[\Lambda_{V(A)} {\cal P}^{x}_{V(A)} \right]$ 
for $x={\mathrm{SMOM}}$ or ${\mathrm{SMOM}}_{\gamma_\mu}$.
Non-zero value of $c_{V-A}$ directly indicates a violation of the equality $Z_V = Z_A$, since the value 
of $c_{V-A}$ turns out to be the ratio of $\frac{Z_V - Z_A}{\left(Z_V + Z_A\right)/2}$. However recall that the vertex functions receive the effects from both the explicit chiral symmetry breaking and 
spontaneous chiral symmetry breaking~\cite{Blum:2001sr}.

%
%
\begin{figure*}
\includegraphics[width=0.7\linewidth,bb=0 0 792 612,clip]{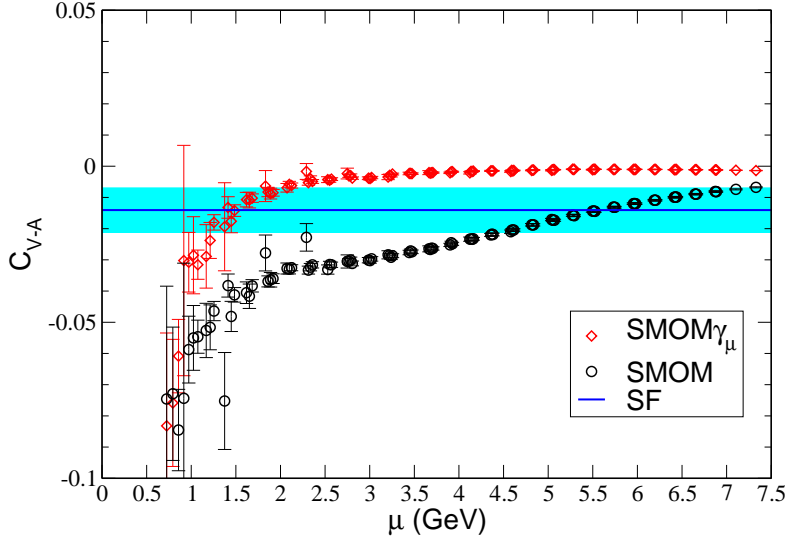}
\caption{
A plot of $c_{V-A}$ as a function of $\mu$ showing that 
the effects of the explicit chiral symmetry breaking tend to be small in high $\mu$ region.
The blue line and shaded band represent the reference value and its uncertainty given by the SF scheme. 
\label{fig:CAV}}
\end{figure*}
The results obtained from the ${\mathrm{SMOM}}$ (black circles) and ${\mathrm{SMOM}}_{\gamma_\mu}$ (red diamonds) schemes are shown in Fig.~\ref{fig:CAV}. The blue line and shaded band represent the reference value and 
its uncertainty given by the SF scheme as
\begin{align}
    c_{V-A}^{\rm SF} = \frac{Z_{V}^{\rm SF}-Z_{A}^{\rm SF}}{\left(Z_{V}^{\rm SF}+Z_{A}^{\rm SF}\right)/2} = -0.0141(71) ,
\end{align}
which is scale independent.
The sizes of $c_{V-A}$ for both results of ${\mathrm{SMOM}}$ and 
${\mathrm{SMOM}}_{\gamma_\mu}$ except for the lower $\mu$ region ($\mu<1.5$ GeV) 
are equally small compared to the SF value $c_{V-A}^{\rm SF}$ as being less than 
about 5\%. The result of ${\mathrm{SMOM}}$ shows slightly stronger dependence 
of the renormalization scale than that of ${\mathrm{SMOM}}_{\gamma_\mu}$, while the 
data of ${\mathrm{SMOM}}$ gradually 
approaches an asymptotic value of ${\mathrm{SMOM}}_{\gamma_\mu}$ at high scale as the scale $\mu$ increases.
The asymptotic value of $c_{V-A}$ obtained in the RI scheme is expected to be the same as the SF value,
which should be independent of the renormalization scheme, when $c_{V-A}$ is evaluated at vanishing quark masses. 
Recall that $c_{V-A}$ evaluated in the RI scheme is calculated at the physical point, where the light quark 
mass is small but finite. This deviation from the SF value is supposed to be contributions from explicit chiral symmetry breaking due to the finite quark mass. On the other hand, the strong scale dependence observed in low $\mu$ region 
for both ${\mathrm{SMOM}}$ and ${\mathrm{SMOM}}_{\gamma_\mu}$ should be considered as 
the infrared divergent effect due to both the explicit chiral symmetry breaking and 
spontaneous chiral symmetry breaking~\cite{Blum:2001sr}. 
These observations therefore indicate that our action indeed achieves ${\cal O}(a)$ improvement
and then discretization uncertainties are expected to be suppressed being less than a few \% according 
to the value of $c_{V-A}^{\rm SF}$. Thus, the framework of the Rome-Southampton method 
to evaluate the renormalization constants is applicable even for Wilson-type fermions being 
highly improved in this study.

\subsubsection{Residual matching scale dependence and systematic uncertainties}
\begin{figure*}
 \includegraphics[width=1.0\linewidth,bb=0 0 792 612,clip]{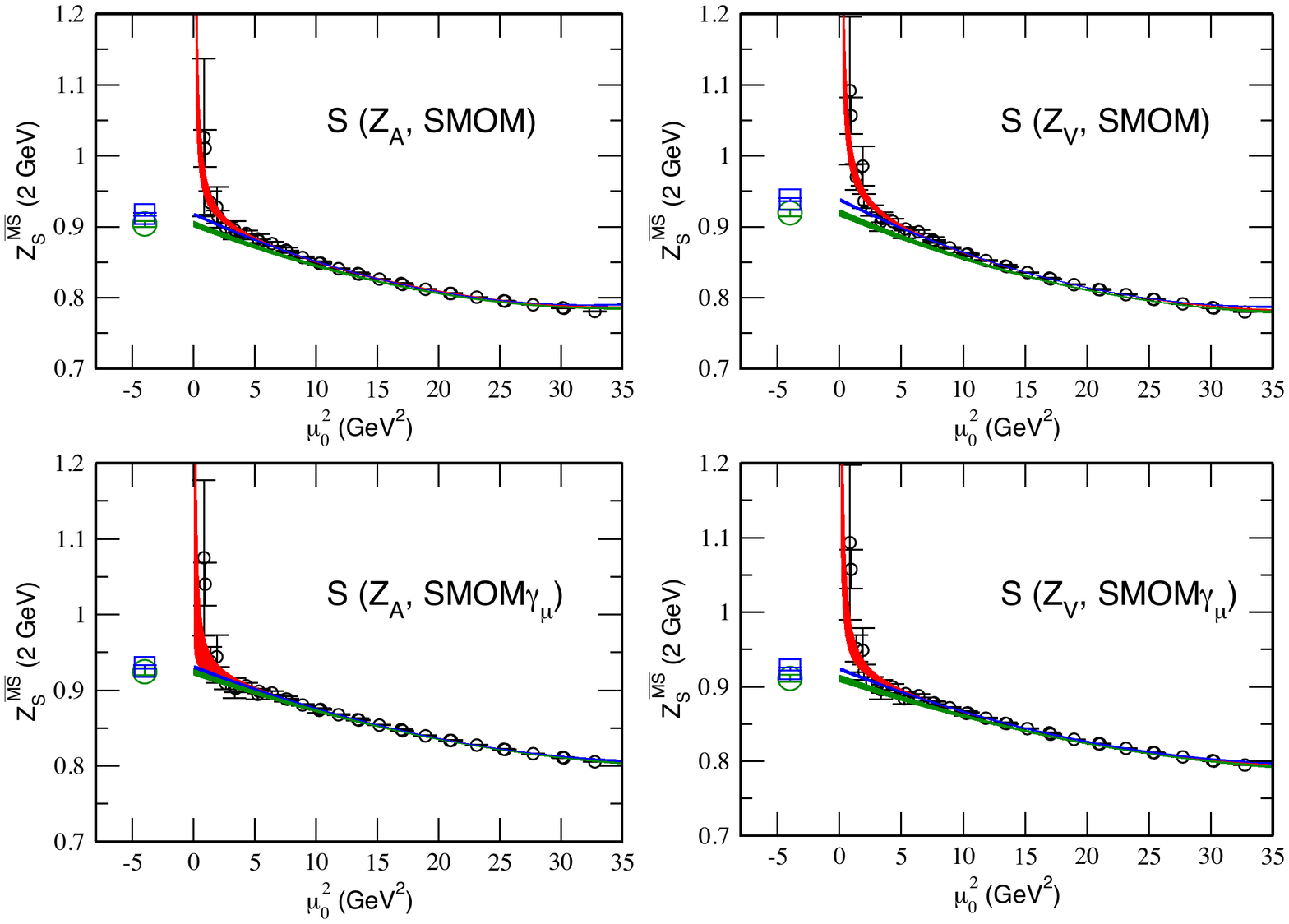}
 \caption{
The renormalization constant for the scalar current operator evaluated 
in the $\overline{\mathrm{MS}}$ scheme at a scale of 2 GeV as a function of the square of the matching scale $\mu_0$. Similar results are obtained for the four combinations of \{SF input, scheme\}=\{$Z_V$, SMOM\} (right upper panel), \{$Z_V$, SMOM$_{\gamma_\mu}$\} (right lower panel), \{$Z_A$, SMOM\} (left upper panel),  \{$Z_A$, SMOM$_{\gamma_\mu}$\} (left lower panel). In each panel, red and blue curves with error bands 
represent two fit results obtained from the global fit and the IR-truncated fit
with $k_{\rm max}=2$, while green curve is given by removing the divergent contribution from the global fit result (red curve). The blue circle and green squared symbols are the $\mu_0$-independent contributions obtained from both the global and IR-truncated fit results.
\label{fig:fitted_redults_ren_S}}
\end{figure*}

\begin{figure*}
 \includegraphics[width=1.0\linewidth,bb=0 0 792 612,clip]{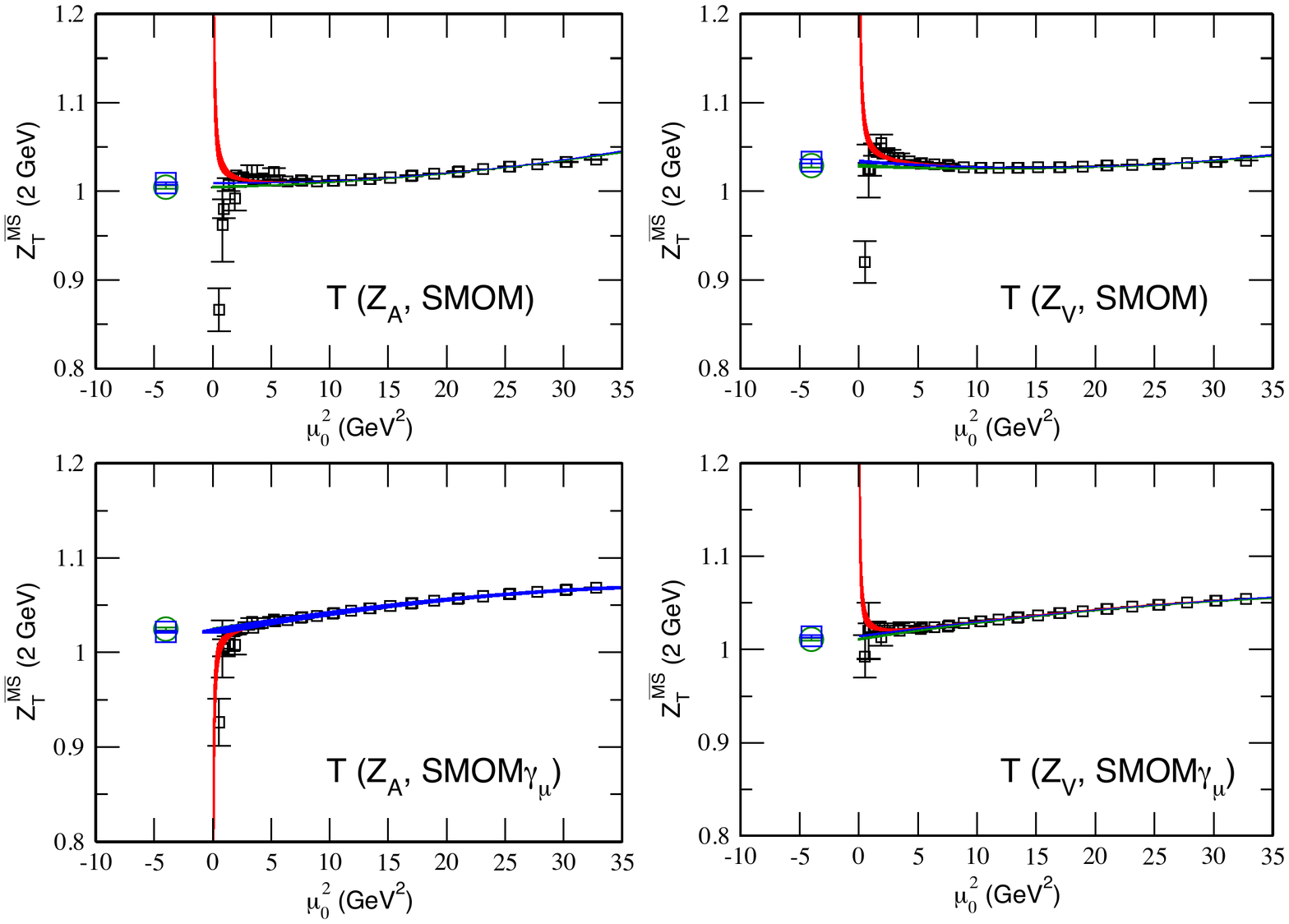}
 \caption{
The renormalization constant for the tensor current operator evaluated 
in the $\overline{\mathrm{MS}}$ scheme at a scale of 2 GeV as a function of the square 
of the matching scale $\mu_0$. The rest is the same as in Fig.~\ref{fig:fitted_redults_ren_S}.
\label{fig:fitted_redults_ren_T}}
\end{figure*}

We show the renormalization constants for the scalar and tensor current operators evaluated 
in the $\overline{\mathrm{MS}}$ scheme at the scale of 2 GeV in Figs.~\ref{fig:fitted_redults_ren_S} 
and \ref{fig:fitted_redults_ren_T}. Similar results are obtained for the four 
combinations of \{SF input, scheme\}. 
As discussed in Sec.\ref{residual_scale_dependence}, these figures clearly 
show the residual dependence 
of the choice of the matching scale $\mu_0$ in the final results 
of $Z_O^{\mathrm{\overline{MS}}} (\mu)$ 
at the renormalization scale $\mu=2$ GeV ($O=S$ or $T$), which are given by Eq.~(\ref{eq:eq_matching}) with the 
input values of $Z_O^{\mathrm{RI}}(\mu_0)$. 
We therefore evaluate the $\mu_0$-independent value 
of $Z_O^{\mathrm{\overline{MS}}} (2\;\mbox{GeV})$
by the analysis described in Sec.~\ref{methods_ren}.

In each panel of Figs.~\ref{fig:fitted_redults_ren_S} and \ref{fig:fitted_redults_ren_T},
red and blue curves with error bands represent two fit results obtained from the global fit~(\ref{eq:ren_fit_global}) 
and the IR-truncated fit~(\ref{eq:ren_fit_IRtru}) with $k_{\rm max}=2$, 
while green curve with error bands is given by removing the divergent contribution from the global fit result. 
The intersections of the green and blue curves with the $y$-axis correspond to
the $\mu_0$-independent contributions obtained from both the global 
and IR-truncated fit results. For the sake of clarity, 
we plot their central values and statistical errors denoted as the circle and 
squared symbols, which are slightly shifted from the $y$-axis in each panel of Figs.~\ref{fig:fitted_redults_ren_S} and \ref{fig:fitted_redults_ren_T}.
Both results obtained from two different fit procedures, where 
the unwanted infrared divergence is treated in a different way, 
coincide each other with the optimal choices of the fit range 
of $[\mu_{\rm min}$:$\mu_{\rm max}]$ for each fit as will be discussed as follows.

%
%

%
%

\begin{figure*}
 \includegraphics[width=1\linewidth,bb=0 0 792 612,clip]{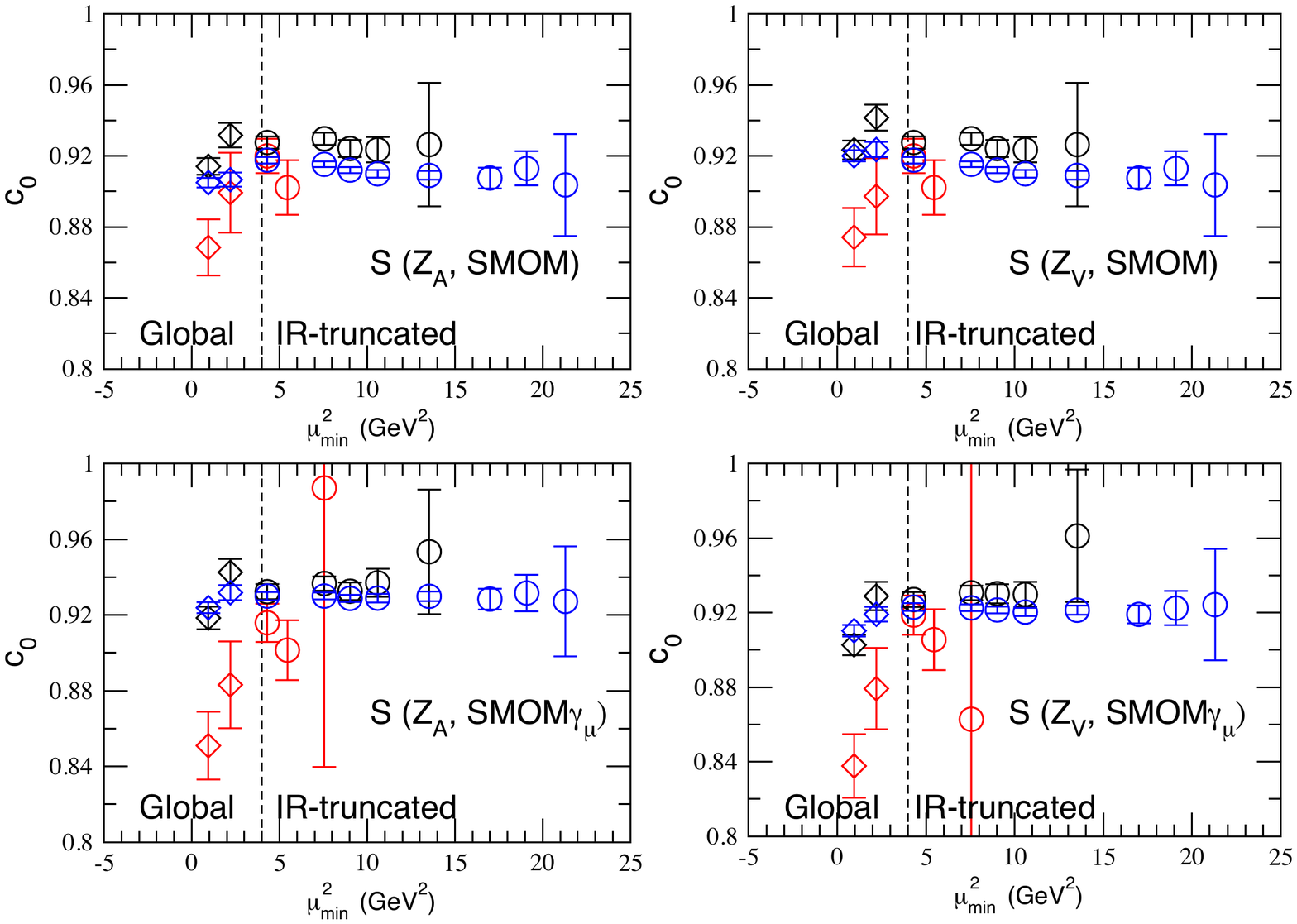}
 \caption{
 Stability test on the $\mu_0$-independent contribution $c_0$ for the scalar channel
 with variation of $\mu_{\mathrm{min}}$ under the choices 
 of $\mu_{\mathrm{max}}=$3 (red),
 4 (black) and 5 (blue) GeV.
 Diamond symbols appearing in the left side of the vertical dashed line are given by 
 the global fit, while circle symbols appearing in the right side of the vertical dashed line
 are given by the IR-truncated fit results. The choice of $k_{\mathrm{max}}=2$ is 
 adopted for the polynomial terms in both functional forms used in the global fit and the IR-truncated fit. 
 \label{fig:mumin_stab_ren_S}}
\end{figure*}

\begin{figure*}
 \includegraphics[width=1\linewidth,bb=0 0 792 612,clip]{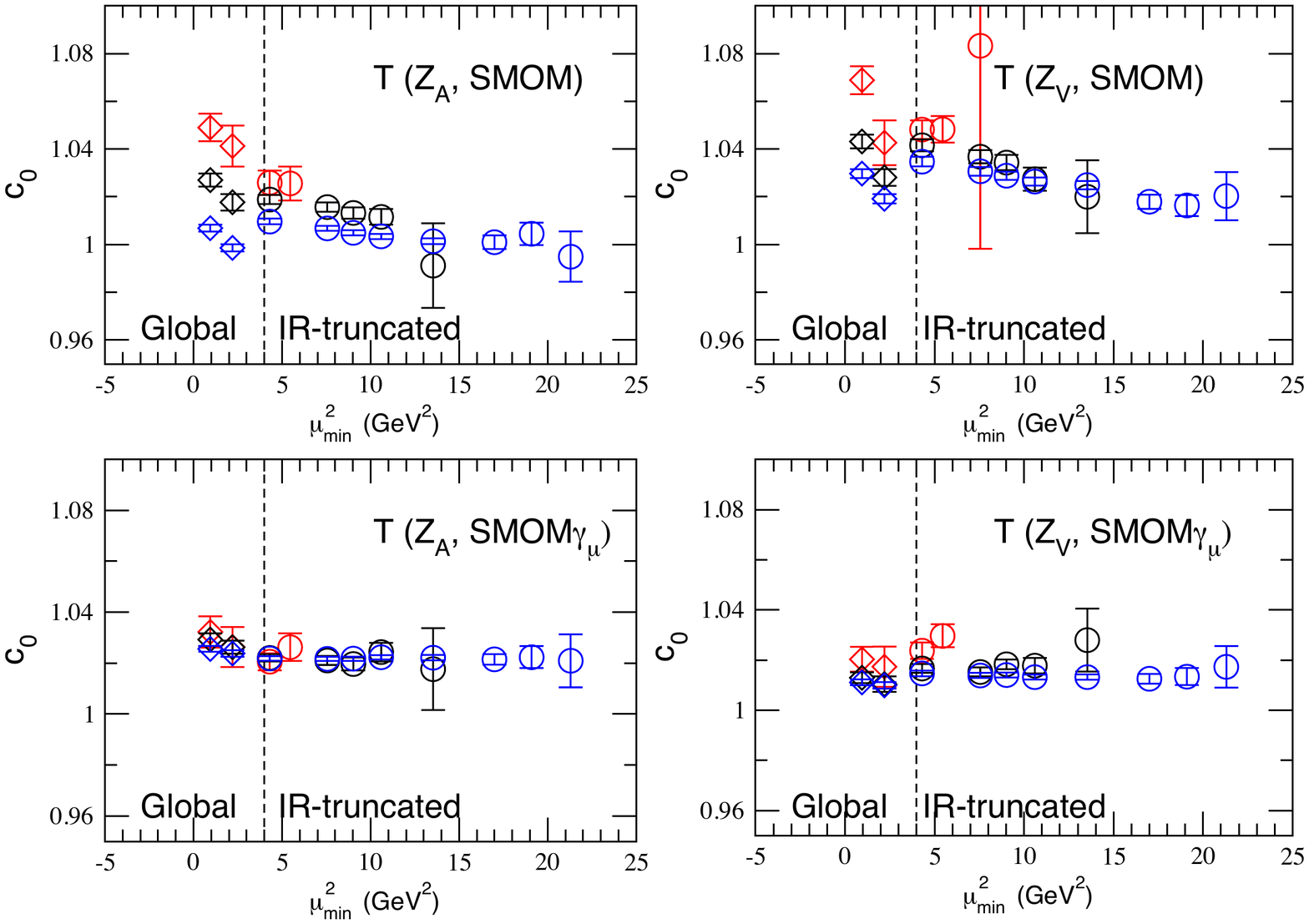}
 \caption{
 Stability test on the $\mu_0$-independent contribution $c_0$ for the tensor channel
 with variation of $\mu_{\mathrm{min}}$ under the choices 
 of $\mu_{\mathrm{max}}=3$ (red),
 4 (black) and 5 (blue) GeV.
 The rest is the same as in Fig.~\ref{fig:mumin_stab_ren_S}.
 \label{fig:mumin_stab_ren_T}}
\end{figure*}

First, we check the stability of $c_0$, which is the $\mu_0$-independent term obtained from either
the global fit with Eq.~(\ref{eq:ren_fit_global}) or the IR-truncated fit with Eq.~(\ref{eq:ren_fit_IRtru}), 
with variation of $\mu_{\mathrm{min}}$ under the condition of $k_{\mathrm{max}}=2$ with three values of $\mu_{\mathrm{max}}=3$, 4 and 5 GeV as shown in Figs.~\ref{fig:mumin_stab_ren_S} and \ref{fig:mumin_stab_ren_T}. 
Figure~\ref{fig:mumin_stab_ren_S} shows the results for $Z_S^{\overline{\mathrm{MS}}}(2\;\mbox{GeV})$ in the cases of the four combinations 
of \{SF input, scheme\} as a function of $\mu_{\mathrm{min}}$, 
while Fig.~\ref{fig:mumin_stab_ren_T} presents the results for $Z_T^{\overline{\mathrm{MS}}}(2\;\mbox{GeV})$. 
The diamond and circle symbols are obtained from the global fit and the IR-truncated fit, respectively.

The former fit results are restricted in the range of $\mu_{\mathrm{min}}\le 2$ GeV (left side of vertical dashed line), while the latter fit results are given with the condition of $\mu_{\mathrm{min}}> 2$ GeV (right side of vertical dashed line).
The red, black and blue symbols represent the results given with the fixed value of 
$\mu_{\mathrm{max}}=$3, 4, and 5 GeV, respectively. 
At a glance, the $c_0$ value is sensitive to the choice of $\mu_{\mathrm{min}}$ 
in the case of $\mu_{\mathrm{max}}=3$ GeV with $k_{\mathrm{max}}=2$.
However, the larger values of $\mu_{\mathrm{max}}$ tend to ensure the stability 
of the $c_0$ value
since the $\mu_0$-dependences of $Z_{S}^{\overline{\mathrm{MS}}}(2\;\mbox{GeV})$ and 
$Z_{T}^{\overline{\mathrm{MS}}}(2\;\mbox{GeV})$ at higher $\mu_0$ can be well described by 
a quadratic function of $(a\mu_0)^{2}$. Furthermore, the resulting $c_0$ value does not change so much
even changing from $\mu_{\mathrm{max}}=4$ GeV (black symbols) 
to $\mu_{\mathrm{max}}=5$ GeV (blue symbols). 
This suggests that the choice of $k_{\mathrm{max}}=2$ is 
enough to cope with the $\mu_0$-dependence appearing at higher $\mu_0$, 
which is caused by the lattice discretization artifacts, up to $\mu_0=5$ GeV. 
Once the ultraviolet behavior is well controlled by the polynomial terms
with the optimal choice of $k_{\mathrm{max}}$, both fit procedures 
lead to compatible results, especially for the ${\mathrm{SMOM}}_{\gamma_\mu}$ cases 
as shown in Figs.~\ref{fig:mumin_stab_ren_S} and \ref{fig:mumin_stab_ren_T}.
We therefore choose the RI/SMOM$_{\gamma_\mu}$ scheme and the SF input of 
$Z_V$ for evaluating the final results of $Z_{S}^{\overline{\mathrm{MS}}}(2\;\mbox{GeV})$ 
and $Z_{T}^{\overline{\mathrm{MS}}}(2\;\mbox{GeV})$, 
and also quote the systematic uncertainties (denoted as a subscript of ``scheme") 
which are determined from the maximum difference among the four 
combinations of \{SF input, scheme\}.

%
%

%
%

According to the aforementioned discussion, we consider three types of the choices of 
$k_{\mathrm{max}}$ and the fit range of $[\mu_{\mathrm{min}}:\mu_{\mathrm{max}}]$ 
as a candidate of the best choice for the global fit.
Our best estimate of $Z_{S}^{\overline{\mathrm{MS}}}(2\;\mbox{GeV})$ 
and $Z_{T}^{\overline{\mathrm{MS}}}(2\;\mbox{GeV})$ obtained from the combination of
\{$Z_V$, SMOM$_{\gamma_\mu}$\} are summarized in Tab.~\ref{tab:candidate_ren}. 
The first error denotes the statistical error, while the second error represents
the systematic error determined from the maximum difference among the four 
combinations of \{SF input, scheme\}.
The third error represents the systematic error determined by the size of the deviation 
from the IR-truncated fit result with the same choice of $k_{\mathrm{max}}$
(denoted as a subscript of ``model").  For the tensor case, all results obtained 
from three different types of the global fit are consistent with each other 
within their statistical accuracy, while a mutual consistency in the scalar case is assured 
if the systematic uncertainties are taken into account.
The results from the Type C case have smaller uncertainties on two systematic
errors in both the scalar and tensor channels. We therefore choose the results obtained
from the Type C case as our central values of 
$Z_{S}^{\overline{\mathrm{MS}}}(2\;\mbox{GeV})$ 
and $Z_{T}^{\overline{\mathrm{MS}}}(2\;\mbox{GeV})$ and quote
the third systematic error that is determined from the maximum difference among
above three results (denoted as a subscript of ``fit") according
to our guidelines in the determination of the $\mu_0$-independent term 
as discussed in Sec.~\ref{methods_ren}.

%
%
 \begin{table}[ht]
     \caption{
     Results of the renormalization constants 
     $Z_S^{\overline{\mathrm{MS}}}(2\;\mbox{GeV})$ and
     $Z_T^{\overline{\mathrm{MS}}}(2\;\mbox{GeV})$ 
     with the RI/SMOM$_{\gamma_\mu}$ intermediate scheme. 
     The obtained values are evaluated by
     three types of the global fit that satisfy all our 
     guidelines to select the appropriate fit range of 
     $[\mu_{\mathrm{min}}:\mu_{\mathrm{max}}]$ with 
     a smaller value of $k_{\rm max}$ as discussed in Sec.~\ref{methods_ren}.
     The first error denotes the statistical one, while the second and
     third errors represent the systematic uncertainties in the determination
     of the $\mu_0$-independent contribution as described in the text. 
     }
     \label{tab:candidate_ren}
     \renewcommand{\arraystretch}{1.5}
     \begin{ruledtabular}
     \begin{tabular}{ccccccc}
        Type & Fit range [GeV] & $k_{\rm max}$ 
       & $Z_S^{\overline{\mathrm{MS}}}(2\;\mbox{GeV})$ 
       & $\chi^2/{\mathrm{dof}}$ 
       & $Z_T^{\overline{\mathrm{MS}}}(2\;\mbox{GeV})$ 
       & $\chi^2/{\mathrm{dof}}$ \\
       \hline
A  &   [1:4] & 1 &  0.8966(23)(136)(156) & 1.05(34)
				 &  1.0143(9)(186)(20)   & 0.34(13) \\
B  &   [1:4] & 2 & 0.9027(55)(207)(243)      & 1.03(37) 
                 &  1.0129(21)(303)(37)  & 0.33(14) \\
C  &   [1:5] & 2 &   0.9103(31)(136)(127) & 0.92(31) 
                 &   1.0111(12)(186)(36)  & 0.30(13) \\
   \end{tabular}
   \end{ruledtabular}
\end{table}

Hence, our final results of the renormalization constants for the scalar and tensor current operators are
%
%
\begin{eqnarray}
    Z_S^{\overline{\mathrm{MS}}}(2\;\mbox
    {GeV}) & = & 0.9103(31)_{\mathrm{stat}}(136)_{\mathrm{scheme}}(127)_{\mathrm{model}}(137)_{\mathrm{fit}}\nonumber\\
    & =&  0.9103(31)_{\mathrm{stat}}(231)_{\mathrm{syst}}\\
    Z_T^{\overline{\mathrm{MS}}}(2\;\mbox
    {GeV})& = & 
      1.0111(12)_{\mathrm{stat}}(186)_{\mathrm{scheme}}(36)_{\mathrm{model}}(32)_{\mathrm{fit}}\nonumber\\
    & = &
      1.0111(12)_{\mathrm{stat}}(192)_{\mathrm{syst}} .
\end{eqnarray}

On these renormalization constants, the error budget is compiled in Tab.~\ref{tab:err_src_ren}. 
All errors of the renormalization constants become small enough 
to be comparable to the precision that is reached for the bare couplings with the statistical errors at the 1-2\% level as tabulated in Tab.~\ref{tab:bare_couplings}, since we use 
the RI/SMOM$_{(\gamma_\mu)}$ intermediate scheme that can considerably reduce the 
systematic errors compared to the case of the RI/MOM scheme.

%
%
 \begin{table}[ht]
     \caption{     
     The error budget for $Z_S^{\overline{\mathrm{MS}}}(2\;\mbox{GeV})$ 
     and $Z_T^{\overline{\mathrm{MS}}}(2\;\mbox{GeV})$ 
     with the RI/SMOM$_{\gamma_\mu}$ intermediate scheme.      
}
     \label{tab:err_src_ren}
     \renewcommand{\arraystretch}{1.5}
     \begin{ruledtabular}
     \begin{tabular}{llcc}
        & & Scalar ($Z_S$) & Tensor ($Z_T$) \\
       \hline
       \multicolumn{2}{l}{Statistical:}     & 0.34\% &  0.12\% \\
       Systematical: & Choice of scheme (``scheme")    & 1.49\% &   1.84\% \\
       & Choice of fit model  (``model") & 1.40\% &   0.36\% \\
       & Variation of fit (``fit") &   1.50\% &   0.32\% \\
       \hline
       \multicolumn{2}{l}{Total:}           &   2.56\% &   1.90\% \\
     \end{tabular}
     \end{ruledtabular}
   \end{table}
%

\subsection{Renormalized coupling}

The renormalized value of the coupling can be evaluated by combining the value of the bare coupling with the renormalization constant as $Z_{O}\times g^{\rm bare}_{O}$. The renormalized couplings $g_{O}$ for $O=V, A, S, T$ 
obtained from both $L=128$ and $L=64$ calculations are summarized in Tab.~\ref{tab:ren_coup_AVST}. 
The scalar and tensor couplings ($g_S$ and $g_T$) are renormalized in the $\overline{\mathrm{MS}}$ scheme 
at the renormalization scale of 2 GeV, while the vector and axial-vector couplings are renormalized with
$Z_{V}=0.95153(76)$ and $Z_{A}=0.9650(68)$ obtained with the SF scheme. 
The jackknife method is used to estimate the statistical error. For the scalar and tensor couplings, 
both the statistical and systematic errors are quoted. The systematic error stems from the determination of the renormalization constant in the Rome-Southampton method as discussed in Sec.~\ref{methods_ren}.
%
%
 \begin{table}[ht]
     \caption{      
     Results of the renormalized couplings $g_{O}$ for $O=V, A, S, T$ obtained 
     from both $128^4$ and $64^4$
     lattices with different choices of the source-sink separation ($t_{\mathrm{sep}}$) 
     and the smearing type.
     The scalar and tensor couplings ($g_S$ and $g_T$) are renormalized 
     in the $\overline{\mathrm{MS}}$ scheme 
     at the renormalization scale of $\mu=2$ GeV. The second errors on $g_S$ and $g_T$
     are given by the total systematic uncertainties for the determination of the
     renormalization constants. 
     \label{tab:ren_coup_AVST}}
     \renewcommand{\arraystretch}{1.3}
	\begin{ruledtabular}
     \begin{tabular}{ccccccc}
       $L$&$t_{\mathrm{sep}}$ & Smearing-type 
       & $g_V$& $g_A$ & $g_S$& $g_T$\\
       \hline
       128 &10 & Exp.(I) 
       & 0.997(1)& 1.276(26) & 0.545(180)$_{\rm stat}$(13)$_{\rm syst}$&   1.120(14)$_{\rm stat}$(21)$_{\rm syst}$\\
               &12 &                      
               & 0.997(2)& 1.278(24) & 0.743(134)$_{\rm stat}$(17)$_{\rm syst}$&   1.036(22)$_{\rm stat}$(20)$_{\rm syst}$ \\
               &14 &                      
               & 1.000(2)& 1.275(38) & 0.950(321)$_{\rm stat}$(22)$_{\rm syst}$&    0.988(37)$_{\rm stat}$(19)$_{\rm syst}$\\
               &16 &                      
               & 0.997(4)& 1.309(60) & 0.859(306)$_{\rm stat}$(20)$_{\rm syst}$&   1.019(57)$_{\rm stat}$(19)$_{\rm syst}$\\
               \cline{2-7}
               &13 & Gauss     
               & 0.998(1)& 1.209(34) & 0.713(140)$_{\rm stat}$(17)$_{\rm syst}$&   1.044(20)$_{\rm stat}$(20)$_{\rm syst}$  \\

               &16 &          
               & 0.993(2)& 1.267(24) & 0.381(296)$_{\rm stat}$(9)$_{\rm syst}$&   0.988(55)$_{\rm stat}$(19)$_{\rm syst}$  \\
       \hline
       64  &11 & Exp.(I)  
       & 0.999(7)& 1.230(18) & 1.313(380)$_{\rm stat}$(31)$_{\rm syst}$&   1.063(21)$_{\rm stat}$(20)$_{\rm syst}$\\
               &14 &                       
               & 1.010(7)& 1.299(35) & 1.055(360)$_{\rm stat}$(25)$_{\rm syst}$&   1.013(27)$_{\rm stat}$(19)$_{\rm syst}$\\
               \cline{2-7}
               &12 & Exp.(II) 
               & 1.011(13)& 1.246(12)& 1.003(116)$_{\rm stat}$(23)$_{\rm syst}$&   1.027(9)$_{\rm stat}$(20)$_{\rm syst}$\\

               &14 &                       
               & 0.997(4)& 1.292(15) & 0.972(153)$_{\rm stat}$(23)$_{\rm syst}$&   1.054(9)$_{\rm stat}$(20)$_{\rm syst}$\\
               &16 &                      
               & 0.998(11)& 1.263(37) & 1.398(397)$_{\rm stat}$(33)$_{\rm syst}$&   0.966(32)$_{\rm stat}$(18)$_{\rm syst}$\\
     \end{tabular}
     \end{ruledtabular}
\end{table}

\subsubsection{Axial-vector couplings}

\begin{figure*}
\includegraphics[width=0.7\linewidth,bb=0 0 792 612,clip]{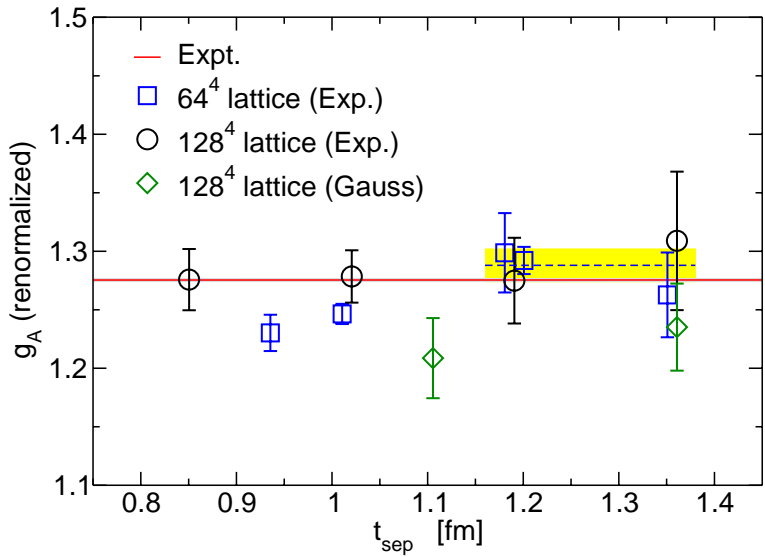}
\caption{ 
    $t_{\mathrm{sep}}$ dependence of the renormalized values of $g_A$. 
    The horizontal axis denotes the source-sink separation 
    $t_{\mathrm{sep}}$ in physical unit. 
    Square symbols for the $L=64$ calculations have been slightly 
    shifted in horizontal direction if they need to avoid overlap. 
    The solid line denotes the experimental value, while
    the dotted line represents the average value, and shaded band displays
    the fit range and one standard deviation. 
\label{fig:ren_axi}}
\end{figure*}

Figure~\ref{fig:ren_axi} shows $t_{\mathrm{sep}}$ dependence on the renormalized
value of the axial-vector coupling $g_A$. In the case of the exponential 
smeared source, our results of $g_A$
obtained with $t_{\mathrm{sep}}=\{12, 14, 16\}$ for the $L=128$ lattice
and with $t_{\mathrm{sep}}=\{14, 16\}$ for the $L=64$ lattice agree well
with the experimental value, $1.2754(13)$~\cite{ParticleDataGroup:2020ssz}, 
while only one result obtained with $t_{\mathrm{sep}}=16$ for the Gaussian 
smeared source can reproduce the experimental value. 
The latter smearing results are only obtained with two different values of $t_{\mathrm{sep}}$, so that the details of $t_{\mathrm{sep}}$ dependence are not clear. 
{However, at least, the condition of $t_{{\mathrm{sep}}}\ge14$, corresponding to $t_{{\mathrm{sep}}}\gtrsim1.2\ \mathrm{fm}$, 
is large enough to eliminate the 
excited-state contaminations regardless of the smearing type that we used in this study. 
In addition, it is certainly confirmed that systematic uncertainties stemming from the
finite-size effect are negligible at the level of the statistical precision of 1.9\%
in two large spatial extents of about 10 and 5 fm. }

Taking into account these observations, we determine the central value of $g_A$ from several combinations that can be made by selecting the data set and imposing restrictions on $t_{\mathrm{sep}}$. In Tab.~\ref{tab:ren_coup_AVST_comb}, we list four different combined results by using only data calculated with the exponential smeared source. The listed mean values are first computed by a weighted average of the selected data with a covariance matrix accounting for correlation on each lattice ensemble. As for the combined analysis with both $128^4$ and $64^4$ lattices, we simply take an average of two combined results obtained from both $128^4$ and $64^4$ lattices with respect to their statistical uncertainties, which are evaluated by the jackknife method~\footnote{We also perform the
super-jackknife analysis~\cite{{CP-PACS:2001vqx},{LHPC:2010jcs}}, 
but the results mostly remain same within their errors since the $L=128$ and $L=64$ lattice ensembles are independent.}.
Although inclusion of the data obtained with $t_{\mathrm{sep}}=12$ in the $L=64$ calculations slightly reduces the mean value, all four results are in good agreement 
with each other. 
We prefer to choose the combined value of $g_A$ with $t_{\mathrm{sep}}=\{14, 16\}$ 
from both $L=128$ and $L=64$ calculations as our best estimate of $g_A$. 

Our final result of $g_A$ in this study is 
\begin{align}
\label{eq:ren_axi}
g_A & = 1.288(14)_{\rm stat}(9)_{Z_A}, 
\end{align}
where we also include a systematic error stemming from the error of $Z^{\mathrm{SF}}_A$. 
This result reproduces the experimental value at the precision level of 1\%.

%
%
\begin{table}
     \caption{      
     Results of the renormalized couplings, $g_A$, $g_S$ and $g_T$ from the combined analysis
     with different data selections for lattice volumes and $t_{\mathrm{sep}}$.  
     \label{tab:ren_coup_AVST_comb}}
     \begin{ruledtabular}
     \begin{tabular}{lccccc}
	 data set & $t_{\mathrm{sep}}$ & \# of data & $g_A$ & $g_S$ & $g_T$ \cr
	 \hline
     $64^4$ & $\{14, 16\}$ &	3 &1.288(16) 
     & $1.021(142)_{\mathrm{stat}}(24)_{\mathrm{syst}}$ 
     &   $1.042(9)_{\mathrm{stat}}(20)_{\mathrm{syst}}$ \cr
     $128^4$ (only Exp.)& $\{12, 14, 16\}$&  3 &1.280(24) 
     & $0.767(141)_{\mathrm{stat}}(18)_{\mathrm{syst}}$
     &   $1.020(22)_{\mathrm{stat}}(20)_{\mathrm{syst}}$ \cr
     $64^4+128^4$ (only Exp.)  & $\{14, 16\}$ & 5 & 1.288(14) 
     & $0.894(100)_{\mathrm{stat}}(21)_{\mathrm{syst}}$ 
     &   $1.038(9)_{\mathrm{stat}}(20)_{\mathrm{syst}}$	\cr
     $64^4+128^4$ (only Exp.)  & $\{12, 14, 16\}$& 7 & 1.267(12) 
     & $0.927(83)_{\mathrm{stat}}(22)_{\mathrm{syst}}$ 
     &   $1.036(6)_{\mathrm{stat}}(20)_{\mathrm{syst}}$ \cr
     \end{tabular}
     \end{ruledtabular}
\end{table}

%
%

\subsubsection{Scalar and tensor couplings}
 
\begin{figure*}
\includegraphics[width=0.49\linewidth,bb=0 0 792 612,clip]{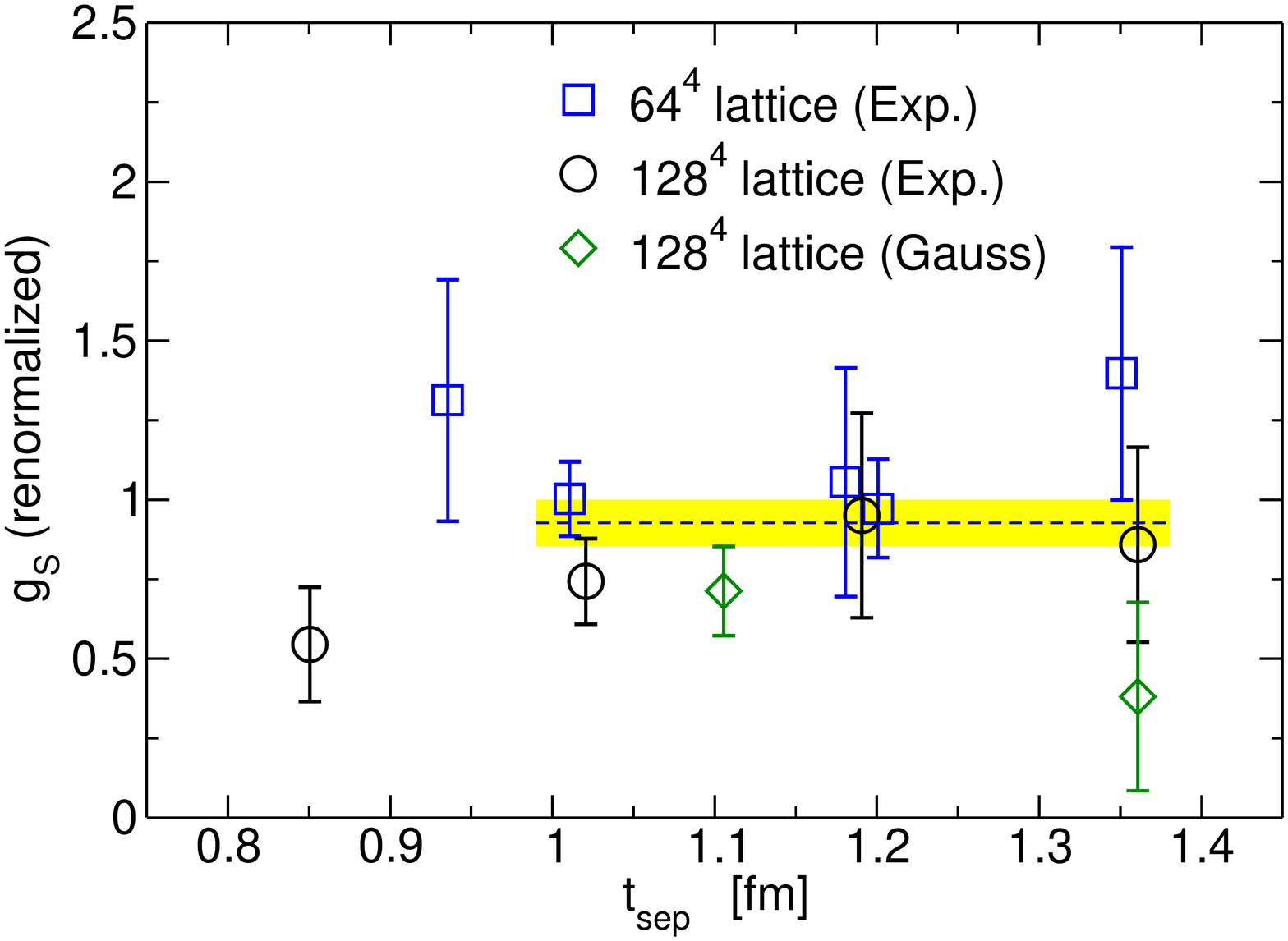}
\includegraphics[width=0.49\linewidth,bb=0 0 792 612,clip]{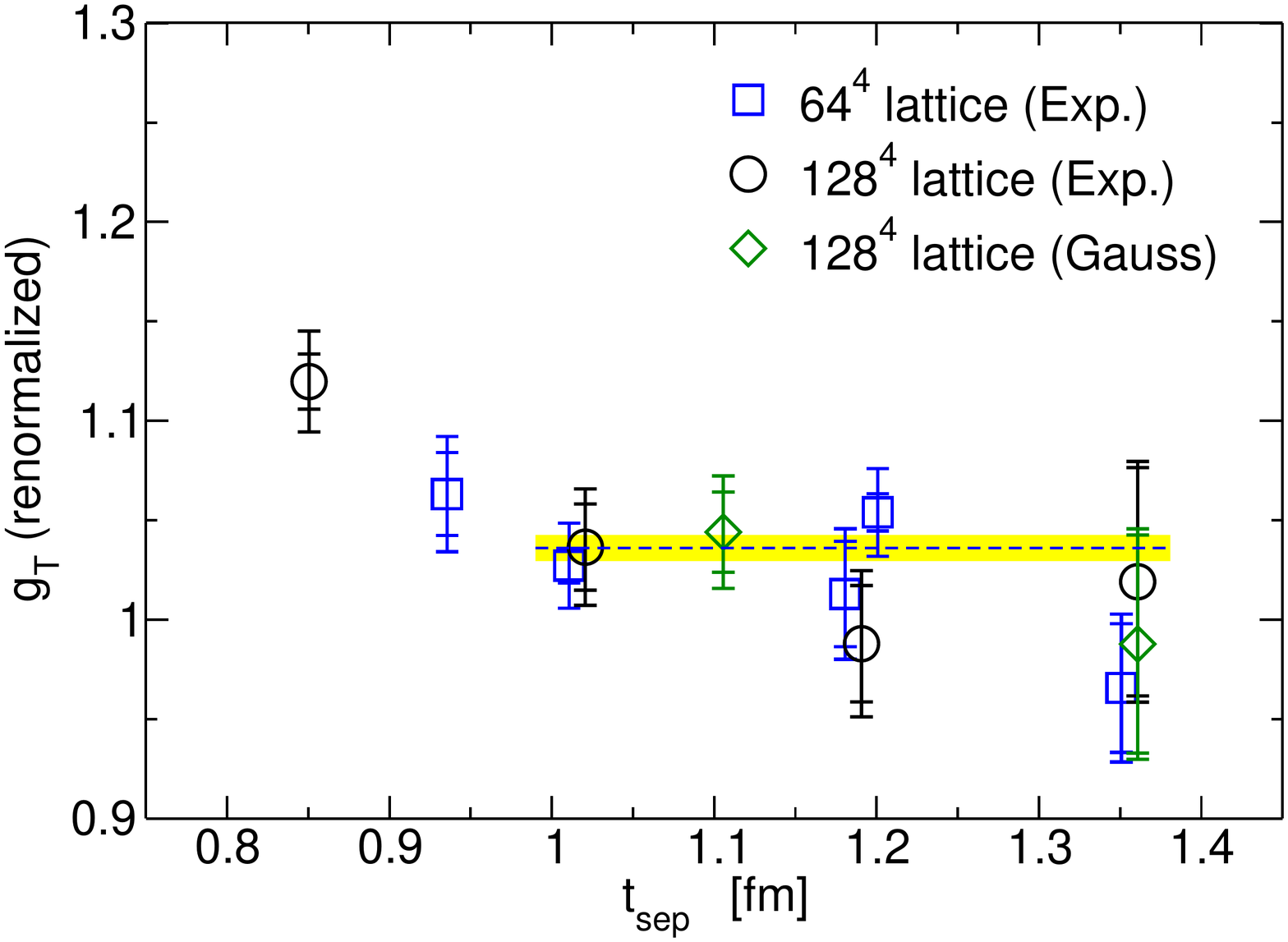}
\caption{ 
    $t_{\mathrm{sep}}$ dependence of the renormalized values of $g_S$ (left) and $g_T$ (right). 
    The horizontal axis denotes the source-sink separation $t_{\mathrm{sep}}$ in physical unit. 
    The inner and outer error bars represent their statistical and total uncertainties, respectively.
    The rest is the same as in Fig.~\ref{fig:ren_axi}.
\label{fig:ren_scaten}}
\end{figure*}

In Fig.~\ref{fig:ren_scaten}, we show the results for the renormalized values of
$g_S$ (left) and $g_T$ (right), which are obtained with several different choices 
of $t_{\mathrm{sep}}$ on two lattice volumes. In both cases of $g_S$ and $g_T$,
there is no distinct $t_{\mathrm{sep}}$ dependence in the region of 
$t_{\mathrm{sep}}\ge 12$, corresponding to $t_{\mathrm{sep}}\gtrsim 1.0$ fm~\footnote{
Here, it is worth mentioning that the choice of the wider range of $t_{\mathrm{sep}}$ than that of $g_A$ 
is based on the fact that two results from the gaussian smeared source 
with $t_{\mathrm{sep}}=\{13, 16\}$ are consistent within the statistical errors 
for $g_S$ and $g_T$.}.
This observation ensures that systematic uncertainties stemming from 
the excited-state contaminations are well under control for $t_{\mathrm{sep}}\ge 12$ with both smeared methods.
Therefore, similar to the case of $g_A$, we get the combined results obtained from 
four types of data selection as summarized in Tab.~\ref{tab:ren_coup_AVST_comb}.
All four results of $g_T$ are fairly consistent with each other, while the scalar case  shows a rough consistency within the larger statistical errors. 
We then choose the combined values of $g_S$ and $g_T$ with $t_{\mathrm{sep}}=\{
12, 14, 16\}$ from both $L=128$ and $L=64$ calculations. 
As will be discussed in Appendix~\ref{app:summation_method}, the values of $g_S$ and $g_T$ 
determined by the summation method with data sets of $t_{\mathrm{sep}}=\{12, 14, 16\}$ are barely consistent with the above estimations, though the results from the summation method are much larger uncertainties than those of the plateau method. Furthermore, the two-state fit analysis, which is only applied to $g_T$, does not lead to a major discrepancy beyond the statistical precision.

Our best estimates are 
%
%
\begin{align}
    \label{eq:ren_scaten}
    g_S & = 0.927(83)_{\rm stat}(22)_{Z_S} & {\rm and }& & 
    g_T & = 1.036(6)_{\rm stat}(20)_{Z_T} ,
\end{align}
where the second errors are given by the total systematic uncertainties for 
the determination of the renormalization constants. 

\begin{figure*}
\includegraphics[width=0.49\textwidth,bb=0 0 612 612,clip]{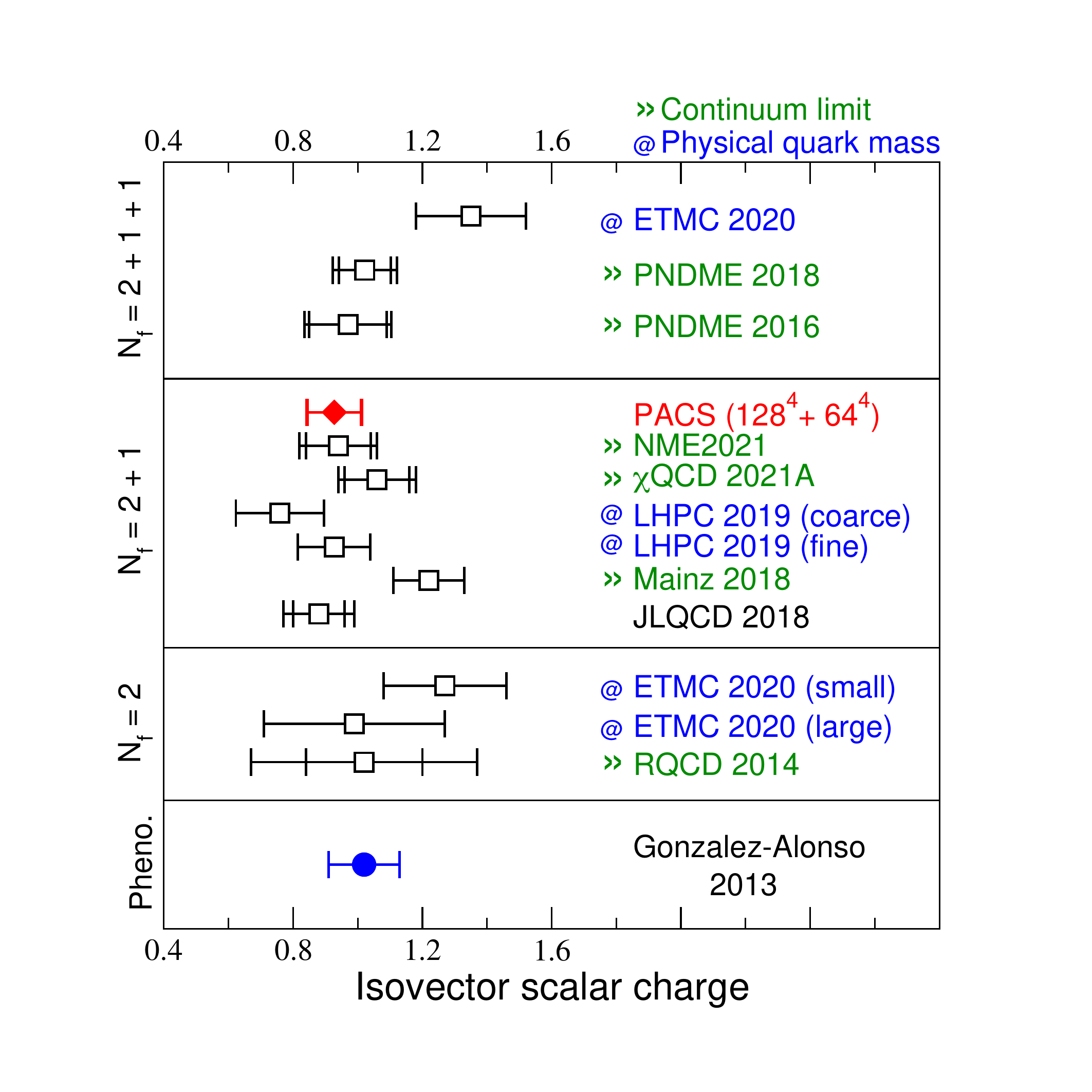}
\includegraphics[width=0.49\textwidth,bb=0 0 612 612,clip]{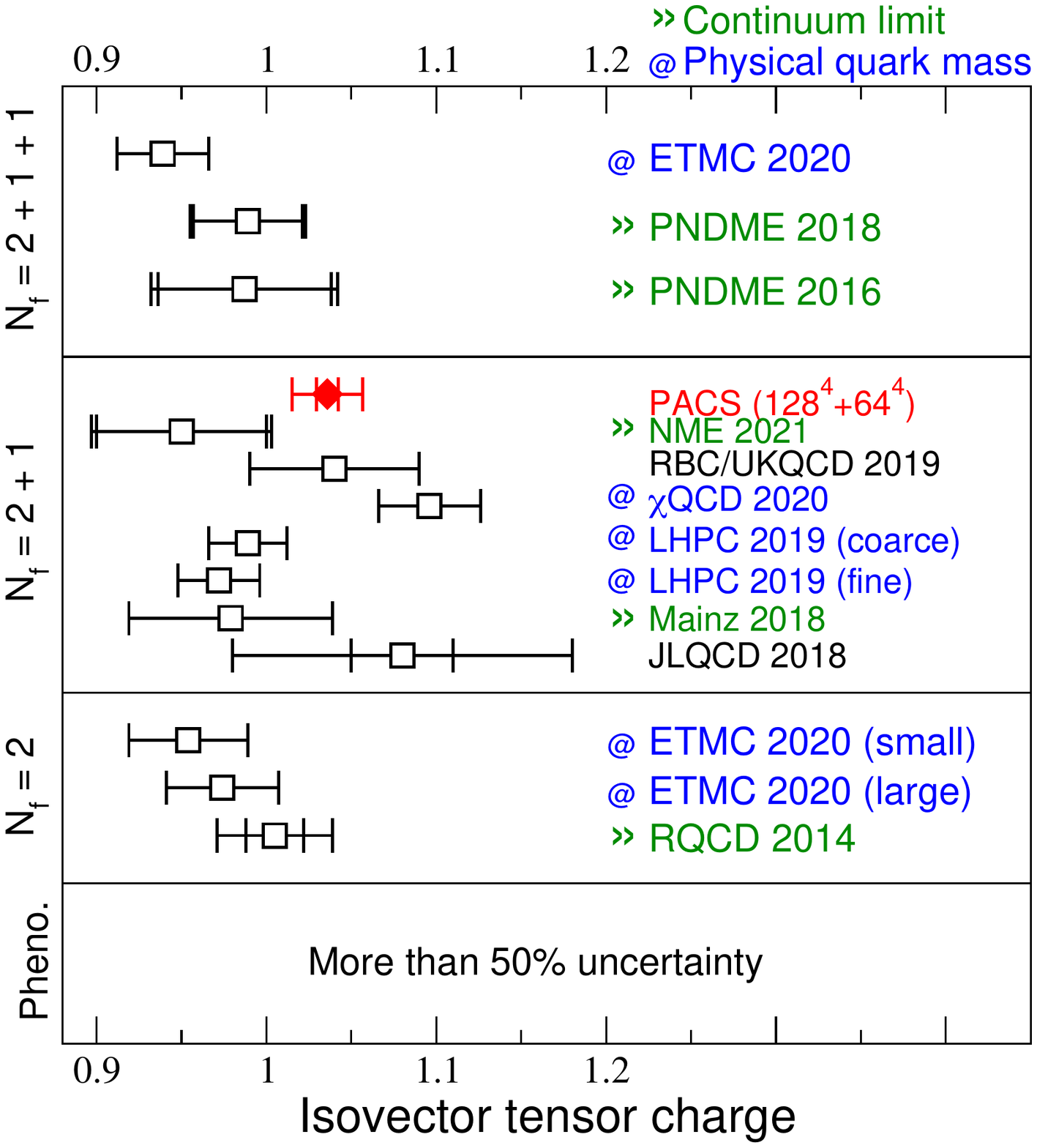}
 \caption{  
  Comparison of our results (red diamonds) with the other lattice results (black squares)~\cite{{Aoki:2021kgd},{Park:2021ypf},{Horkel:2020hpi},{Gupta:2018qil},{Bhattacharya:2016zcn},{Alexandrou:2019brg},{Hasan:2019noy},{Ottnad:2018fri},{Yamanaka:2018uud},{Bali:2014nma}} 
  and the phenomenological value (blue circle)~\cite{Gonzalez-Alonso:2013ura} 
for $g_S$ (left panel) and $g_T$ (right panel). 
The inner error bars represent the statistical uncertainties, while the outer ones represent the total uncertainties given 
by adding the statistical and systematic errors in quadrature. Blue labels indicate that the analysis includes 
the data from lattice QCD simulation near the physical point, while green labels indicate that the continuum extrapolation
is achieved.
}
 \label{fig:comparison}
\end{figure*}

We finally compare our results of the renormalized values of $g_S$ (left) and $g_T$ (right) together 
with results from the recent lattice QCD calculations ~\cite{{Aoki:2021kgd},{Park:2021ypf},{Horkel:2020hpi},{Gupta:2018qil},{Bhattacharya:2016zcn},{Alexandrou:2019brg},{Hasan:2019noy},{Ottnad:2018fri},{Yamanaka:2018uud},{Bali:2014nma}} in Fig.~\ref{fig:comparison}. 
Remark that our results are obtained solely from the physical point simulations which suffer from large statistical fluctuations, while the other lattice results are given by the combined data that includes the data taken from simulations at the heavier pion masses. However, the statistical and total errors on our results are comparable to the other lattice results, since we use the AMA method for the bare matrix elements and the RI/SMOM$_{(\gamma_\mu)}$ scheme 
for the renormalization, both of which reduce the statistical and systematic uncertainties on the final results. 
As for $g_S$, our result is consistent with the trend of the other results. On the other hand, our result of $g_T$ 
locates slightly higher than other continuum results (green labels), though this discrepancy would be caused by the discretization uncertainty that is not yet accounted in our calculations.

%
%
%

%
%

\section{Summary}
 
We have calculated the renormalized values of the nucleon isovector-couplings 
in the axial-vector, scalar and tensor channels using 2+1 flavor lattice QCD with physical light quarks. 
The calculations are carried out with the gauge configurations generated by the PACS Collaboration 
with the stout-smeared $O(a)$ improved Wilson quark action 
and Iwasaki gauge action~\cite{Iwasaki:1983iya} at a single lattice spacing of $a\approx0.085$ fm
on the two lattice volumes (linear spatial extents of 10.9 fm and 5.5 fm).
In order to achieve high-precision and high-accuracy determination, we employ the AMA technique 
which can reduce the statistical error significantly, and the RI/SMOM$_{(\gamma_\mu)}$ scheme 
which keeps the systematic error under control in the determination of 
$Z_S^{\overline{\mathrm{MS}}}(2\;\mbox{GeV})$ and $Z_T^{\overline{\mathrm{MS}}}(2\;\mbox{GeV})$.

We have confirmed that systematic uncertainties stemming from the finite-size effects are negligible 
in two large spatial extents of about 10 and 5 fm.
The effects of the excited-state contamination have also been investigated
by varying $t_{\mathrm{sep}}$ from 0.85 to 1.36 fm in the standard plateau method.
It was found that the condition of $t_{{\mathrm{sep}}}\gtrsim1.0\ \mathrm{fm}$ for $g_S$ and $g_T$ 
is large enough to eliminate the excited-state contaminations thanks to the elaborated tuning of the sink and source functions,
while the condition of $t_{{\mathrm{sep}}}\gtrsim1.2\ \mathrm{fm}$ is required for $g_A$. 
Taking account of the uncertainties stemming from the excited-state contamination, our best estimates for
the nucleon isovector couplings, $g_A$, $g_S$ and $g_T$
are obtained as follows:
\begin{align}
\label{eq:ren_all}
g_A & = 1.288(14)(9) \\
g_S & = 0.927(83)(22)  \\
g_T & =   1.036(6)(20) ,
\end{align}
where the first error is statistical, while the second one
represents a systematic error given by the total uncertainties in the
determination of the renormalization constant. 
Our result of $g_A$ is fairly consistent with the experimental result at a percent level of accuracy ($\Delta g_A/g_A\sim 1\%$).
For $g_S$ and $g_T$, our results are consistent with those of the FLAG average~\cite{Aoki:2021kgd},
though the discretization uncertainty is not yet accounted in this study.
It is worth remarking that our preliminary result of $g_A/g_V$ obtained at a second, finer lattice spacing ($a\approx0.063$ fm)
shows a very small discretization error on $g_A/g_V$ at the lattice spacing of $a\approx0.085$ fm~\cite{Tsuji:2022knz}.
The overall uncertainties on our results of $g_S$ and $g_T$ determined directly at the physical point
reach $\Delta g_S/g_S\sim 8\%$ and $\Delta g_T/g_T\sim 2\%$, which are smaller than the FLAG average 
for both $N_f = 2 + 1 + 1$ ($\Delta g_S/g_S\sim 10\%$ and $\Delta g_T/g_T\sim 3\%$) and 
$N_f=2+1$ ($\Delta g_S/g_S\sim 12\%$ and $\Delta g_T/g_T\sim 6\%$)~\cite{Aoki:2021kgd}.

We continue our research to evaluate the systematic uncertainties of $g_S$ and $g_T$ 
due to the discretization error, using the PACS10 gauge configurations at the finer lattice spacing 
of $a\approx0.063$ fm at the physical point.
We also plan to extend to investigate individual components of $g_O^u$, $g_O^d$ and $g_O^s$ for $O=A$, $S$, $T$,
which demand calculations of the disconnected contribution of the nucleon 3-point correlation functions. 
Further study is now in progress~\cite{Tsuji:2022knz}.

\begin{acknowledgments}
{We would like to thank members of the PACS collaboration for useful discussions. We also thank Y. Namekawa for his careful reading of the manuscript.
R.~T. is supported by the RIKEN Junior Research Associate Program.
R.~T. and N.~T. acknowledge the support from Graduate Program on Physics for the Universe (GP-PU) of Tohoku University.
Numerical calculations in this work were performed on Oakforest-PACS in Joint Center for Advanced High Performance Computing (JCAHPC) and Cygnus in Center for Computational Sciences at University of Tsukuba under Multidisciplinary Cooperative Research Program of Center for Computational Sciences, University of Tsukuba, and Wisteria/BDEC-01 in the Information Technology Center, The University of Tokyo. This research also used computational resources through the HPCI System Research Projects (Project ID: hp170022, hp180051, hp180072, hp180126, hp190025, hp190081, hp200062, hp200188, hp210088, hp220050) provided by Information Technology Center of the University of Tokyo and RIKEN Center for Computational Science (R-CCS). The  calculation employed OpenQCD system(http://luscher.web.cern.ch/luscher/openQCD/). 
This work is supported by the JLDG constructed over the SINET5 of NII.
This work was also supported in part by Grants-in-Aid for Scientific Research from the Ministry of Education, Culture, Sports, Science and Technology (Nos. 18K03605, 19H01892, 22K03612).

}
\end{acknowledgments}

 
\appendix
\section{Comparison between the RI/MOM and RI/SMOM$_{(\gamma_\mu)}$ schemes}
\label{app:mom_and_smom}
It was reported in Ref.~\cite{Aoki:2007xm} that the RI/SMOM$_{(\gamma_\mu)}$ scheme
significantly reduces sensitivity to the unwanted infrared divergence, which stems from 
the nonperturbative effects such as the spontaneous chiral symmetry breaking.
It is simply because the nonexceptional momenta ($p_1^2=p_2^2=q^2$) can avoid contributions from 
the nonperturbative effects due to the nonvanishing value of $q^2$.
Furthermore, the better convergence of the perturbative series for the matching
from RI/SMOM$_{(\gamma_\mu)}$ to $\overline{\mathrm{MS}}$
 was observed 
for the case of the mass renormalization $Z_m$~\cite{Sturm:2009kb} and it has been confirmed 
that this trend is maintained up to the two-loop level~\cite{{Almeida:2010ns}, {Gorbahn:2010bf}}.
Recall that $Z_m=1/Z_P$ is satisfied in the RI/SMOM$_{(\gamma_\mu)}$ scheme~\cite{Sturm:2009kb},
while the good chiral property leads to the relation of $Z_S=Z_P$ in the massless limit.
In this context, the RI/SMOM$_{(\gamma_\mu)}$ scheme is expected to provide
better estimation of the renormalization constant for the scalar channel $Z_S$
with reduced systematic uncertainties. 

In this calculation, we use the quark propagator calculated with 
a single source location $x_0=(0,0,0,0)$, while the SMOM$_{(\gamma_\mu)}$ results are obtained with 
four different source locations as described in Sec.~\ref{Sec.5:renorm}.

\begin{figure*}[b]
 \includegraphics[width=1.0\linewidth,bb=0 0 792 612,clip]{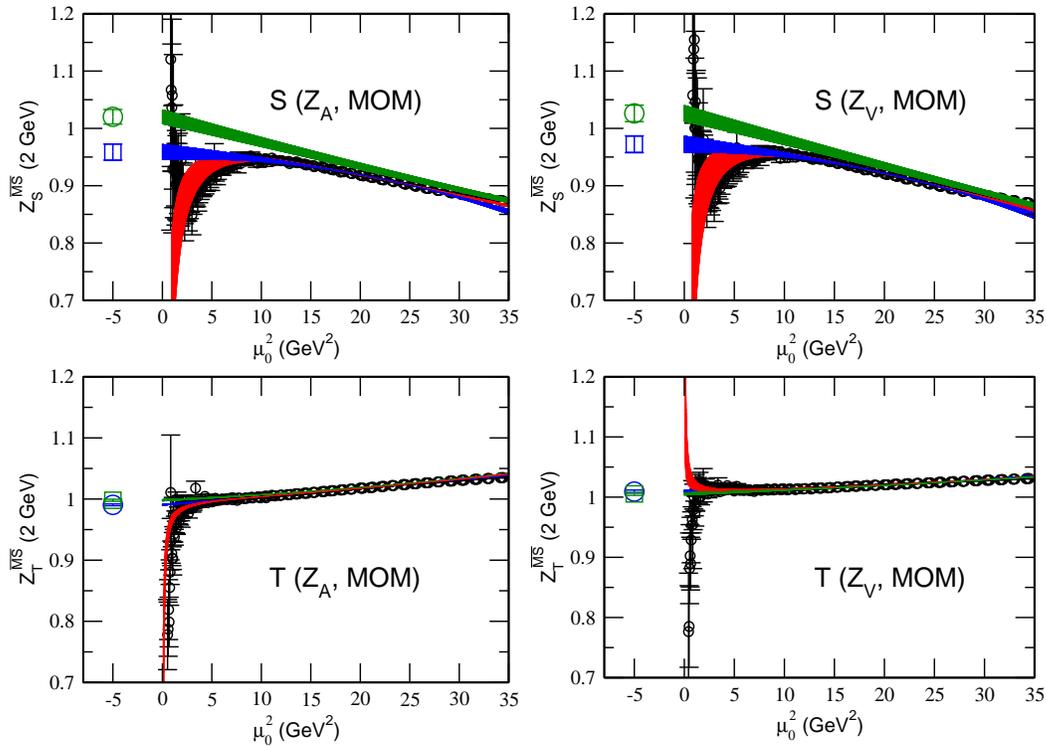}
 \caption{
 The renormalization constants of $Z_{S}^{\overline{\mathrm{MS}}}(2\;\mbox{GeV})$ 
  (upper panels) and  $Z_{T}^{\overline{\mathrm{MS}}}(2\;\mbox{GeV})$ (lower panels) obtained from the RI/MOM scheme
  as a function of the square of the matching scale $\mu_0$. 
  The SF input of $Z_A$ is used for the left panels, while the SF input of $Z_V$ is used for the right panels.  
  The rest is the same as in Figs.~\ref{fig:fitted_redults_ren_S} and \ref{fig:fitted_redults_ren_T}.
 \label{fig:mom_vs_smom}}
\end{figure*}

Figure~\ref{fig:mom_vs_smom} shows the renormalization constants 
$Z_S^{\overline{\mathrm{MS}}}(2\;\mbox{GeV})$ (upper panels) and
$Z_T^{\overline{\mathrm{MS}}}(2\;\mbox{GeV})$ (lower panels) obtained from 
the RI/MOM intermediate scheme as a function of the square of the matching scale $\mu_0$. 
The SF input of $Z_A$ is adopted in the left panels, while $Z_V$ is used for the right panels.  
In each panel of Fig.~\ref{fig:mom_vs_smom}, red and blue curves with error bands represent two fit
results from the global fit (\ref{eq:ren_fit_global}) with $\mu_{\mathrm{min}}=1$ GeV
and the IR-truncated fit (\ref{eq:ren_fit_IRtru}) with $\mu_{\mathrm{min}}=2$ GeV under the conditions of 
$k_{\mathrm{max}}=2$ and $\mu_{\mathrm{max}}=5$ GeV, 
while green curve with error bands is given by removing the divergent contribution from the global fit result.
The intersections of the green and blue curves with the $y$-axis correspond to the
$\mu_0$-independent contributions in the global fit (\ref{eq:ren_fit_global})
and the IR-truncated fit (\ref{eq:ren_fit_IRtru}). 
Their central values and statistical errors denoted as the circle and squared symbols
are slightly shifted from the $y$-axis in each panel of Fig.\ref{fig:mom_vs_smom} 
similar to those shown in Figs.~\ref{fig:fitted_redults_ren_S} and \ref{fig:fitted_redults_ren_T}.

As clearly displayed in the upper panels of Fig.~\ref{fig:mom_vs_smom}, 
the $\mu_0$-independent value of $Z_S^{\overline{\mathrm{MS}}}(2\;\mbox{GeV})$ is severely infrared sensitive compared 
to the case of the RI/SMOM$_{(\gamma_\mu)}$ scheme. On the other hand, as for the tensor channel displayed in the 
lower panels of Fig.~\ref{fig:mom_vs_smom}, the $\mu_0$-independent value of $Z_S^{\overline{\mathrm{MS}}}(2\;\mbox{GeV})$
obtained from the RI/MOM scheme equally shows less sensitivity to the unwanted 
infrared divergence as well as the RI/SMOM$_{(\gamma_\mu)}$ scheme, since the tensor channel 
is not strongly affected by the spontaneous chiral symmetry breaking.  

Finally, similar to the case of SMOM$_{(\gamma_\mu)}$, we choose the global fit results for the central value of 
$Z_S^{\overline{\mathrm{MS}}}(2\;\mbox{GeV})$ and $Z_T^{\overline{\mathrm{MS}}}(2\;\mbox{GeV})$ 
obtained from the RI/MOM intermediate scheme as summarized in Tab.~\ref{tab:val_mom}.
The first, second and third errors are the statistical one and the systematic ones associated 
with ``scheme" and ``model".  The third error caused by choice of fit model is 
the largest of the three errors and its size is about a few times its own statistical accuracy.
Especially, this largest error on $Z_S^{\overline{\mathrm{MS}}}(2\;\mbox{GeV})$ from
the MOM case is much larger than that of the SMOM$_{(\gamma_\mu)}$ case due to
more significant sensitivity to the unwanted infrared divergence. Therefore, we confirm that
the RI/SMOM$_{(\gamma_\mu)}$ scheme can provide better estimation of the renormalization
constant with reduced sensitivity to the unwanted infrared divergence compared to the RI/MOM scheme.

%
%
%

%
%
 \begin{table}[ht]
     \caption{
     Results of the renormalization constants $Z_S^{\overline{\mathrm{MS}}}(2\;\mbox{GeV})$ and
     $Z_T^{\overline{\mathrm{MS}}}(2\;\mbox{GeV})$ from the RI/MOM scheme
     using the global fit of the Type C.
     }
     \label{tab:val_mom}
     \renewcommand{\arraystretch}{1.5}
     \begin{ruledtabular}
     \begin{tabular}{cccccccc}
        Scheme & Fit range [GeV] & $k_{\rm max}$ 
       & $Z_S^{\overline{\mathrm{MS}}}(2\;\mbox{GeV})$ 
       & $\chi^2/{\mathrm{dof}}$ 
       & $Z_T^{\overline{\mathrm{MS}}}(2\;\mbox{GeV})$ 
       & $\chi^2/{\mathrm{dof}}$ \\
       \hline
MOM  &   [1:5] & 2 & 0.9723(148)(136)(539)&1.13(52) & 1.0092(25)(38)(62) & 0.45(13)\\
\end{tabular}
\end{ruledtabular}
\end{table}
%

\section{Truncation errors of the perturbative series in matching and evolution}
\label{app:uncertainties_pert}
As described in Sec.~\ref{Renormalization}, we first determine the renormalization constant $Z_O^{\mathrm{RI}}$ 
for the $O$ channel in one of three RI schemes in a fully nonperturbative way. 
The obtained value of $Z_O^{\mathrm{RI}}$ is converted and evolved to the $\overline{\mathrm{MS}}$ scheme
at the renormalization scale of 2 GeV with the help of perturbation theory.
Therefore, it is inevitable that there are truncation errors of the perturbative series in matching
and evolution. To examine the size of such systematic uncertainties, we compare the values of 
$Z_O^{\mathrm{RI}}$ using three different types of perturbative conversion as listed in Tab.~\ref{tab:pert_level}. 

For the case of the RI/SMOM$_{\gamma_\mu}$ intermediate scheme, we evaluate 
$Z_S^{\overline{\mathrm{MS}}}(2\;\mbox{GeV})$ and $Z_T^{\overline{\mathrm{MS}}}(2\;\mbox{GeV})$
with the leading order (LO), the next-to-leading order (NLO) and NNLO conversion factors as tabulated in Tab.~\ref{tab:pert_assumption_results}.
The first, second, third and fourth errors represent the statistical one and the systematic ones associated with ``scheme", ``model" and ``fit", respectively. As mentioned in Sec.~\ref{error_pt}, we assume 
that the systematic error associated with ``scheme" can be regarded as possible truncation errors 
in perturbation matching and evolution. 

We indeed observe that the difference between the NLO and NNLO results is sufficiently small compared to 
the difference between the LO and NNLO results in Tab.~\ref{tab:pert_level}. 
For the case of the scalar, the size of the truncation error from NLO to NNLO in perturbation theory is
comparable to the systematic error associated with ``scheme", while such perturbation truncation 
error is negligible for the case of the tensor in comparison to any other systematic uncertainties.

%
%
 \begin{table}[ht]
     \caption{
     List of three types of perturbative conversion to $\overline{\mathrm{MS}}$ at the scale of 2 GeV.    
     }
     \label{tab:pert_level}
     \begin{ruledtabular}
     \begin{tabular}{lccc}
        & LO & NLO & NNLO \\
       \hline
       Scheme matching from RI/SMOM$_{(\gamma_\mu)}$ to $\overline{\mathrm{MS}}$             & tree & 1-loop& 2-loop  \\
       Scale evolution in $\overline{\mathrm{MS}}$ (anomalous dim.)   &  1-loop& 2-loop& 3-loop	\\
       Scale evolution in $\overline{\mathrm{MS}}$ (running coupling $\alpha_{s}$) & 4-loop& 4-loop& 4-loop 	\\
     \end{tabular}
     \end{ruledtabular}
   \end{table}

%
%
 \begin{table}[ht]
     \caption{     
     Results of the renormalization constants 
     $Z_S^{\overline{\mathrm{MS}}}(2\;\mbox{GeV})$ and
     $Z_T^{\overline{\mathrm{MS}}}(2\;\mbox{GeV})$ from the RI/SMOM$_{\gamma_\mu}$ 
     with the LO, NLO and NNLO conversion factors using the global fit of the Type C. 
     The first, second, third and fourth errors
     represent the statistical one and the systematic ones associated with ``scheme", ``model" and ``fit", respectively. 
     }.
     \renewcommand{\arraystretch}{1.5}
     \label{tab:pert_assumption_results}
     \begin{ruledtabular}
     \begin{tabular}{lcccc}
        & FIT range [GeV] & $k_{\mathrm{max}}$ & $Z_S^{\overline{\mathrm{MS}}}(2\;\mbox{GeV})$ &  
        $Z_T^{\overline{\mathrm{MS}}}(2\;\mbox{GeV})$ \\
        \hline
         LO  & [1:5] &  2 & 0.8982(27)(431)(358)(154)&  0.9726(11)(497)(193)(33) \\
          NLO & [1:5] &  2 & 0.9045(30)(130)(1)(130)& 1.0109(12)(265)(53)(31)  \\
	   NNLO &   [1:5] & 2 & 0.9103(31)(53)(73)(137)  & 1.0111(12)(149)(20)(32)   \\
     \end{tabular}
     \end{ruledtabular}
   \end{table}
%

\section{The QCD running of $\alpha_s$ and Perturbative matching for $Z_O$}
\label{app:pert_matching}
The QCD running coupling $\alpha_s(\mu)$ is determined by the renormalization group equation. To calculate the coupling constant $\alpha_s=g^2/4\pi$ at any scale, we used the four-loop ($\mathrm{NNNLO}$) running formula:
\begin{align}
    \frac{d a_s}{d \mathrm{ln}\mu^2}
    & =
    - \sum^{\infty}_{n=0}\beta_n(a_s)^{n+2}\nonumber \\
    & =
    - \beta_0 a_s^2 - \beta_1 a_s^3 - \beta_2 a_s^4 - \beta_3 a_s^5 + \mathcal{O}(a_s^6) ,
\end{align}
where $a_s=\alpha_s/4\pi$ using the four-loop QCD beta function which was calculated in Ref.~\cite{Ritbergen1997}.

For a numerical implementation, the methodology  described in Ref.~\cite{Aoki:2007xm} is employed in this work. The initial condition of the running coupling is given at the scale of $\mu=M_Z$ with the world-average $Z$-boson mass of 91.1876(21) GeV~\cite{ParticleDataGroup:2020ssz} as below, 
\begin{align}
    \alpha^{(5)}_s(M_Z)=0.1176\pm0.0011 ,
\end{align}
 where the number in superscript brackets indicates the number of dynamical quarks.
 While the scale in $\alpha_s$ runs across the $m_b$ and $m_c$ threshold, where each flavor quark is decoupled at its mass scale, the following matching conditions are imposed as
 \begin{align}
     \alpha^{(5)}_s(m_b) = \alpha^{(4)}_s(m_b)\quad\mathrm{and}\quad
     \alpha^{(4)}_s(m_c) = \alpha^{(3)}_s(m_c) .
 \end{align}
Once $\alpha^{(3)}_s(m_c)$ is computed, we can calculate the value of the running coupling $\alpha_s(\mu)$ at any scale in the 3-flavor theory. For example, in this study, 
we use 
\begin{align}
    \alpha^{(3)}_s (2\ \mathrm{GeV}) = 0.2904.
\end{align}

After performing the conversion from the intermediate renormalization scheme to the $\overline{\mathrm{MS}}$ scheme, we can evaluate the renormalization constant for the quark bilinear operator $Z_{O}(\mu)$ at any scale. The renormalization group equation is given as 
\begin{align}
    \frac{\partial \mathrm{ln} Z_{O}}{\partial \mathrm{ln} \mu^2} & = \gamma_{O}=-\sum_{n=0}\gamma^{(n)}_O\left( \frac{\alpha_s}{4\pi} \right)^{n+1},
\end{align}
where $\gamma^{(n)}$ represents the $n$-loop anomalous dimension. For the case of $O=S$ and $T$, these are evaluated up to two-loop order~\cite{Almeida:2010ns}. 
In order to compare with the experimental values or other lattice results, 
we choose the renormalization scale $\mu=2$ GeV, and then use the evolution factor 
from the matching scale $\mu_0$ to 2 GeV,
$R_O(2\ \mathrm{GeV},\mu_0)=Z_O^{\mathrm{\overline{MS}}}(2\ \mathrm{GeV})/Z_O^{\mathrm{\overline{MS}}}(\mu_0)$, which is given as the NNLO accuracy:
\begin{align}
    \frac{Z^{\overline{\mathrm{MS}}}_O(2\ \mathrm{GeV})}{Z^{\overline{\mathrm{MS}}}_O(\mu_0)}
    &=
    \left( \frac{\alpha_s(2\ \mathrm{GeV})}{\alpha_s(\mu_0)} \right)^{\frac{\gamma_O^{(0)}}{\beta_0}}
    \mathrm{exp}
    \left(
    \frac{\alpha_s(2\ \mathrm{GeV})}{4\pi} - \frac{\alpha_s(\mu_0)}{4\pi}
    \right)^{\left\{ \frac{\gamma_O^{(1)}}{\beta_0} - \frac{\gamma_O^{(0)}\beta_1}{\beta_0^2} \right\}}\nonumber\\
    & \times
    \mathrm{exp}
    \left\{
    \left(
    \frac{\alpha_s(2\ \mathrm{GeV})}{4\pi}
    \right)^2-
    \left(
        \frac{\alpha_s(\mu_0)}{4\pi}
    \right)^2
    \right\}^{\frac{1}{2}
    \left\{ \frac{\gamma_O^{(2)}}{\beta_0} - \frac{\gamma_O^{(1)}\beta_1}{\beta_0^2}
    - \frac{\gamma_O^{(0)}\beta_2}{\beta_0^2} + \frac{\gamma_O^{(0)}\beta_1^2}{\beta_0^3} \right\}
    } \times \mathrm{exp}\{\mathcal{O}(\alpha_s^3)\}.
\end{align}

%

\section{Assessment of excited-state contamination}
\label{app:summation_method}

\subsection{Summation method}

In this study, we mainly adopted the standard plateau method to determine the bare coupling directly from the ratio (\ref{Eq:ratio_plateau}) as described in Sec.~\ref{Sec2:ratio_method}. In order to eliminate the systematic uncertainties
stemming from the excited-state contaminations, one should calculate the ratio  (\ref{Eq:ratio_plateau}) with several choices of $t_{\mathrm{sep}}$, and then make
sure whether the evaluated value of the bare coupling does not change with a variation
of $t_{\mathrm{sep}}$ within a certain precision. Alternatively, there is 
another method called the summation method, which was proposed to extract the bare coupling from the ratio (\ref{Eq:ratio_plateau})
with a better control of the excited-state contamination~\cite{Maiani:1987by}. 
In this method, the ratio of $R(t_{\mathrm{op}}, t_{\mathrm{sep}})$ 
defined in Eq.~(\ref{Eq:ratio_plateau}) is summed over the range of  $t_{\mathrm{sink}} > t_{\mathrm{op}} > t_{\mathrm{src}}$ as
\begin{align}
    S(t_{\mathrm{sep}})
    & =
    \sum^{t_{\mathrm{sink}}-1}_{t_{\mathrm{op}}=t_{\mathrm{src}}+1} R(t_{\mathrm{op}}, t_{\mathrm{sep}}) \nonumber \\
    & =
    C +
    g_{O} \cdot t_{\mathrm{sep}}
    + \mathcal{O}\left( e^{-\Delta E t_{\mathrm{sep}}} \right),
    \label{eq:summation_method}
\end{align}
where $C$ is a constant being independent of $t_{\mathrm{sep}}$. The bare coupling $g_{O}$ can be read off from the slope of $S(t_{\mathrm{sep}})$ with respect to $t_{\mathrm{sep}}$. When more than 3 sets of $t_{\mathrm{sep}}$ are carried out in 
the plateau method, one can also perform the summation method as well.

The summation method takes an advantage for reducing the excited-state contamination, compared to the plateau method that receives ${\cal O}(e^{-\Delta Et_{\mathrm{sep}}/2})$ at most (even when $t_{\mathrm{op}}$ is chosen to be right in the middle of the nucleon source and sink operators).
However, in general, the summation method requires {\it various choices} of the source-sink separation $t_{\mathrm{sep}}$ to determine the value of $g_{O}$
by a linear fit as a function of $t_{\mathrm{sep}}$ with {\it smaller uncertainties}.
We calculate the nucleon 3-point functions using the sequential source method with a fixed source location~\cite{{Martinelli:1988rr},{Sasaki:2003jh}}. 
Therefore, the number of available data of $t_{\mathrm{sep}}$ is quite limited in this study.

The summation method is then applied for both $128^4$ and $64^4$ lattices 
with four data
sets of $t_{\mathrm{sep}}=\{10,12,14,16\}$ with the smearing-type of Exp.(I)
for the former and three data sets of $t_{\mathrm{sep}}=\{12,14,16\}$ 
with the smearing-type of Exp.(II) for the latter. We then use an uncorrelated linear fit to determine the coupling constant from the slope of the linear dependence of $t_{\mathrm{sep}}$ in Eq.~(\ref{eq:summation_method}).
Because of a small number of choices with respect to $t_{\mathrm{sep}}$, 
the resulting values of the bare coupling as tabulated in Tab.~\ref{tab:bare_couplings_summation}, receive much larger statistical uncertainties than the corresponding values determined by the plateau method. 
Indeed, no statistically significant results for the scalar channel were obtained from the summation method with both $128^4$ and $64^4$ lattices. 

The consistency between the $128^4$ and $64^4$ results is clearly seen in the case of the vector channel. These
values are fully consistent with the results obtained by the plateau method as listed in Table~\ref{tab:bare_couplings}.
On the other hand, for the axial-vector and tensor channels, there are some discrepancies between the $128^4$ and $64^4$ results 
beyond the statistical errors. It is worth remembering that the plateau method 
does not show any difference between the $128^4$ and $64^4$ results of $g_A$ and $g_T$ with high precision. This observation suggest that the discrepancies found in the case of $g_A$ and $g_T$ are mainly caused by rather large uncertainties in the summation method due to fewer choices of $t_{\mathrm{sep}}$. Indeed, in the case of the axial-vector channel, the trend of two data in the $L=64$ calculations at $t_{\mathrm{sep}}=12$ and 14, whose statistical errors are much smaller than that of $t_{\mathrm{sep}}=16$, has a strong influence on the results 
of the linear fit in the summation method with fewer choices of $t_{\mathrm{sep}}$.
Nevertheless, the maximum difference between the plateau and summation methods is less than twice of the standard
deviation. In this sense, we confirm that the bare couplings of $g_V$, $g_A$ and $g_T$
obtained from both methods are barely consistent with each other.

\subsection{Two-state fit analysis}

Although the two analyses with the plateau method and summation method do not lead to major discrepancies in all channels, 
one may still have concerns about the tensor coupling, which exhibits curvature beyond the statistical errors in Fig.~\ref{fig:128_bare_all} and
Fig.~\ref{fig:64_bare_all} unlike the other couplings.
Therefore, we additionally perform the two-state fit analysis to
extract the bare value of $g_T$ from $R(t_{\rm op},t_{\rm sep})$ with the following functional form:
\begin{equation}
    R(t^\prime_{\rm op}, t_{\rm sep}) = g_T + A \cosh \left[
    \Delta E\left(t^\prime_{\rm op}-\frac{t_{\rm sep}}{2}\right)
    \right]
\end{equation}
where $t^\prime_{\rm{op}}=t_{\rm{op}}-t_{\rm{src}}$. In this formula, the excited-state contribution remains
at most as the size of $A$ at the center of the source and sink operators $t^\prime_{\rm{op}}=t_{\rm{sep}}/2$. 
Therefore, the value of $A$ corresponds to a typical size of the systematic uncertainty associated with
the excited-state contamination in the plateau method. 

The two-state fit results for the bare couplings $g_{T}$ are summarized in Table.~\ref{tab:bare_couplings_two_state}.
Although the results of $g_T^{\rm{bare}}$ from the two-state fit analysis are slightly smaller than those of the plateau method, 
the size of the corresponding correction due to non-zero value of $A$ is a few percent level in each case,
but statistically not so significant. If we choose data sets of $t_{\rm{sep}}=\{12,14,16\}$ from both $L=128$ and $L=64$ calculations,
we evaluate the average value of the renormalized coupling $g_T$ 
from the combined analysis using the super-jackknife method as
\begin{equation}
g_T=0.999(45)
\end{equation}
which is consistent with that of the plateau method within the statistical error. 
This observation indicates that systematic uncertainties stemming from the excited-state contamination are well under control within the limits of the statistical precision thanks to the optimal choice of the smearing parameters used in this study.

%
%
\begin{table}[ht]
     \caption{
     Summation method results
     of the bare couplings $g_{O}$ for $O=V, A, S, T$ obtained from both $128^4$ and $64^4$ lattices
     with available source-sink separations.
     The smearing-type is chosen to be Exp.(I) for the $L=128$ calculation and Exp.(II) for the $L=64$ calculation. 
     }
     \label{tab:bare_couplings_summation}
     \renewcommand{\arraystretch}{1.3}
     \begin{ruledtabular}
     \begin{tabular}{ccccccc}
        $L$ &  Smearing-type & $t_{\mathrm{sep}}$ &  $g_V^{\mathrm{bare}}$& $g_A^{\mathrm{bare}}$& $g_S^{\mathrm{bare}}$& $g_T^{\mathrm{bare}}$\\
       \hline
       128 & Exp.(I) & \{10,12,14,16\} & 1.052(5)& 1.298(144)& 1.138(887)& 0.872(88)\\
       \hline
       64  & Exp.(II) & \{12,14,16\}& 1.034(34)& 1.517(89)& 1.030(882)& 1.073(57)\\
      \hline
      
      \end{tabular}
     \end{ruledtabular}
   \end{table}
%

%
%
\begin{table}[ht]
     \caption{
     Two-state fit results
     of the bare couplings $g_{T}$ obtained from both $128^4$ and  $64^4$ lattices with smearing-types of 
     Exp.(I) and Exp.(II).
     } 
     \label{tab:bare_couplings_two_state}
     \begin{ruledtabular}
     \begin{tabular}{cccccccc}
\hline 
$L$ & $t_{\rm sep}$ & Smearing-type & $g_T^{\rm bare}$ & $A$ & $\Delta E$ & fit-range & $\chi^2/{\rm{dof}}$\cr
\hline
128& 10 & Exp.(I)  & 1.040(36)  & 0.025(17) & 0.55(12) & [1, 9] & 0.94 \cr
   & 12 &          & 1.015(49)  & 0.016(49) & 0.59(63) & [3, 9] & 1.91 \cr
   & 14 &          & 0.953(54)  & 0.004(19) & 0.89(113)& [4,10] & 0.45 \cr
   & 16 &          & 1.020(63)  & 0.0003(36) & 0.84(179) & [3,13] & 0.91\cr
\hline
64 & 11 & Exp.(I)  & 1.037(25)  & 0.0095(93) & 0.74(21)& [2, 9] & 0.78 \cr
   & 14 &          & 0.934(84)  & 0.051(71) & 0.35(18) & [2,11] & 1.19 \cr  
\cline{2-8} 
   & 12 & Exp.(II) & 0.975(31)  & 0.035(28) & 0.47(15) & [3, 9] & 0.51 \cr
   & 14 &          & 0.999(23)  & 0.020(15) & 0.46(12) & [3,11] & 1.29 \cr
   & 16 &          & 0.818(178) & 0.012(16) & 0.24(13) & [2,14] & 0.65 \cr
\hline
 \end{tabular}
 \end{ruledtabular}
\end{table}

\clearpage

\end{document}